\documentclass[pra,aps,preprint,amsmath,amssymb,showkeys]{revtex4}
\usepackage{lineno,hyperref}
\usepackage{graphicx}
\usepackage{epsfig}
\usepackage{epstopdf}

\include{graphics}

\begin{document}


\title{Fast two-beam collisions in a linear optical medium 
with weak cubic loss in spatial dimension higher than 1}








\author{Avner Peleg$^{1}$, Toan T. Huynh$^{2,3}$, and Quan M. Nguyen$^{4}$}

\affiliation{$^{1}$Department of Mathematics, Azrieli College of Engineering,   Jerusalem 9371207, Israel}

\affiliation{$^{2}$Department of Mathematics, University of Science, 
Vietnam National University-HCMC, Ho Chi Minh City, Vietnam}

\affiliation{$^{3}$Department of Mathematics, University of Medicine and Pharmacy at Ho Chi Minh City, Ho Chi Minh City, Vietnam}

\affiliation{$^{4}$Department of Mathematics, International University, 
Vietnam National University-HCMC, Ho Chi Minh City, Vietnam}

\date{\today}

\begin{abstract} 
We study the dynamics of fast two-beam collisions in linear 
optical media with weak cubic loss in spatial dimension higher than 1.  
For this purpose, we extend the perturbation theory that was developed 
for analyzing two-pulse collisions in spatial dimension 1 
to spatial dimension 2. We use the extended two-dimensional version 
of the perturbation theory to show that the collision leads to 
a change in the beam shapes in the direction transverse to the 
relative velocity vector. Furthermore, we show that in the important case 
of a separable initial condition for both beams, the longitudinal part 
in the expression for the amplitude shift is universal, while the 
transverse part is not universal.
Additionally, we demonstrate that the same behavior holds for 
collisions between pulsed optical beams in spatial dimension 3. 
We check these predictions of the perturbation theory along 
with other predictions concerning the effects on the collision 
of partial beam overlap and anisotropy in the initial condition 
by extensive numerical simulations with the weakly perturbed 
linear propagation model in spatial dimensions 2 and 3. The agreement 
between the perturbation theory and the simulations is very good.   
Therefore, our study significantly extends and generalizes the results 
of previous works, which were limited to spatial dimension 1.  
\end{abstract}


\maketitle


\section{Introduction}
\label{Introduction}

Linear evolution models have an important role in many 
areas of science. Examples include the linear diffusion 
equation \cite{Van_Kampen}, the linear wave equation \cite{Whitham99}, 
the linear propagation equation \cite{Ishimaru2017,Siegman86,Kogelnik66}, 
and the linear Schr\"odinger equation \cite{Merzbacher98}. In many cases, 
the physical systems that are described by these linear 
evolution models include weak nonlinear dissipation \cite{Van_Kampen,Agrawal2007a}. 
As a result, the latter physical systems are more accurately 
described by perturbed linear evolution models with weak nonlinear 
dissipation. The presence of the nonlinear dissipation induces 
new physical effects, which do not exist in the unperturbed 
linear physical systems. A major example is the change in the 
pulse mass or energy during fast collisions between pulses 
of the linear propagation model \cite{PNH2017B,QMN2018,NHP2020}. 
Since the pulses of the linear physical systems (and also of 
their weakly perturbed counterparts) are not shape preserving, 
one does not expect to observe simple dynamics in these collisions. 
As a result, one also does not expect to be able to make simple 
general statements about the collision-induced effects.

In two previous works \cite{PNH2017B,NHP2020}, we showed that 
the opposite is in fact true for fast two-pulse collisions. 
The latter are collisions, in which the collision length, i.e., 
the length of the interval where the two pulses overlap, 
is much smaller than all the other length scales in the problem 
\cite{fast_collisions}. In Refs. \cite{PNH2017B,NHP2020}, 
we showed that the amplitude shifts in fast two-pulse collisions 
in linear physical systems, weakly perturbed by nonlinear dissipation,  
exhibit simple soliton-like behavior. The behavior was demonstrated 
for the following two central cases: (1) systems described by the 
linear propagation equation with weak cubic loss; (2) systems 
described by the linear diffusion-advection equation with weak 
quadratic loss. We first developed a perturbation method for 
analyzing the fast two-pulse collision dynamics in these weakly 
perturbed linear systems. We then used the perturbation method 
to show that in both cases, the expressions for the collision-induced 
amplitude shifts in the presence of weak nonlinear loss have the 
same simple form as the expression for the amplitude shift in a fast 
collision between two solitons of the nonlinear Schr\"odinger equation 
in the presence of weak cubic loss. Furthermore, in Ref. \cite{NHP2020}, 
we showed that the expressions for the amplitude shifts are universal 
in the sense that they are independent of the details of the  
initial pulse shapes. In addition, we found that within the leading order 
of the perturbation theory, the pulse shapes are not changed by 
the collision. The perturbation theory predictions for the 
collision-induced amplitude shifts were verified by extensive 
numerical simulations with the two perturbed 
linear evolution models for a variety of initial pulse 
shapes \cite{PNH2017B,NHP2020}. Additionally, in Ref. \cite{QMN2018}, 
we showed that the amplitude shift in a fast two-pulse collision in 
systems described by the linear propagation model with high-order 
nonlinear loss can be calculated by the same perturbation method 
that was developed in Ref. \cite{PNH2017B}.

The three studies in Refs. \cite{PNH2017B,QMN2018,NHP2020} were 
limited to spatial dimension 1 \cite{spatial_dimension}. As a result, 
these studies did not consider important collisional effects, 
which exist only is spatial dimension higher than 1, such as anisotropy 
and partial pulse overlap. Additionally, it is not clear if the 
simple (universal) dependence of the expressions for the collision-induced 
amplitude shifts on the physical parameters that was found in Refs. 
\cite{PNH2017B} and \cite{NHP2020} remains valid in the high-dimensional 
problem. It is also unclear if the pulse shapes remain unchanged 
in fast collisions in the presence of cubic (or quadratic) loss in the 
high-dimensional problem. Thus, all the key aspects of the fast two-pulse 
collision problem, which are associated with the collision dynamics in 
spatial dimension higher than 1, were not addressed in previous studies.

In the current paper, we address the important aspects of the 
high-dimensional fast two-pulse collision problem that were 
mentioned in the preceding paragraph. For this purpose, we 
first develop a perturbation method, which generalizes the 
perturbation method that was introduced in Refs. 
\cite{PNH2017B,NHP2020} for the one-dimensional problem in 
three major ways. (1) It extends the perturbative calculation 
from spatial dimension 1 to spatial dimension 2, and enables 
further extension of the calculation to a general spatial dimension 
in a straightforward manner. (2) It provides a perturbative calculation of 
the collision-induced dynamics of the beam shape both inside 
and outside of the collision interval. In contrast, the perturbative 
calculation of Refs. \cite{PNH2017B,NHP2020} was limited to 
the collision interval only. (3) It enables the discovery 
and analysis of several collision-induced effects, 
which exist only in the high-dimensional problem.

We use the generalized version of the perturbation method to 
derive formulas for the collision-induced changes in the beam 
shapes and amplitudes in spatial dimension 2. We find that for 
a general initial condition, the collision leads to a change in 
the beam shape in the direction transverse to the relative 
velocity vector between the beam centers. Additionally, 
we find that in the important case of an initial condition 
that is separable for both beams, the beam shape in the longitudinal 
direction is not changed by the collision within the leading order 
of the perturbation theory. Furthermore, we show that for a separable 
initial condition, the longitudinal part in the expression for 
the amplitude shift is universal, while the transverse part 
is not universal and is proportional to the integral of the product of 
the beam intensities with respect to the transverse coordinate. 
We also show that the same behavior of the longitudinal and transverse parts in the 
expression for the collision-induced amplitude shift exists  
in collisions between pulsed-beams in spatial dimension 3.

We check these predictions of the perturbation theory 
together with other predictions concerning the effects 
of partial beam overlap and anisotropy in the initial condition 
by extensive numerical simulations with the perturbed linear 
propagation model in spatial dimensions 2 and 3. The simulations 
in spatial dimension 2 are carried out for four different two-beam 
collision setups. These setups demonstrate the following four major effects 
and properties of the collision that either exist only in 
spatial dimension higher than 1, or are qualitatively 
different from their one-dimensional counterparts. 
(1) The universality of the longitudinal part in the expression 
for the collision-induced amplitude shift. 
(2) The effect of partial beam overlap. 
(3) The effect of anisotropy in the initial condition. 
(4) The collision-induced change in the beam shape in the transverse direction. 
The prediction for universal behavior of the longitudinal part in the expression 
for the amplitude shift is also checked in spatial dimension 3 by 
numerical simulations of collisions between pulsed optical beams. 
In all the simulation setups we obtain very good agreement between 
the perturbation theory and the numerical simulations. 
Therefore, the simulations validate the theoretical predictions 
for the four high-dimensional effects and properties 
of the collision and show that the extended perturbation 
approach can indeed be used for analyzing the effects of fast 
two-beam collisions in spatial dimension higher than 1.

In a related work, we studied the dynamics of fast 
two-pulse collisions in systems described by linear 
diffusion-advection models with weak quadratic loss 
in spatial dimension higher than 1 \cite{PHN2020}.  
We first developed a perturbation method for analyzing 
the collision dynamics, which is similar to the one 
introduced in the current paper. Using this perturbation 
method and numerical simulations, we showed that the 
collision-induced changes in pulse shapes and amplitudes in 
these systems exhibit similar behavior to the one found 
in the current paper \cite{PHN2020}. Thus, the current 
paper and the related study of Ref. \cite{PHN2020}      
significantly extend and generalize the results of the 
previous works in Refs. \cite{PNH2017B,QMN2018,NHP2020}, 
which were limited to spatial dimension 1.  
We also comment that detailed analytic results on collisions 
between pulse solutions of linear or nonlinear evolution models 
in the presence of nonlinear dissipation in spatial dimension 
higher than 1 are quite scarce. Therefore, the current work 
and the work in Ref. \cite{PHN2020} also significantly extend 
the understanding of the more general high-dimensional problem 
of two-pulse collisions in the presence of nonlinear dissipation.

We choose to study two-beam collisions in the presence of cubic loss, 
since cubic loss is important in many optical systems, 
and is therefore a major example for nonlinear dissipative perturbations. 
The optical medium's cubic loss typically arises due to two-photon 
absorption (2PA) \cite{Agrawal2007a,Dekker2007,Borghi2017,Boyd2008}. 
Propagation of optical pulses and optical beams in the presence 
of cubic loss has been studied in many earlier works, both in weakly 
perturbed linear media \cite{PNH2017B,NHP2020,Perry97,Liang2005,Cohen2005b,Cohen2004}, 
and in nonlinear media \cite{Malomed89,Stegeman89,Silberberg90,Aceves92,
Kivshar95,Tsoy2001,Silberberg2008,PCDN2009,PNC2010,Gaeta2012,PC2018,PC2020}.
The subject attracted renewed attention in recent years due to the importance of 2PA 
in silicon nanowaveguides, which are expected to play a major role in many applications 
in optoelectronic devices \cite{Agrawal2007a,Dekker2007,Borghi2017,Gaeta2008,Soref2006}.          
In the current paper, we assume that the optical medium is 
weakly nonlinear and neglect the effects of cubic (Kerr) nonlinearity. 
We emphasize that this assumption was successfully used in previous 
experimental and theoretical works, see, e.g., Refs. 
\cite{Perry97,Liang2005,Cohen2005b,Cohen2004}. For the same reason, 
we also neglect the effects of high-order nonlinear loss
on the collision. We remark that the latter effects can be described 
by the same perturbation method that is introduced in the current paper 
(see also Ref. \cite{QMN2018}, where the calculation 
was carried out for spatial dimension 1).

The rest of the paper is organized as follows. In Section \ref{lp_2D}, 
we present the extended perturbation method for calculating the 
amplitude and beam shape dynamics in fast collisions between beams 
of the linear propagation model in spatial dimension 2. 
In Section \ref{simu}, we present the perturbation theory 
predictions and the results of numerical simulations with 
the weakly perturbed linear propagation model for four major 
collision setups. These setups demonstrate four major effects 
and properties of the collision that exist only in spatial 
dimension higher than 1. In Section \ref{lp_3D}, we present 
the main predictions of the perturbation theory for collisions 
between pulsed-beams of the linear propagation equation in spatial 
dimension 3. We also present a comparison between the perturbation 
theory prediction for the collision-induced amplitude shift and 
the results of numerical simulations with the weakly perturbed linear 
propagation model. In Section \ref{conclusions}, we summarize our 
conclusions. The five Appendixes contain calculations that support 
the material in the main body of the paper.

\section{The perturbation theory for fast two-beam collisions 
in spatial dimension 2}
\label{lp_2D}

\subsection{Introduction}
\label{lp_model} 
We consider fast collisions between two optical beams in a 
three-dimensional linear optical medium with weak cubic loss. 
We assume that the beams propagate along the $z$ 
axis with beam-steering in the $xy$ plane, and that 
the propagation is accurately described by the paraxial approximation 
\cite{Ishimaru2017,Siegman86,Kogelnik66}. For each given value of $z$, 
the distribution of the electric field is a function of $x$ and $y$. 
Therefore, we can think about the $z$ coordinate as a dynamical 
coordinate, and about the $x$ and $y$ coordinates as the actual 
spatial coordinates, which help describe the distribution 
of the electric field for each value of $z$. We refer to the 
dimension of the space, in which the distribution of the 
electric field is described (for a given $z$) as the spatial dimension. 
Thus, in the current problem, the spatial dimension is 2 and the 
total dimension is 3. The propagation is described by a 
$(2+1)$-dimensional propagation model, where the 2 in this terminology 
corresponds to the spatial dimension, and the 1 is the dimension of 
the dynamical axis (the $z$ axis).

We take into account the effects of isotropic diffraction 
and weak cubic loss, as well as the velocity difference between 
the beam centers, which is controlled by beam-steering 
\cite{McManamon2009,Carlson88,Brandl2013,Gomez2015,Oh2016,
Kivshar2012,Kartashov2009,Lederer2002}. 
For simplicity and without loss of generality, we assume that the relative velocity 
vector between the beam centers lies along the $x$ axis. This choice enables 
one to obtain closed formulas for the collision-induced changes in beam shapes 
and amplitudes, and in this manner, enables a significantly deeper insight into 
the collision dynamics. Furthermore, in Appendix \ref{appendE}, we show that the choice 
of the relative velocity vector along the $x$ axis does not change the value of 
the collision-induced amplitude shift obtained by our perturbation approach. 
That is, the latter value is invariant under rotations of the $x$ and $y$ axes. 
Thus, the dynamics of the two-beam collision is described by the following 
weakly perturbed linear propagation model:        
\begin{eqnarray}&&
\!\!\!\!\!\!\!
i\partial_z\psi_{1} + \partial_{x}^{2}\psi_{1}
+\partial_{y}^{2}\psi_{1}=
-i\epsilon_{3}|\psi_{1}|^2\psi_{1}
-2i\epsilon_{3}|\psi_{2}|^2\psi_{1},
\nonumber \\&&
\!\!\!\!\!\!\!
i\partial_z\psi_{2}+id_{11}\partial_{x}\psi _{2}
+\partial_{x}^2\psi_{2} + \partial_{y}^{2}\psi_{2} =
-i\epsilon_{3}|\psi_{2}|^2\psi_{2}
-2i\epsilon_{3}|\psi_{1}|^2\psi_{2}.   
\!\!\!\!\!\!\!\!
\label{lp1}
\end{eqnarray}                                    
In Eq. (\ref{lp1}), $\psi_{j}$ with $j=1,2$ are proportional to the electric 
fields of the beams, and $x$, $y$, and $z$ are the spatial coordinates 
\cite{Dimensions1}. In addition, $d_{11}$ is the coefficient related to  
the velocity difference between the beam centers (the beam-steering 
coefficient), and $\epsilon_{3}$ is the cubic loss coefficient, 
which satisfies $0<\epsilon_{3} \ll 1$. The terms $\partial_{x}^{2}\psi_{j}$ 
and $\partial_{y}^{2}\psi_{j}$ on the left hand side of Eq. (\ref{lp1}) 
describe the effects of isotropic diffraction, while $id_{11}\partial_{x}\psi_{2}$ 
is related to the velocity difference between the beam centers. 
The first and second terms on the right hand side 
of Eq. (\ref{lp1}) describe intra-beam and inter-beam effects due to cubic loss.  
In the current paper, we do not take into account the effects of linear loss, 
since these effects do not change the form of the expressions for the collision-induced 
changes in beam amplitudes and shapes. Furthermore, the simple effects of linear loss 
on amplitude dynamics can be incorporated into the analysis in exactly the same manner 
as was done in Refs. \cite{PNH2017B,NHP2020} for spatial dimension 1 (see also Appendix \ref{appendB}). 
We remark that the same perturbed linear propagation model (with some changes 
in the physical variables) also describes the dynamics of a fast collision between 
two pulsed-beams in a two-dimensional linear optical medium (e.g., a planar waveguide) 
with weak cubic loss.  In this case, the coordinate $x$ is replaced 
by the time variable $t$, the term $id_{11}\partial_{t}\psi_{2}$ describes the effects 
of the group velocity difference, and the terms $\partial_{t}^{2}\psi_{j}$ describe 
the effects of second-order dispersion. The more general case of fast collisions between 
pulsed-beams in a three-dimensional medium (i.e., in spatial dimension 3 
and total dimension 4) is studied in section \ref{lp_3D}.

We consider fast collisions between beams with generic initial shapes and 
with tails that decay sufficiently fast, such that the values of the integrals 
$\int_{-\infty}^{\infty} dx \int_{-\infty}^{\infty} dy |\psi_{j}(x,y,0)|^{2}$ 
are finite. We assume that the initial beams can be characterized by the following 
parameters. (1) The initial amplitudes $A_{j}(0)$. 
(2) The initial beam widths, i.e., the widths of the maxima of  $|\psi_{j}(x,y,0)|$, 
which can be expressed in terms of the widths along the $x$ and $y$ axes, 
$W_{j0}^{(x)}$ and $W_{j0}^{(y)}$, respectively. (3) The initial positions of the 
beam centers, i.e., the locations of the maxima of $|\psi_{j}(x,y,0)|$, which are 
denoted by $(x_{j0}, y_{j0})$. (4) The initial phases $\alpha_{j0}$. 
Therefore, the initial electric fields of the optical beams can be 
expressed as: 
\begin{eqnarray} &&
\psi_{j}(x,y,0)=A_{j}(0)h_{j}(x,y)\exp(i\alpha_{j0}),  
\label{lp_IC1}
\end{eqnarray} 
where $h_{j}(x,y)$ is real-valued. Note that for brevity of notation, we did not 
write the dependence of the function $h_{j}(x,y)$ on the beam parameters explicitly. 
We are also interested in the important case, where the initial 
electric fields of both beams are separable, i.e., where each of 
the functions $\psi_{j}(x,y,0)$ can be expressed as a product 
of a function of $x$ and a function of $y$ \cite{laser_modes}. 
In this case, the initial electric fields are given by:
\begin{eqnarray} &&
\!\!\!\!\!\!
\psi_{j}(x,y,0)=A_{j}(0)h_{j}^{(x)}[(x-x_{j0})/W_{j0}^{(x)}]
h_{j}^{(y)}[(y-y_{j0})/W_{j0}^{(y)}]\exp(i\alpha_{j0}). 
\nonumber \\&& 
\label{lp_IC2}
\end{eqnarray}       
In what follows, we will also consider cases where the initial electric field is 
separable for one beam and nonseparable for the other beam.

In the current paper, we study the collision-induced dynamics of complete 
fast collisions. The complete collision assumption means that the beams are 
well-separated before and after the collision. More specifically, in these collisions, 
the values of the $x$ coordinate of the beam centers at $z=0$ and 
at the final propagation distance $z_{f}$, $x_{j0}$ and $x_{j}(z_f)$, satisfy 
$|x_{20}-x_{10}| \gg W_{10}^{(x)}+W_{20}^{(x)}$ and 
$|x_{2}(z_f)-x_{1}(z_f)| \gg W_{1}^{(x)}(z_f)+W_{2}^{(x)}(z_f)$, 
where $W_{j}^{(x)}(z_f)$ are the beam widths in the $x$ direction at $z=z_{f}$. 
To obtain the condition for a fast collision, we first define the collision length 
$\Delta z_{c}$, as the distance along which the beam widths in the $x$ direction 
overlap. From this definition it follows that 
$\Delta z_{c}=2(W_{10}^{(x)}+W_{20}^{(x)})/|d_{11}|$. 
For a fast collision, we require that $\Delta z_{c}$ would be much smaller 
than the smallest diffraction length in the problem. We note that the 
diffraction lengths of the $j$th beam in the $x$ and $y$ directions are 
$z_{Dj}^{(x)}=W_{j0}^{(x)2}/2$ and $z_{Dj}^{(y)}=W_{j0}^{(y)2}/2$,  
respectively. Thus, the smallest diffraction length $z_{D}^{(min)}$
is $z_{D}^{(min)}=\min \left\{ z_{D1}^{(x)}, z_{D2}^{(x)}, 
z_{D1}^{(y)}, z_{D2}^{(y)}\right\}$. Requiring that $\Delta z_{c} \ll z_{D}^{(min)}$ 
and using the definition of $\Delta z_{c}$, we obtain that the condition for a fast 
collision can be expressed as $2(W_{10}^{(x)}+W_{20}^{(x)}) \ll |d_{11}| z_{D}^{(min)}$.

\subsection{Calculation of the collision-induced changes in the beam  
shape and amplitude for a general initial condition} 
\label{lp_general_IC}

\subsubsection{Introduction}     
The perturbation method that we present here generalizes the perturbation 
method presented in Refs. \cite{PNH2017B,NHP2020} in three major ways. 
First, it extends the calculation from spatial dimension 1 to spatial dimension 2 \cite{nD}. 
Second, it provides a perturbative calculation and 
analytic expressions for the collision-induced change 
in the beam shape both in the collision interval and away from the collision interval, 
whereas the calculation of the change in the beam shape in Refs. \cite{PNH2017B,NHP2020} 
was limited to the collision interval only. Third, it helps uncover several collision-induced 
effects, which exist only in spatial dimension higher than 1.  In the first step in 
the perturbative calculation, we look for a solution of Eq. (\ref{lp1}) in the form: 
\begin{eqnarray}&&
\psi_{j}(x,y,z)=\psi_{j0}(x,y,z)+\phi_{j}(x,y,z), 
\label{lp2}
\end{eqnarray}           
where $j=1,2$, $\psi_{j0}$ are the solutions of the weakly perturbed linear propagation 
equations without the inter-beam interaction terms, and $\phi_{j}$ describe corrections to 
the $\psi_{j0}$ due to the effects of inter-beam interaction on the collision. 
By their definition, the $\psi_{j0}$ satisfy the following two equations:  
\begin{eqnarray}&&
i\partial_z\psi_{10} + \partial_{x}^{2}\psi_{10}
+ \partial_{y}^{2}\psi_{10}=
-i\epsilon_{3}|\psi_{10}|^2\psi_{10},
\label{lp3}
\end{eqnarray}         
and
\begin{eqnarray}&&
i\partial_z\psi_{20}+id_{11}\partial_{x}\psi _{20}
+\partial_{x}^2\psi_{20} + \partial_{y}^{2}\psi_{20} =
-i\epsilon_{3}|\psi_{20}|^2\psi_{20}.   
\label{lp4}
\end{eqnarray}                                    
Substituting the ansatz (\ref{lp2}) into Eq. (\ref{lp1}) 
and using Eqs. (\ref{lp3}) and  (\ref{lp4}), we obtain equations for the $\phi_{j}$. 
We concentrate on the calculation of $\phi_{1}$, since the calculation of 
$\phi_{2}$ is similar. The equation for $\phi_{1}$ in the 
leading order of the perturbative calculation is 
\begin{eqnarray}&&
i\partial_z\phi_{1} + \partial_{x}^{2}\phi_{1}
+ \partial_{y}^{2}\phi_{1}=
-2i\epsilon_{3}|\psi_{20}|^2\psi_{10}. 
\label{lp5}
\end{eqnarray}      
Note that in writing Eq. (\ref{lp5}), we neglected the high-order 
terms containing $\phi_{j}$ on the right hand side of the equation.

In solving the equation for $\phi_{1}$, we distinguish 
between two intervals along the $z$ axis, the collision interval and the 
post-collision interval. These intervals are defined in terms of the collision 
distance $z_{c}$, which is the distance at which the $x$ coordinates of 
the beam centers coincide, i.e., $x_{1}(z_{c})=x_{2}(z_{c})$. 
The collision interval is the small interval 
$z_{c} - \Delta z_{c}/2 \le z \le z_{c} + \Delta z_{c}/2$ centered 
about $z_{c}$, in which the two beams are overlapping. 
The post-collision interval is the interval $z > z_{c} + \Delta z_{c}/2$, 
in which the beams are no longer overlapping.

\subsubsection{Collision-induced effects in the collision interval}     

We substitute $\psi_{j0}(x,y,z)=\Psi_{j0}(x,y,z)\exp[i\chi_{j0}(x,y,z)]$ and 
$\phi_{1}(x,y,z)=\Phi_{1}(x,y,z)\exp[i\chi_{10}(x,y,z)]$, where $\Psi_{j0}$ and 
$\chi_{j0}$ are real-valued, into Eq. (\ref{lp5}), and obtain the following equation 
for $\Phi_{1}$:          
\begin{eqnarray} &&
i\partial_{z}\Phi_{1} - \left(\partial_{z}\chi_{10}\right)\Phi_{1}
+\left[\partial_{x}^{2}\Phi_{1}
+2i\left(\partial_{x}\chi_{10}\right)\partial_{x}\Phi_{1}
\right. 
\nonumber \\&&
\left. 
+i\left(\partial_{x}^{2}\chi_{10}\right)\Phi_{1}
-\left(\partial_{x}\chi_{10}\right)^2\Phi_{1}\right] 
+\left[\partial_{y}^{2}\Phi_{1}
+2i\left(\partial_{y}\chi_{10}\right)\partial_{y}\Phi_{1}
\right. 
\nonumber \\&&
\left. 
+i\left(\partial_{y}^{2}\chi_{10}\right)\Phi_{1}
-\left(\partial_{y}\chi_{10}\right)^2\Phi_{1}\right] 
=-2i\epsilon_{3}\Psi_{20}^{2}\Psi_{10}.
\label{lp6}
\end{eqnarray} 
Since the collision length $\Delta z_{c}$ is of order 
$1/|d_{11}|$, the term $i\partial_{z}\Phi_{1}$ is of order 
$|d_{11}| \times O(\Phi_{1})$. Additionally, the term 
$-2i\epsilon_{3}\Psi_{20}^{2}\Psi_{10}$ is of order $\epsilon_{3}$.  
Equating the orders of $i\partial_{z}\Phi_{1}$ and 
$-2i\epsilon_{3}\Psi_{20}^{2}\Psi_{10}$, we find that $\Phi_{1}$ 
is of order $\epsilon_{3}/|d_{11}|$. In addition, $\Phi_{1}$ does not 
contain any fast dependence on $x$ and $y$, and $\chi_{10}$ does 
not contain any fast dependence on $x$, $y$, and $z$. As a result, 
all the other terms in Eq. (\ref{lp6}) are of order $\epsilon_{3}/|d_{11}|$ 
or higher, and can therefore be neglected. It follows that in the leading 
order of the perturbative calculation, the equation for $\Phi_{1}$ is 
\begin{eqnarray} &&
\partial_{z}\Phi_{1}=
-2\epsilon_{3}\Psi_{20}^{2}\Psi_{10}.
\label{lp7}
\end{eqnarray} 
Equation (\ref{lp7}) has the same form as the equation obtained 
for a fast collision between two pulses of the linear propagation equation 
in the presence of weak cubic loss in spatial dimension 1 \cite{PNH2017B,NHP2020}.
It also has the same form as the equation obtained for a fast collision between two 
solitons of the NLS equation in the presence of weak cubic loss 
in spatial dimension 1 \cite{PNC2010}.

We now introduce the following approximations to the solutions 
$\psi_{j0}(x,y,z)$ of Eqs. (\ref{lp3}) and (\ref{lp4}): 
\begin{eqnarray} &&
\psi_{j0}(x,y,z) \simeq A_{j}(z)\tilde\psi_{j0}(x,y,z), 
\label{lp_add1}
\end{eqnarray}    
where $A_{j}(z)$ are the $z$ dependent beam amplitudes, and 
\begin{eqnarray} &&
\tilde\psi_{j0}(x,y,z)=\tilde\Psi_{j0}(x,y,z)\exp[i\chi_{j0}(x,y,z)], 
\label{lp_add4}
\end{eqnarray}    
are the solutions to the unperturbed linear propagation equation 
with the initial condition (\ref{lp_IC1}) with unit amplitude. 
From Eqs. (\ref{lp_add1}) and (\ref{lp_add4}), it follows that 
\begin{eqnarray} &&
\Psi_{j0}(x,y,z) \simeq A_{j}(z)\tilde\Psi_{j0}(x,y,z). 
\label{lp_add2}
\end{eqnarray}          
Using Eqs. (\ref{lp2}) and (\ref{lp_add1}), we find that the total 
electric fields of the beams can be approximated by 
\begin{eqnarray}&&
\psi_{j}(x,y,z) \simeq A_{j}(z)\tilde\psi_{j0}(x,y,z)+\phi_{j}(x,y,z). 
\label{lp_add3}
\end{eqnarray}  
Note that the approximate expressions (\ref{lp_add1}), (\ref{lp_add2}), 
and (\ref{lp_add3}) are used both inside and outside of the collision interval. 
In addition, the dynamics of the $A_{j}(z)$ that is associated with 
single-beam propagation is described in Appendix \ref{appendB}.

The collision-induced amplitude shift of beam 1 is calculated from  
the collision-induced change in $\Phi_{1}$ in the collision interval, 
$\Delta\Phi_{1}(x,y,z_{c})=\Phi_{1}(x,y,z_{c}+\Delta z_{c}/2)-
\Phi_{1}(x,y,z_{c}-\Delta z_{c}/2)$. To calculate $\Delta\Phi_{1}(x,y,z_{c})$, 
we substitute the approximations (\ref{lp_add2}) for the $\Psi_{j0}$ 
into Eq. (\ref{lp7}), and integrate with respect to $z$ over the collision interval. 
This calculation yields   
\begin{eqnarray}&&
\!\!\!\!\!\!
\Delta\Phi_{1}(x,y,z_{c})\!=\!-2\epsilon_{3}
\!\!\int_{z_{c}-\Delta z_{c}/2}^{z_{c}+\Delta z_{c}/2} 
\!\!\!\!\!\!\!\!\!\!\! dz' A_{1}(z') A_{2}^{2}(z')
\nonumber \\&&
\times
\tilde\Psi_{10}(x,y,z')\tilde\Psi_{20}^{2}(x,y,z').
\label{lp8}
\end{eqnarray}  
Note that $\tilde\Psi_{20}$ is the only function in the integrand on 
the right hand side of Eq. (\ref{lp8}) that contains fast variations 
with respect to $z$, which are of order 1. Therefore, we can 
approximate the other functions $A_{1}(z)$, $A_{2}(z)$, and $\tilde\Psi_{10}(x,y,z)$ 
by $A_{1}(z_{c}^{-})$, $A_{2}(z_{c}^{-})$, and $\tilde\Psi_{10}(x,y,z_{c})$, 
where $A_{j}(z_{c}^{-})$ is the limit from the left of $A_{j}$ at $z_{c}$.   
Furthermore, in calculating the integral, we can take into account in an exact manner 
only the fast dependence of $\tilde\Psi_{20}$ on $z$, i.e., the dependence on $z$ 
that is contained in the factors $\tilde x=x-x_{20}-d_{11}z$, and replace $z$ by $z_{c}$ 
everywhere else in the expression for $\tilde\Psi_{20}$. We denote this approximation 
of $\tilde\Psi_{20}(x,y,z)$ by $\bar\Psi_{20}(\tilde x,y,z_{c})$.   
Carrying out all the aforementioned approximations in Eq. (\ref{lp8}), we obtain: 
\begin{eqnarray}&&
\!\!\!\!\!\!\!\!
\Delta\Phi_{1}(x,y,z_{c})\!=\!
-2\epsilon_{3}A_{1}(z_{c}^{-}) A_{2}^{2}(z_{c}^{-})
\tilde\Psi_{10}(x,y,z_{c})
\nonumber \\&&
\times
\!\!\int_{z_{c}-\Delta z_{c}/2}^{z_{c}+\Delta z_{c}/2}  
\!\!\!\!\!\!\!\! dz' 
\bar\Psi_{20}^{2}(x - x_{20}-d_{11}z',y,z_{c}).
\label{lp9}
\end{eqnarray}    
We assume that the integrand on the right hand side of 
Eq. (\ref{lp9}) is sharply peaked at a small interval around 
the collision distance $z_{c}$. Under this assumption, we can 
extend the integral's limits to $-\infty$ and $\infty$. 
We also change the integration variable from 
$z'$ to $\tilde x=x-x_{20}-d_{11}z'$ and obtain: 
\begin{eqnarray} &&
\!\!\!\!\!\!\!
\Delta\Phi_{1}(x,y,z_{c})\!=\!-\frac{2\epsilon_{3}
A_{1}(z_{c}^{-}) A_{2}^{2}(z_{c}^{-})}{|d_{11}|}\tilde\Psi_{10}(x,y,z_{c})
\!\!\!\int_{-\infty}^{\infty} \!\!\!\!\! d\tilde x \bar\Psi_{20}^{2}(\tilde x,y,z_{c}).
\nonumber \\&&
\label{lp10}
\end{eqnarray}
We see that the $y$ dependence of beam 2 at $z=z_{c}$ affects the 
$y$ dependence of $\Delta\Phi_{1}(x,y,z_{c})$, while the $x$ dependence 
of beam 2 does not affect the $x$ dependence of $\Delta\Phi_{1}(x,y,z_{c})$. 
Thus, inside the collision interval, the beam shape in the longitudinal direction 
is preserved, while the beam shape in the transverse direction is changed by the collision. 
We also point out that the collision-induced change in the beam shape 
is an effect that exists only in spatial dimension higher than 1. 
Indeed, it was shown in Refs. \cite{PNH2017B,NHP2020} that in the one-dimensional case, 
the beam shape is preserved by the collision within the leading order 
of the perturbative calculation.

In Appendix \ref{appendA}, we show that the collision-induced amplitude shift 
of beam 1 $\Delta A_{1}^{(c)}$ is related to $\Delta\Phi_{1}(x,y,z_{c})$ by:
\begin{eqnarray}&&
\!\!\!\!\!\!\!\!\!\!\!\!\!\!
\Delta A_{1}^{(c)}=C_{p1}^{-1}
\!\!\int_{-\infty}^{\infty} \!\!\!\!\! dx 
\!\int_{-\infty}^{\infty} \!\!\!\!\! dy
\;\tilde\Psi_{10}(x,y,z_{c})\Delta\Phi_{1}(x,y,z_{c}), 
\label{lp11}
\end{eqnarray}       
where 
\begin{eqnarray}&&
C_{p1}= 
\!\!\int_{-\infty}^{\infty} \!\!\!\!\! dx 
\!\int_{-\infty}^{\infty} \!\!\!\!\! dy
\;\tilde\Psi_{10}^{2}(x,y,0).  
\label{lp12}
\end{eqnarray}       
Substituting Eq. (\ref{lp10}) into  Eq. (\ref{lp11}), we obtain the following   
expression for the collision-induced amplitude shift of beam 1 for the general 
initial condition (\ref{lp_IC1}): 
\begin{eqnarray} &&
\!\!\!\!
\Delta A_{1}^{(c)}=-\frac{2\epsilon_{3}
A_{1}(z_{c}^{-}) A_{2}^{2}(z_{c}^{-})}{C_{p1}|d_{11}|}
\nonumber \\&&
\times
\!\!\int_{-\infty}^{\infty} \!\!\!\!\! dx 
\!\int_{-\infty}^{\infty} \!\!\!\!\! dy
\;\tilde\Psi_{10}^{2}(x,y,z_{c})
\!\int_{-\infty}^{\infty} \!\!\!\!\! d\tilde x \;\bar\Psi_{20}^{2}(\tilde x,y,z_{c}). 
\label{lp13}
\end{eqnarray}

\subsubsection{Dynamics of $\phi_{1}(x,y,z)$ in the post-collision interval}
In the post collision interval, i.e., for $z > z_{c} + \Delta z_{c}/2$, 
the two beams are no longer overlapping. As a result, the inter-beam  
interaction terms $-2i\epsilon_{3}|\psi_{2}|^2\psi_{1}$ and 
$-2i\epsilon_{3}|\psi_{1}|^2\psi_{2}$ are negligible in this interval. 
Therefore, in the leading order of the perturbative calculation, 
the equation describing the dynamics of $\phi_{1}(x,y,z)$ in the 
post-collision interval is the unperturbed linear propagation equation 
\begin{eqnarray}&&
i\partial_z\phi_{1} + \partial_{x}^{2}\phi_{1}
+ \partial_{y}^{2}\phi_{1}=0. 
\label{lp21}
\end{eqnarray}       
To find the initial condition for Eq. (\ref{lp21}), we first note 
that for $|d_{11}| \gg 1$, $\Delta\Phi_{1}(x,y,z_{c})$ can be written as 
\begin{eqnarray}&&
\Delta\Phi_{1}(x,y,z_{c}) \simeq \Phi_{1}(x,y,z_{c}^{+}) - \Phi_{1}(x,y,z_{c}^{-}) 
\simeq \Phi_{1}(x,y,z_{c}^{+}),  
\label{lp22}
\end{eqnarray}       
where $\Phi_{1}(x,y,z_{c}^{+})$ is the limit from the right of $\Phi_{1}(x,y,z)$ 
at $z=z_{c}$. Thus, using the relation $\phi_{1}(x,y,z)=\Phi_{1}(x,y,z)\exp[i\chi_{10}(x,y,z)]$, 
the initial condition for Eq. (\ref{lp21}) is: 
\begin{eqnarray}&&
\phi_{1}(x,y,z_{c}^{+}) = \Phi_{1}(x,y,z_{c}^{+})\exp[i\chi_{10}(x,y,z_{c})],  
\label{lp23}
\end{eqnarray}        
where $\Phi_{1}(x,y,z_{c}^{+})$ is given by Eq. (\ref{lp10}). 
The solution of Eq. (\ref{lp21}) with the IC (\ref{lp23}) 
can be written as 
\begin{eqnarray}&&
\phi_{1}(x,y,z) = {\cal F}^{-1}\left(\hat\phi_{1}(k_{1},k_{2},z_{c}^{+})
\exp[-i(k_{1}^{2} + k_{2}^{2})(z-z_{c})]\right),  
\label{lp24}
\end{eqnarray}         
where $\hat\phi_{1}(k_{1},k_{2},z_{c}^{+})=
{\cal F}\left(\phi_{1}(x,y,z_{c}^{+}) \right)$, 
and ${\cal F}$ and ${\cal F}^{-1}$ are the Fourier transform and the 
inverse Fourier transform with respect to $x$ and $y$. In addition, the 
total electric field of beam 1 in the post-collision interval is given by 
Eq. (\ref{lp_add3}), where $A_{1}(z)$ is given by Eq. (\ref{appB_3}) 
in Appendix \ref{appendB}.

\subsection{Calculation of the collision-induced changes in the beam  
shape and amplitude for a separable initial condition} 
\label{lp_separable_IC}

\subsubsection{Introduction}
We now describe the collision-induced dynamics in the important case, 
where the initial condition for both beams is separable, i.e., it is given 
by Eq. (\ref{lp_IC2}). This case is of special importance for two main reasons. 
First, this initial condition describes the output electric field from many types 
of lasers \cite{Siegman86,Kogelnik66}. Second, in this case, it is possible to 
further simplify the expressions for the collision-induced changes of the beam  
shape and amplitude, and in this manner, obtain deeper insight into the 
collision dynamics.

It is straightforward to show that the solutions of the unperturbed linear 
propagation equation with the separable initial condition (\ref{lp_IC2}) 
and with unit amplitude can be written as 
\begin{eqnarray}&&
\tilde\psi_{j0}(x,y,z)=g_{j}^{(x)}(x,z)g_{j}^{(y)}(y,z)\exp(i\alpha_{j0})=
\nonumber \\&&
G_{j}^{(x)}(x,z)G_{j}^{(y)}(y,z)
\exp\left\{i\left[\chi_{j0}^{(x)}(x,z) + \chi_{j0}^{(y)}(y,z) 
+ \alpha_{j0}\right] \right\},
\label{lp31}
\end{eqnarray}   
where 
\begin{eqnarray}&&
g_{1}^{(x)}(x,z) = (2\pi)^{-1/2}
\!\int_{-\infty}^{\infty} \!\!\!\! d k_{1} \hat f_{1}^{(x)}(k_{1})
\exp[-ik_{1}^{2}z + ik_{1}x],  
\nonumber \\&& 
\!\!\!\!\!\!\!\!\!\!\!\!
g_{2}^{(x)}(x,z) = (2\pi)^{-1/2}
\!\int_{-\infty}^{\infty} \!\!\!\! d k_{1} \hat f_{2}^{(x)}(k_{1})
\exp[-id_{11}k_{1}z -ik_{1}^{2}z + ik_{1}x], 
\label{lp32}
\end{eqnarray}         
\begin{eqnarray}&& 
g_{j}^{(y)}(y,z) = (2\pi)^{-1/2}
\!\int_{-\infty}^{\infty} \!\!\!\! d k_{2} \hat f_{j}^{(y)}(k_{2})
\exp[-ik_{2}^{2}z + ik_{2}y],  
\label{lp33}
\end{eqnarray}         
and $G_{j}^{(x)}$, $G_{j}^{(y)}$, $\chi_{j0}^{(x)}$, and 
$\chi_{j0}^{(y)}$ are real-valued. The functions $\hat f_{j}^{(x)}$ 
and $\hat f_{j}^{(y)}$ in Eqs. (\ref{lp32}) and (\ref{lp33}) 
are defined by: 
\begin{eqnarray}&&
\hat f_{j}^{(x)}(k_{1}) =
{\cal F}\left(h_{j}^{(x)}[(x-x_{j0})/W_{j0}^{(x)}]\right),  
\label{lp34}
\end{eqnarray}           
and 
\begin{eqnarray}&&
\hat f_{j}^{(y)}(k_{2}) =
{\cal F}\left(h_{j}^{(y)}[(y-y_{j0})/W_{j0}^{(y)}]\right). 
\label{lp35}
\end{eqnarray}    
Using Eqs. (\ref{lp_add4}) and (\ref{lp31}), we obtain 
\begin{eqnarray}&&
\tilde\Psi_{j0}(x,y,z)=G_{j}^{(x)}(x,z)G_{j}^{(y)}(y,z),
\label{lp36}
\end{eqnarray}   
and 
\begin{eqnarray}&&
\chi_{j0}(x,y,z) = \chi_{j0}^{(x)}(x,z) + \chi_{j0}^{(y)}(y,z) + \alpha_{j0}.
\label{lp37}
\end{eqnarray}      
In addition, using the conservation of the total energy for the 
unperturbed linear propagation equation, the definitions of 
$G_{j}^{(x)}$ and $G_{j}^{(y)}$, and the initial condition 
(\ref{lp_IC2}), we obtain  
\begin{eqnarray}&&
 \!\!\!\! \!\!\!\!
\int_{-\infty}^{\infty} \!\!\!\! dx \, G_{j}^{(x)2}(x,z)
=\int_{-\infty}^{\infty} \!\!\!\! dx \, G_{j}^{(x)2}(x,0)
=W_{j0}^{(x)}\int_{-\infty}^{\infty} \!\!\!\! ds \, h_{j}^{(x)2}(s)
=W_{j0}^{(x)}  c_{pj}^{(x)},  
\nonumber \\&&
\label{lp38}
\end{eqnarray}         
and 
\begin{eqnarray}&&
\!\!\!\! \!\!\!\!
\int_{-\infty}^{\infty} \!\!\!\! dy \, G_{j}^{(y)2}(y,z)
=\int_{-\infty}^{\infty} \!\!\!\! dy \, G_{j}^{(y)2}(y,0)
=W_{j0}^{(y)}\int_{-\infty}^{\infty} \!\!\!\! ds \, h_{j}^{(y)2}(s)
=W_{j0}^{(y)}  c_{pj}^{(y)},  
\nonumber \\&&
\label{lp39}
\end{eqnarray}                    
where $c_{pj}^{(x)}$ and $c_{pj}^{(y)}$ are constants.

\subsubsection{Collision-induced effects in the collision interval}  

We first obtain the general expression for $\Delta\Phi_{1}(x,y,z_{c})$ 
for an initial condition that is separable for both beams. For this purpose, 
we note that from the definition of $\bar\Psi_{20}(\tilde x,y,z)$ 
it follows that at $z=z_{c}$, $\bar\Psi_{20}(\tilde x,y,z_{c})=\tilde\Psi_{20}(x,y,z_{c})$. 
Using this relation along with Eq. (\ref{lp36}), we obtain 
$\bar\Psi_{20}(\tilde x,y,z_{c})=G_{2}^{(x)}(x,z_{c}) G_{2}^{(y)}(y,z_{c})$. 
It follows that: 
\begin{eqnarray}&&
\int_{-\infty}^{\infty} \!\!\!\!\! d\tilde x \, \bar\Psi_{20}^{2}(\tilde x,y,z_{c})=
G_{2}^{(y)2}(y,z_{c})
\int_{-\infty}^{\infty} \!\!\!\!\! dx \, G_{2}^{(x)2}(x,z_{c}).
\label{lp41}
\end{eqnarray}   
Employing the conservation law (\ref{lp38}) in Eq. (\ref{lp41}), we obtain:       
\begin{eqnarray}&&
\int_{-\infty}^{\infty} \!\!\!\!\! d\tilde x \, \bar\Psi_{20}^{2}(\tilde x,y,z_{c})=
c_{p2}^{(x)} W_{20}^{(x)} G_{2}^{(y)2}(y,z_{c}).
\label{lp42}
\end{eqnarray}               
Substitution of Eq. (\ref{lp42}) into Eq. (\ref{lp10}) yields the following 
expression for $\Delta\Phi_{1}(x,y,z_{c})$, which is valid 
for an initial condition that is separable for beam 2: 
\begin{eqnarray} &&
\!\!\!\!\!\!\!
\Delta\Phi_{1}(x,y,z_{c})\!=\!-\frac{2\epsilon_{3}
A_{1}(z_{c}^{-}) A_{2}^{2}(z_{c}^{-})}{|d_{11}|}
c_{p2}^{(x)} W_{20}^{(x)} G_{2}^{(y)2}(y,z_{c})
\tilde\Psi_{10}(x,y,z_{c}).
\nonumber \\&&
\label{lp43}
\end{eqnarray}   
Equation (\ref{lp43}) is valid for a general initial condition for beam 1. 
When the initial condition for beam 1 is also separable, Eq. (\ref{lp43}) 
takes the form 
\begin{eqnarray} &&
\!\!\!\!\!\!\!
\Delta\Phi_{1}(x,y,z_{c})\!=\!-\frac{2\epsilon_{3}
A_{1}(z_{c}^{-}) A_{2}^{2}(z_{c}^{-})}{|d_{11}|}
\nonumber \\&&
\times c_{p2}^{(x)} W_{20}^{(x)} 
G_{1}^{(x)}(x,z_{c})G_{1}^{(y)}(y,z_{c})
G_{2}^{(y)2}(y,z_{c}). 
\label{lp44}
\end{eqnarray}    
We see that as in the case of the general initial condition (\ref{lp_IC1}), 
the shape of the beam in the longitudinal direction does not change inside 
of the collision interval. Moreover, it follows from Eq. (\ref{lp44}) that for 
a separable initial condition, the shape of the beam in the longitudinal direction 
is not changed by the collision at all, i.e., for any $z>z_{c}$
(within the leading order of the perturbative calculation). Indeed, for 
$|d_{11}| \gg 1$, Eq. (\ref{lp44}) is also the initial condition for the dynamics 
of $\phi_{1}(x,y,z)$ in the post-collision interval. We observe that this initial 
condition is separable. In addition, at $z=z_{c}$ the $x$ dependences of 
$\phi_{1}$ and $\tilde\psi_{10}$ are identical. Since in the post-collision 
region $\phi_{1}$ and $\tilde\psi_{10}$ satisfy the same linear propagation 
equation with separable initial conditions, which have the same dependence on $x$, 
the $x$ dependences of $\phi_{1}$ and $\tilde\psi_{10}$ remain identical for 
any $z>z_{c}$. Thus, for a separable initial condition, the shape of the beam in 
the longitudinal direction is not changed by the collision at all.

We now turn to obtain the expression for $\Delta A_{1}^{(c)}$ for 
an initial condition that is separable for both beams. 
Using the conservation of the total energy and 
Eqs. (\ref{lp38}) and (\ref{lp39}), we obtain   
$C_{p1}=c_{p1}^{(x)}c_{p1}^{(y)}W_{10}^{(x)}W_{10}^{(y)}$. 
In addition, using Eqs. (\ref{lp36}) and (\ref{lp42}), we find 
 \begin{eqnarray} &&
\!\!\int_{-\infty}^{\infty} \!\!\!\!\! dx 
\!\int_{-\infty}^{\infty} \!\!\!\!\! dy
\;\tilde\Psi_{10}^{2}(x,y,z_{c})
\!\int_{-\infty}^{\infty} \!\!\!\!\! d\tilde x \;\bar\Psi_{20}^{2}(\tilde x,y,z_{c})= 
\nonumber \\&&
c_{p1}^{(x)}c_{p2}^{(x)}W_{10}^{(x)}W_{20}^{(x)} 
\!\int_{-\infty}^{\infty} \!\!\!\!\! dy \, 
G_{1}^{(y)2}(y,z_{c})G_{2}^{(y)2}(y,z_{c}). 
\label{lp45}
\end{eqnarray}        
Substituting Eq. (\ref{lp45}) and the expression for $C_{p1}$ 
into Eq. (\ref{lp13}), we obtain the following expression for the 
collision-induced amplitude shift for a separable initial condition: 
\begin{eqnarray} &&
\!\!\!\!
\Delta A_{1}^{(c)}=-\frac{2\epsilon_{3}
A_{1}(z_{c}^{-}) A_{2}^{2}(z_{c}^{-})}{|d_{11}|}
\nonumber \\&&
\times
\frac{c_{p2}^{(x)}W_{20}^{(x)}} 
{c_{p1}^{(y)}W_{10}^{(y)}}
\!\int_{-\infty}^{\infty} \!\!\!\!\! dy \, 
G_{1}^{(y)2}(y,z_{c})G_{2}^{(y)2}(y,z_{c}). 
\label{lp46}
\end{eqnarray}     
Note that the expression for $\Delta A_{1}^{(c)}$ has the form 
\begin{eqnarray} &&
\!\!\!\!
\Delta A_{1}^{(c)}=-(\mbox{overall factor}) \times 
(\mbox{longitudinal factor}) \times (\mbox{transverse factor}), 
\nonumber \\&&
\label{lp47}
\end{eqnarray}  
where the overall factor is equal to $2\epsilon_{3}
A_{1}(z_{c}^{-}) A_{2}^{2}(z_{c}^{-})/|d_{11}|$, 
and the longitudinal factor is $c_{p2}^{(x)}W_{20}^{(x)}$. 
This form of the expression for $\Delta A_{1}^{(c)}$ is 
expected to be valid for a general spatial dimension, 
when the initial condition is separable in the longitudinal direction 
for both beams. It is interesting to note that Eq. (\ref{lp46}) is 
a generalization of the equation obtained for a fast two-pulse collision 
in spatial dimension 1. Indeed, using the notation of the current paper, 
the latter equation, which is Eq. (19) in Ref. \cite{NHP2020}, can be 
written as: 
\begin{eqnarray} &&
\Delta A_{1}^{(c)(1D)}=-\frac{2\epsilon_{3}
A_{1}(z_{c}^{-}) A_{2}^{2}(z_{c}^{-})}{|d_{11}|}
c_{p2}^{(x)}W_{20}^{(x)}. 
\label{lp48}
\end{eqnarray}   
We observe that the overall and longitudinal factors in the equation 
for the amplitude shift in spatial dimension 1 have the same form as 
the overall and longitudinal factors in spatial dimension 2, 
while the transverse factor in the one-dimensional case 
is equal to 1. We also observe that the longitudinal factor 
$c_{p2}^{(x)}W_{20}^{(x)}$ is universal in the sense that it does 
not depend on the exact details of the initial pulse shapes and on the 
collision distance $z_{c}$. In contrast, the transverse factor is not 
universal since it does depend on the details of the initial pulse shapes 
and on the collision distance. Therefore, the universality of the expression 
for $\Delta A_{1}^{(c)}$ in the one-dimensional case, which was first demonstrated 
in Ref. \cite{NHP2020}, is extended to spatial dimension 2 
(and to spatial dimension $n$), but in a somewhat restricted manner. 
That is, in the two-dimensional (and the $n$-dimensional) case, 
only the overall and longitudinal parts of the expression for $\Delta A_{1}^{(c)}$ 
are universal, and this is true when the initial condition is separable in the 
longitudinal direction for both beams.

\subsubsection{Dynamics of $\phi_{1}(x,y,z)$ in the post-collision interval}

We now turn to analyze the dynamics of $\phi_{1}(x,y,z)$ in the 
post-collision interval. This analysis is especially interesting for two 
main reasons. First, we showed in subsection \ref{lp_general_IC} 
that the collision induces a change of the beam shape in the transverse 
direction. Even though this effect exists for a general 
initial condition, its simplest and clearest demonstration 
is realized in the case of an initial condition that is separable 
for both beams. Furthermore, since in both experiments and simulations  
the change in the beam shape is measured in the post-collision 
interval, we must analyze the evolution of the beam shape in 
this interval. Second, we claimed in section 2.3.2 that for a separable 
initial condition, the shape of the beam in the longitudinal direction does not 
change at all due to the collision. This claim can be directly proved by analyzing 
the dynamics of $\phi_{1}(x,y,z)$ in the post-collision interval.

In the post-collision interval, $\phi_{1}$ satisfies the unperturbed linear 
propagation equation (\ref{lp21}). Using Eqs. (\ref{lp23}), (\ref{lp36}), 
and (\ref{lp44}), we find that the initial condition for Eq. (\ref{lp21}) is 
\begin{eqnarray} &&
\!\!\!\!\!\!\!
\phi_{1}(x,y,z_{c}^{+}) =
-\frac{2\epsilon_{3}
A_{1}(z_{c}^{-}) A_{2}^{2}(z_{c}^{-})}{|d_{11}|}
\nonumber \\&&
\times c_{p2}^{(x)} W_{20}^{(x)} 
g_{1}^{(x)}(x,z_{c})g_{1}^{(y)}(y,z_{c})
G_{2}^{(y)2}(y,z_{c})\exp(i\alpha_{10}). 
\label{lp51}
\end{eqnarray}                                                       
This initial condition can be written as: 
\begin{eqnarray} &&
\phi_{1}(x,y,z_{c}^{+})=
-\tilde a_{1}(z_{c}^{-}) 
g_{1}^{(x)}(x,z_{c})g_{12}^{(y)}(y,z_{c})\exp(i\alpha_{10}),
\label{lp52}
\end{eqnarray}        
where 
\begin{eqnarray} &&
\tilde a_{1}(z_{c}^{-})=
2\epsilon_{3}A_{1}(z_{c}^{-}) A_{2}^{2}(z_{c}^{-}) 
c_{p2}^{(x)} W_{20}^{(x)}/|d_{11}|, 
\label{lp53}
\end{eqnarray}        
and 
\begin{eqnarray} &&
g_{12}^{(y)}(y,z_{c})= 
g_{1}^{(y)}(y,z_{c})G_{2}^{(y)2}(y,z_{c}).
\label{lp54}
\end{eqnarray}        
The Fourier transform of the initial condition (\ref{lp52}) is 
\begin{eqnarray} &&
\hat \phi_{1}(k_{1},k_{2},z_{c}^{+})=
-\tilde a_{1}(z_{c}^{-}) \hat g_{1}^{(x)}(k_{1},z_{c}) 
\hat g_{12}^{(y)}(k_{2},z_{c})\exp(i\alpha_{10}),
\label{lp55}
\end{eqnarray}         
where $\hat g_{1}^{(x)}$ and $\hat g_{12}^{(y)}$ are the Fourier 
transforms of $g_{1}^{(x)}$ and $g_{12}^{(y)}$ with respect to 
$x$ and $y$, respectively. Substituting Eq. (\ref{lp55}) into Eq. (\ref{lp24}), 
we obtain: 
\begin{eqnarray}&&
\phi_{1}(x,y,z)=
-\tilde a_{1}(z_{c}^{-}) 
{\cal F}^{-1}\left(\hat g_{1}^{(x)}(k_{1},z_{c})
\exp[-i k_{1}^{2}(z-z_{c})]\right)
\nonumber \\&&
\times  {\cal F}^{-1}\left(\hat g_{12}^{(y)}(k_{2},z_{c})
\exp[-i k_{2}^{2}(z-z_{c})]\right)\exp(i\alpha_{10}). 
\label{lp56}
\end{eqnarray}              
Note that when the initial condition for beam 1 is separable, 
$\hat g_{1}^{(x)}(k_{1},z_{c})\exp[-i k_{1}^{2}(z-z_{c})]$ 
is equal to $\hat g_{1}^{(x)}(k_{1},z)$: 
\begin{eqnarray}&&
\hat g_{1}^{(x)}(k_{1},z_{c})\exp[-i k_{1}^{2}(z-z_{c})]= 
\hat g_{1}^{(x)}(k_{1},0)\exp(-i k_{1}^{2}z_{c})
\exp[-i k_{1}^{2}(z-z_{c})]=
\nonumber \\&&
\hat g_{1}^{(x)}(k_{1},0)\exp(-i k_{1}^{2}z)=
\hat g_{1}^{(x)}(k_{1},z). 
\label{lp57}
\end{eqnarray}              
Substituting this relation into Eq. (\ref{lp56}), we obtain the 
expression for $\phi_{1}(x,y,z)$ in the post-collision interval 
for a separable initial condition: 
\begin{eqnarray}&&
\phi_{1}(x,y,z)=
-\tilde a_{1}(z_{c}^{-}) g_{1}^{(x)}(x,z)
\nonumber \\&&
\times  {\cal F}^{-1}\left(\hat g_{12}^{(y)}(k_{2},z_{c})
\exp[-i k_{2}^{2}(z-z_{c})]\right)\exp(i\alpha_{10}). 
\label{lp58}
\end{eqnarray}       
We see that when the initial condition is separable for both beams, 
the $x$ dependences of $\phi_{1}(x,y,z)$ and $\tilde\psi_{10}(x,y,z)$ 
are identical for $z>z_{c}$. Therefore, as argued in subsection 2.3.2, 
the shape of the beam in the longitudinal direction does not change at 
all by the collision. Furthermore, the calculation of the modified beam 
shape in the transverse direction amounts to the calculation of the 
inverse Fourier transform of $\hat g_{12}^{(y)}(k_{2},z_{c})
\exp[-i k_{2}^{2}(z-z_{c})]$.

\section{Perturbative calculation and numerical simulations 
for new collisional effects in spatial dimension 2}
\label{simu} 

\subsection{Introduction}
\label{simu_intro}
We now use the perturbation method of subsections \ref{lp_general_IC} 
and \ref{lp_separable_IC} along with numerical simulations with Eq. (\ref{lp1})  
to demonstrate four important effects and properties of the collision, 
which either exist only in spatial dimension higher than 1, 
or are qualitatively different from their one-dimensional counterparts.    
These four effects and properties are: 
(1) universality of the longitudinal part in the expression 
for the collision-induced amplitude shift, 
(2) the effect of partial beam overlap, 
(3) the effect of anisotropy in the initial condition, 
(4) the collision-induced change in the beam shape in 
the transverse direction. 
For each effect or property, we first use the perturbation theory to obtain 
explicit formulas, which demonstrate the collisional effect or property. 
Since these formulas are only approximate expressions, which are based 
on a number of simplifying assumptions of the perturbative calculation, 
it is important to check their validity by numerical simulations with the 
perturbed linear propagation equation (\ref{lp1}). 
Therefore, in the current subsection, we also take on this 
important numerical investigation, by carrying out extensive numerical simulations 
with Eq. (\ref{lp1}), and by comparing the simulations results with the 
approximate predictions of the perturbation theory for each of the four 
collisional effects and properties. We solve Eq. (\ref{lp1}) numerically by the split-step method 
with periodic boundary conditions \cite{Agrawal2001,Yang2010}.

\subsection{Universality of the longitudinal part in the 
expression for the amplitude shift} 
\label{simu_universality}

In subsection \ref{lp_separable_IC}, we showed that for a separable initial condition, 
the longitudinal factor in the expression for the collision-induced amplitude shift is 
universal in the sense that it does not depend on the exact details of the initial beam 
shapes. In contrast, the transverse factor is not universal, since it does depend on the 
details of the beam shapes and on the collision distance. Thus, according to the 
perturbative calculation, the universality of the expression for $\Delta A_{1}^{(c)}$ 
is extended from spatial dimension 1 to higher spatial dimensions, 
but in a somewhat restricted manner.

In the current subsection, we demonstrate the universality of the longitudinal part 
in the expression for the collision-induced amplitude shift. For this purpose, we first obtain 
explicit expressions for $\Delta A_{1}^{(c)}$ for two initial beam shapes that 
have widely different dependences on the $x$ coordinate. Moreover, we verify the 
validity of the expressions for $\Delta A_{1}^{(c)}$ by extensive numerical simulations 
with Eq. (\ref{lp1}). This numerical investigation is especially important, since it shows 
that the approximations used in the perturbative calculation are indeed valid for widely 
different beam shapes. In this manner, the extensive numerical simulations with 
Eq. (\ref{lp1}) help validate the universal nature of the longitudinal part 
in the expression for $\Delta A_{1}^{(c)}$. 
The initial $x$ dependence for the first beam type that we consider is Gaussian, 
i.e., it is rapidly decreasing with increasing distance from the beam center.                      
In contrast, the initial $x$ dependence for the second beam type that 
we consider is given by a Cauchy-Lorentz distribution, i.e., it decreases 
slowly (as a power-law) with increasing distance from the beam center. 
The initial beam profile in the transverse direction is taken as Gaussian, 
as this choice enables the explicit calculation of the integral with respect 
to $y$ on the right hand side of Eq. (\ref{lp46}). Therefore, the two 
initial conditions for the two-beam collision problem are 
\begin{eqnarray}&&
\psi_{1}(x,y,0)=A_{1}(0)\exp \left[-\frac{x^{2}}{2W^{(x)2}_{10}}
-\frac{y^{2}}{2W^{(y)2}_{10}} + i\alpha_{10} \right],
\nonumber \\&&
\psi_{2}(x,y,0)=A_{2}(0)\exp \left[-\frac{(x-x_{20})^{2}}{2W^{(x)2}_{20}}
-\frac{y^{2}}{2W^{(y)2}_{20}} + i\alpha_{20} \right], 
\label{lp61}
\end{eqnarray}  
for Gaussian beams, and      
 \begin{eqnarray}&&
\psi_{1}(x,y,0)=A_{1}(0)\left[1 + \frac{2x^{4}}{W^{(x)4}_{10}} \right]^{-1}
\exp \left[-\frac{y^{2}}{2W^{(y)2}_{10}} + i\alpha_{10} \right],
\nonumber \\&&
 \psi_{2}(x,y,0)=A_{2}(0)\left[1 + \frac{2(x-x_{20})^{4}}{W^{(x)4}_{20}} \right]^{-1}\exp \left[-\frac{y^{2}}{2W^{(y)2}_{20}} + i\alpha_{20} \right],
\label{lp62}
\end{eqnarray}    
for Cauchy-Lorentz-Gaussian beams.

Let us obtain the expression for the collision-induced amplitude shift 
for the initial conditions (\ref{lp61}) and (\ref{lp62}). From Eq. (\ref{appD_10}) 
in Appendix \ref{appendD} it follows that for both initial conditions 
\begin{eqnarray}&&
G_{j}^{(y)}(y,z_{c})=
\frac{W_{j0}^{(y)}}
{(W_{j0}^{(y)4} + 4z_{c}^{2})^{1/4}}
\exp\left[-\frac{W_{j0}^{(y)2}y^{2}}{2(W_{j0}^{(y)4} + 4z_{c}^{2})} \right], 
\label{lp63}
\end{eqnarray}  
where $j=1,2$. In addition, $c_{p1}^{(y)}=\pi^{1/2}$ for both initial conditions, 
$c_{p2}^{(x)}=\pi^{1/2}$ for Gaussian beams, and $c_{p2}^{(x)}=3\pi/2^{11/4}$
for Cauchy-Lorentz-Gaussian beams. 
We now substitute Eq. (\ref{lp63}) along with the values of $c_{p1}^{(y)}$ 
and $c_{p2}^{(x)}$ into Eq. (\ref{lp46}), and perform the integration 
with respect to $y$. This calculation yields the following expression for 
$\Delta A_{1}^{(c)}$: 
\begin{eqnarray}&&
\Delta A_{1}^{(c)}=
-\frac{2b\epsilon_{3} A_{1}(z_{c}^{-})A_{2}^{2}(z_{c}^{-})}{|d_{11}|}
\nonumber \\&&
\times 
\frac{W_{10}^{(y)}W_{20}^{(x)}W_{20}^{(y)2}}
{(W_{10}^{(y)2} + W_{20}^{(y)2})^{1/2}
(W_{10}^{(y)2}W_{20}^{(y)2} + 4z_{c}^{2})^{1/2}},
\label{lp64}     
\end{eqnarray} 
where $b=\pi^{1/2}$ for Gaussian beams, and $b=3\pi/2^{11/4}$ for 
Cauchy-Lorentz-Gaussian beams. The longitudinal part in the expression 
for $\Delta A_{1}^{(c)}$, $c_{p2}^{(x)}W_{20}^{(x)}=bW_{20}^{(x)}$, is clearly universal. 
In contrast, the transverse part in the expression, which is given by: 
\begin{eqnarray} &&
\!\!\!\!\!\!\!\!\!
\mbox{transverse factor}=
\frac{W_{10}^{(y)}W_{20}^{(y)2}}
{(W_{10}^{(y)2} + W_{20}^{(y)2})^{1/2}
(W_{10}^{(y)2}W_{20}^{(y)2} + 4z_{c}^{2})^{1/2}}, 
\label{lp65}
\end{eqnarray}                                   
depends on $z_{c}$, and does not seem to have a simple universal form. 
One aspect of the nonuniversal nature of the expression for $\Delta A_{1}^{(c)}$ 
in spatial dimension 2 is the deviation of the dependence on $|d_{11}|$ from the $1/|d_{11}|$ 
scaling, which is observed in the one-dimensional case \cite{PNH2017B,NHP2020}, 
and also in fast collisions between NLS solitons in the presence of nonlinear dissipation 
in spatial dimension 1 \cite{PNC2010,PC2012}. Note that the collision distance $z_{c}$ satisfies 
$z_{c}=(x_{10}-x_{20})/d_{11}$. Therefore, the deviation of the $|d_{11}|$ 
dependence of $\Delta A_{1}^{(c)}$ from the $1/|d_{11}|$ scaling is due to 
the term $4(x_{10}-x_{20})^2/d_{11}^{2}$ inside the factor 
$(W_{10}^{(y)2}W_{20}^{(y)2} + 4(x_{10}-x_{20})^2/d_{11}^{2})^{1/2}$ 
on the right hand side of Eq. (\ref{lp64}). 
It is useful to define the quantity $\Delta A_{1}^{(c)(s)}$, 
which is the approximate expression for the amplitude shift that is obtained 
from the full expression by neglecting the $(x_{10}-x_{20})^2/d_{11}^{2}$ 
term. Carrying out the latter approximation, we find:                        
\begin{eqnarray}&&
\Delta A_{1}^{(c)(s)}=
-\frac{2b\epsilon_{3} A_{1}(z_{c}^{-})A_{2}^{2}(z_{c}^{-})}{|d_{11}|}
\frac{W_{20}^{(x)}W_{20}^{(y)}}
{(W_{10}^{(y)2} + W_{20}^{(y)2})^{1/2}}. 
\label{lp66}     
\end{eqnarray}                                                                                                       
Therefore, the difference $|\Delta A_{1}^{(c)} - \Delta A_{1}^{(c)(s)}|$
is a measure for the deviation of the $d_{11}$ dependence of $\Delta A_{1}^{(c)}$ 
from the $1/|d_{11}|$ scaling observed in the one-dimensional case. 
Since in a complete collision $|x_{20}-x_{10}| \gg 1$,  
the term $4(x_{10}-x_{20})^2/d_{11}^{2}$ is not necessarily 
small for intermediate values of $|d_{11}|$. As a result, the deviation from the 
$1/|d_{11}|$ scaling might be significant even for intermediate $|d_{11}|$ values.

To check the perturbation theory predictions for universality of the longitudinal 
part in the expression for $\Delta A_{1}^{(c)}$, we carry out numerical simulations 
with Eq. (\ref{lp1}) with the two initial conditions (\ref{lp61}) and (\ref{lp62}), 
which possess widely different initial beam profiles in the longitudinal direction. 
The extensive simulations with these initial conditions provide a careful test for the 
validity of the perturbation theory approximations for widely different beam shapes, 
and in this manner, help confirm the universality of the longitudinal part 
in the expression for $\Delta A_{1}^{(c)}$. We carry out the simulations for 
$d_{11}$ values in the intervals $4 \le |d_{11}| \le 60$. For concreteness, 
we present the results of the simulations with $\epsilon_{3}=0.01$.     
The parameter values of the initial conditions (\ref{lp61}) and (\ref{lp62}) are 
$A_{j}(0)=1$, $\alpha_{j0}=0$, $x_{20}=\pm 20$, $W_{10}^{(x)}=3$, $W_{10}^{(y)}=5$, 
$W_{20}^{(x)}=4$, and $W_{20}^{(y)}=6$. The final propagation distance 
is $z_{f}=2z_{c}=-2x_{20}/d_{11}$. The values of $x_{20}$ and $z_{f}$ 
ensure that the centers of the two beams are well separated at $z=0$ and at $z=z_{f}$. 
We point out that results similar to the ones described 
in the current subsection are obtained in simulations with other parameter values. 
For each initial condition, we compare the dependence of $\Delta A_{1}^{(c)}$ 
on $d_{11}$ obtained in the simulations with the perturbation theory prediction 
of Eq. (\ref{lp64}), and with the more crude approximation $\Delta A_{1}^{(c)(s)}$  
of Eq. (\ref{lp66}). We also discuss the behavior of the relative errors 
in the approximation of $\Delta A_{1}^{(c)}$ (in percentage), which are defined 
by $E_{r}^{(1)}=|\Delta A_{1}^{(c)(num)}-\Delta A_{1}^{(c)(th)}|
\times 100/|\Delta A_{1}^{(c)(th)}|$ and 
$E_{r}^{(2)}=|\Delta A_{1}^{(c)(num)}-\Delta A_{1}^{(c)(s)}|
\times 100/|\Delta A_{1}^{(c)(s)}|$, respectively.

\begin{figure}[ptb]
\begin{center}
\epsfxsize=10cm  \epsffile{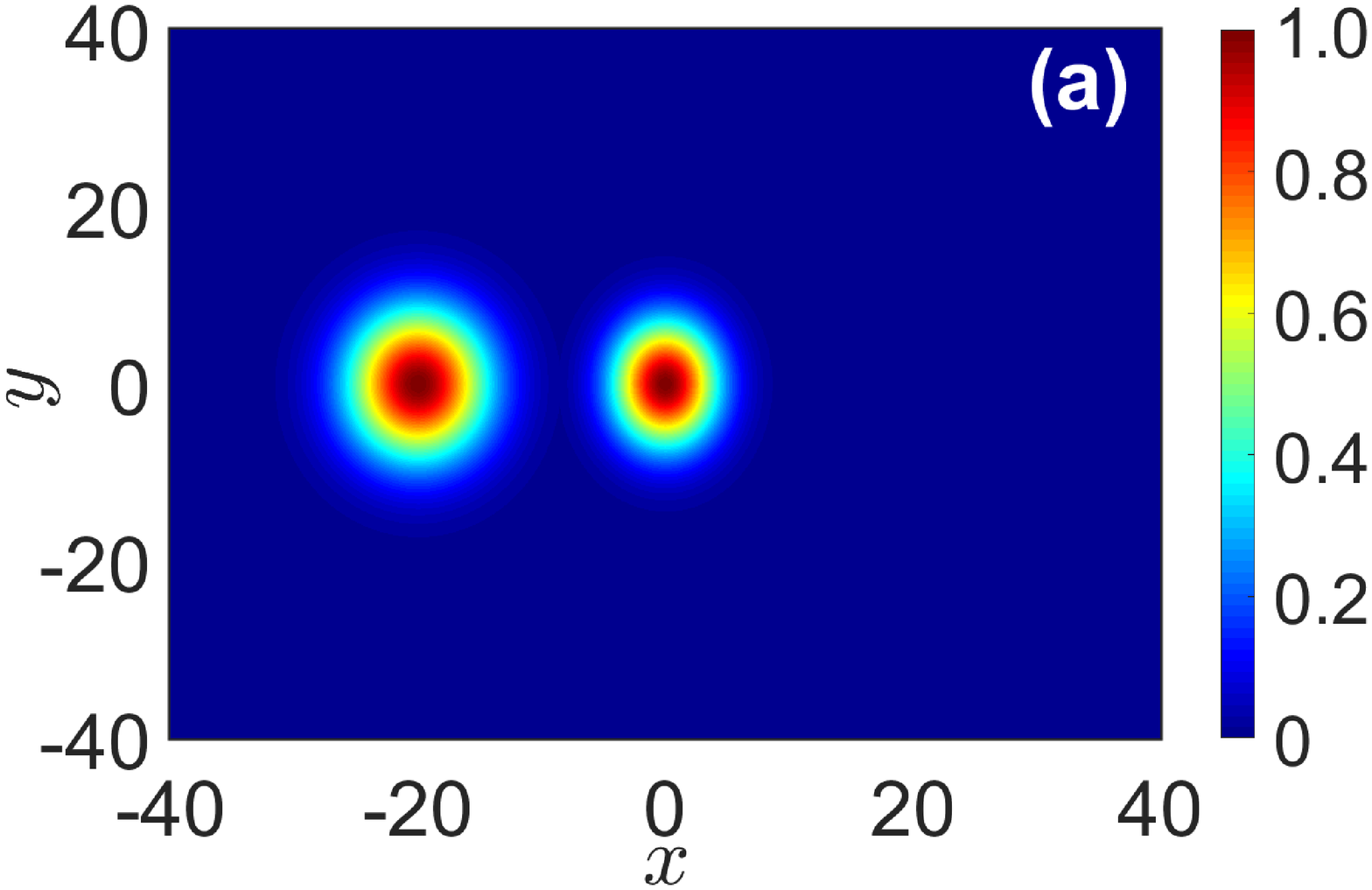} \\
\epsfxsize=10cm  \epsffile{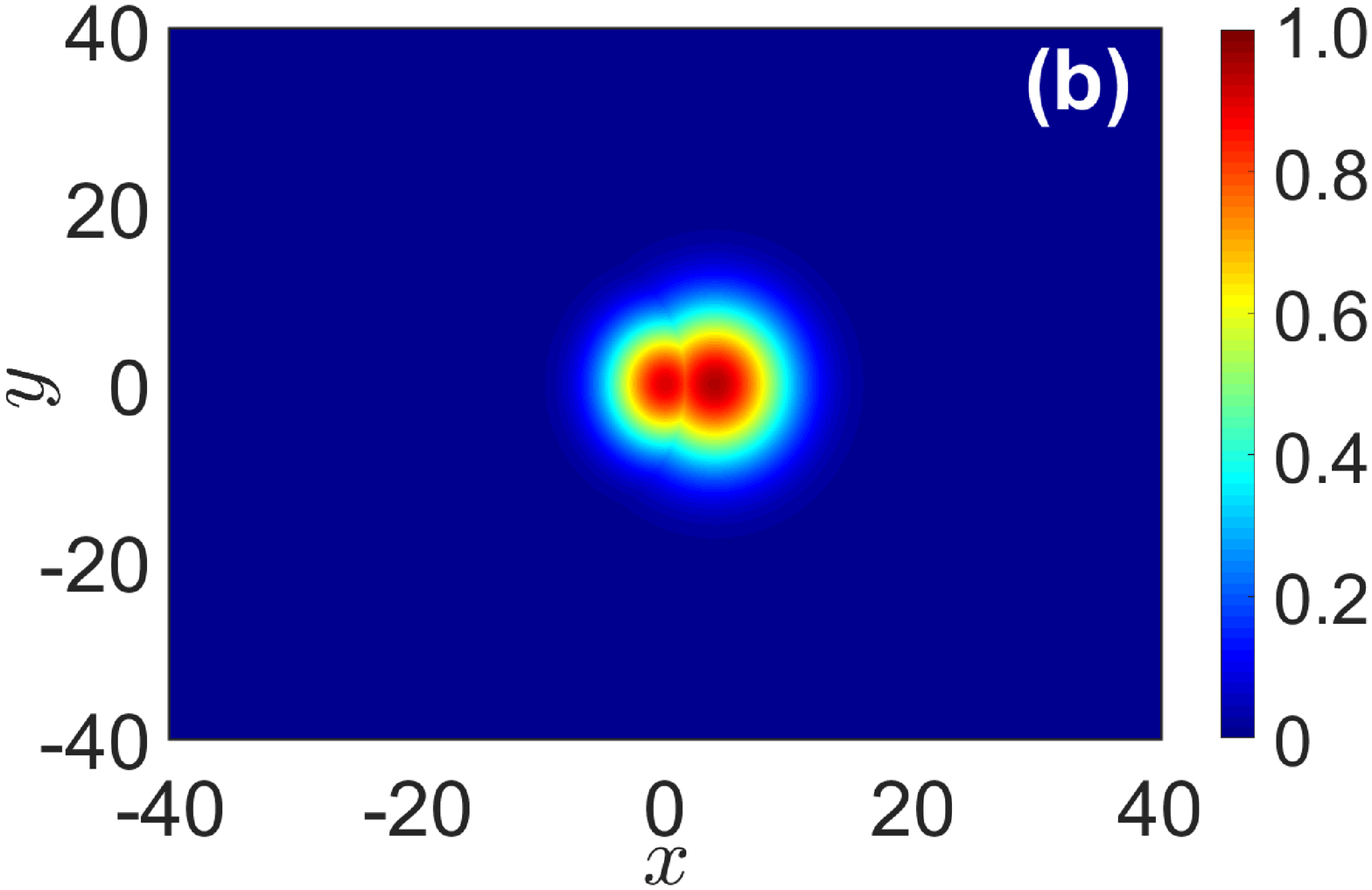} \\
\epsfxsize=10cm  \epsffile{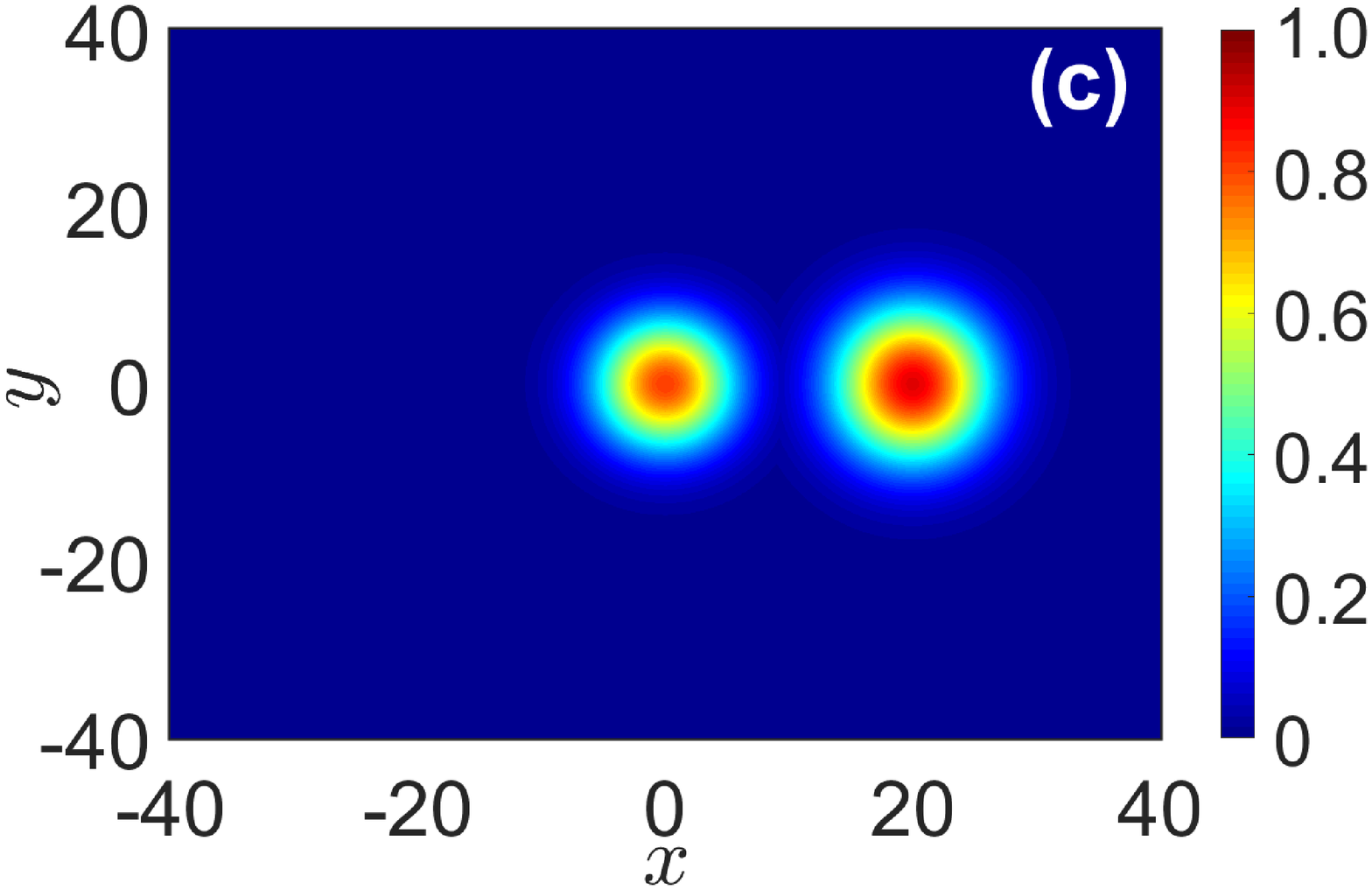} 
\end{center}
\caption{(Color online) 
Contour plots of the beam shapes $|\psi_{j}(x,y,z)|$ at $z=0$ (a), 
$z=z_{i}=2.4$ (b), and $z=z_{f}=4$ (c) in a fast collision between 
two Gaussian beams with parameter values $\epsilon_{3}=0.01$ and $d_{11}=10$.
The plots represent the beam shapes obtained by numerical solution 
of Eq. (\ref{lp1}) with the initial condition (\ref{lp61}).}
\label{fig1}
\end{figure}

\begin{figure}[ptb]
\begin{center}
\epsfxsize=10cm \epsffile{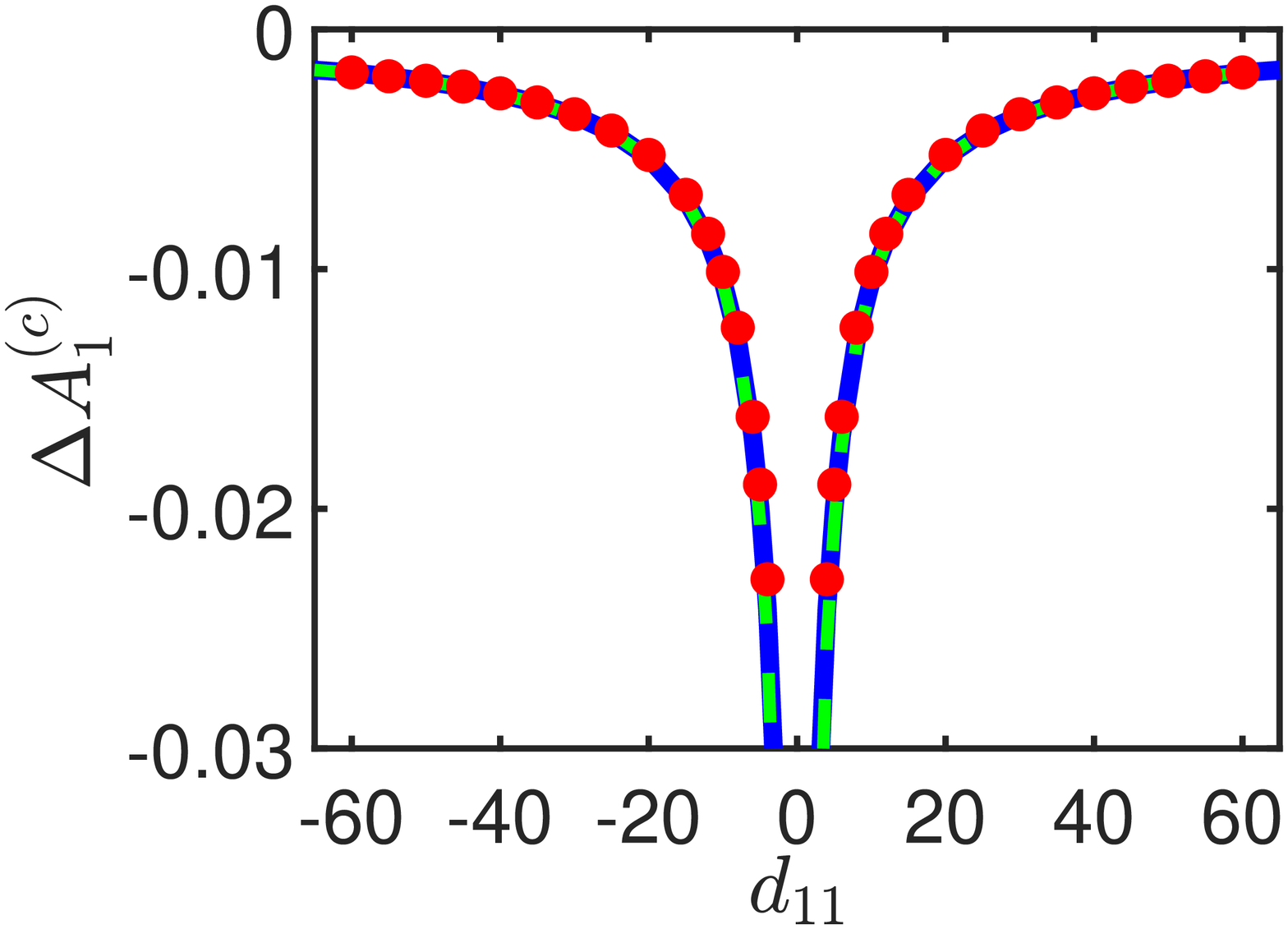}
\end{center}
\caption{(Color online) 
Dependence of the collision-induced amplitude shift of beam 1 
$\Delta A_{1}^{(c)}$ on the beam-steering coefficient $d_{11}$ 
in a fast collision between two Gaussian beams for 
$\epsilon_{3}=0.01$. The red circles represent 
the result obtained by numerical simulations with Eq. (\ref{lp1}) 
with the initial condition (\ref{lp61}). 
The solid blue and dashed green curves represent the theoretical   
predictions of Eqs. (\ref{lp64}) and (\ref{lp66}), respectively.}
\label{fig2}
\end{figure}

We start by discussing the results of the simulations for Gaussian beams, 
which represent beams with rapidly decaying tails. 
The initial beam shapes $|\psi_{j}(x,y,0)|$, and the beam 
shapes $|\psi_{j}(x,y,z)|$ obtained in the simulation with $d_{11}=10$
at the intermediate distance $z_{i}=2.4>z_{c}$ \cite{zi_values}, and at 
the final distance $z_{f}=4$ are shown in Fig. \ref{fig1}. We observe that 
the beams undergo broadening due to diffraction without developing 
significant tails. In addition, the maximum values of $|\psi_{j}(x,y,z)|$ 
decrease with increasing $z$, mainly due to diffraction. The dependence of 
$\Delta A_{1}^{(c)}$ on $d_{11}$ that is obtained in the simulations is 
shown in Fig. \ref{fig2} along with the analytic prediction of Eq. (\ref{lp64}) 
and the more crude approximation $\Delta A_{1}^{(c)(s)}$ of Eq. (\ref{lp66}). 
We observe that despite the diffraction-induced broadening of the beams, 
the agreement between the simulations result and the analytic prediction 
of Eq. (\ref{lp64}) is very good. In particular, the relative error $E_{r}^{(1)}$ 
is less than $3.5\%$ for $10 \le |d_{11}| \le 60$ and less than 
$5.1\%$ for $ 4 \le |d_{11}| <10$. We also observe that the two theoretical 
curves for $\Delta A_{1}^{(c)}$ and $\Delta A_{1}^{(c)(s)}$ are close 
to each other. As a result, the relative error $E_{r}^{(2)}$ is only somewhat 
larger than $E_{r}^{(1)}$. More specifically, the value of $E_{r}^{(2)}$ is 
smaller than $4.4\%$ for $ 10 \le |d_{11}| \le 60$ and smaller than 
$10.0\%$ for $ 4 \le |d_{11}| <10$. This means that the deviation of the 
$d_{11}$ dependence of $\Delta A_{1}^{(c)}$ from the $1/|d_{11}|$ scaling 
is not significant for $|d_{11}| \ge 10$. However, it should be noted that the 
latter result is due to the choice of the values of $W_{10}^{(y)}$ and 
$W_{20}^{(y)}$, i.e., for smaller values of these parameters we can observe 
significantly larger deviations from the $1/|d_{11}|$ scaling.

\begin{figure}[ptb]
\begin{center}
\epsfxsize=10cm  \epsffile{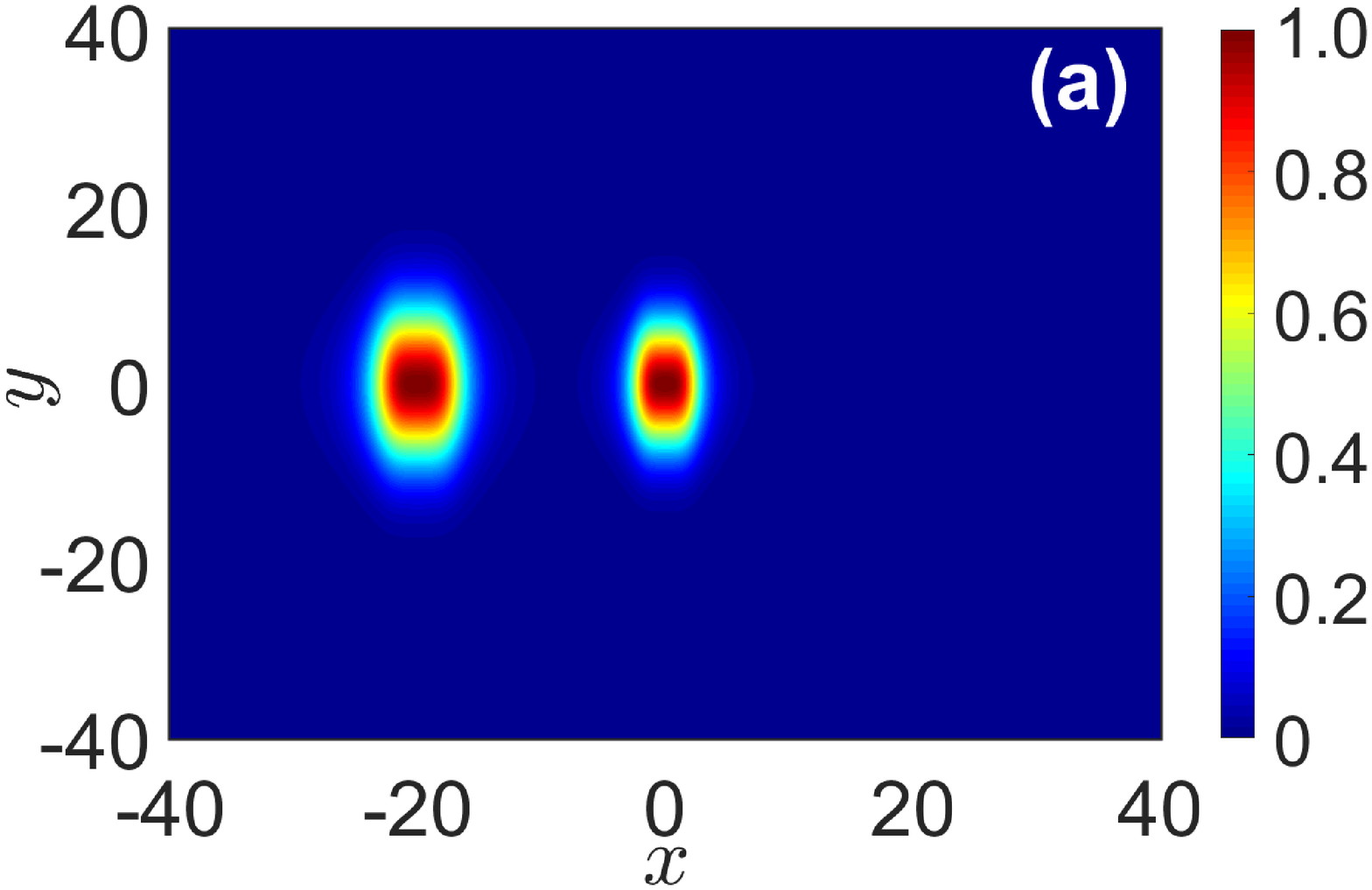} \\
\epsfxsize=10cm  \epsffile{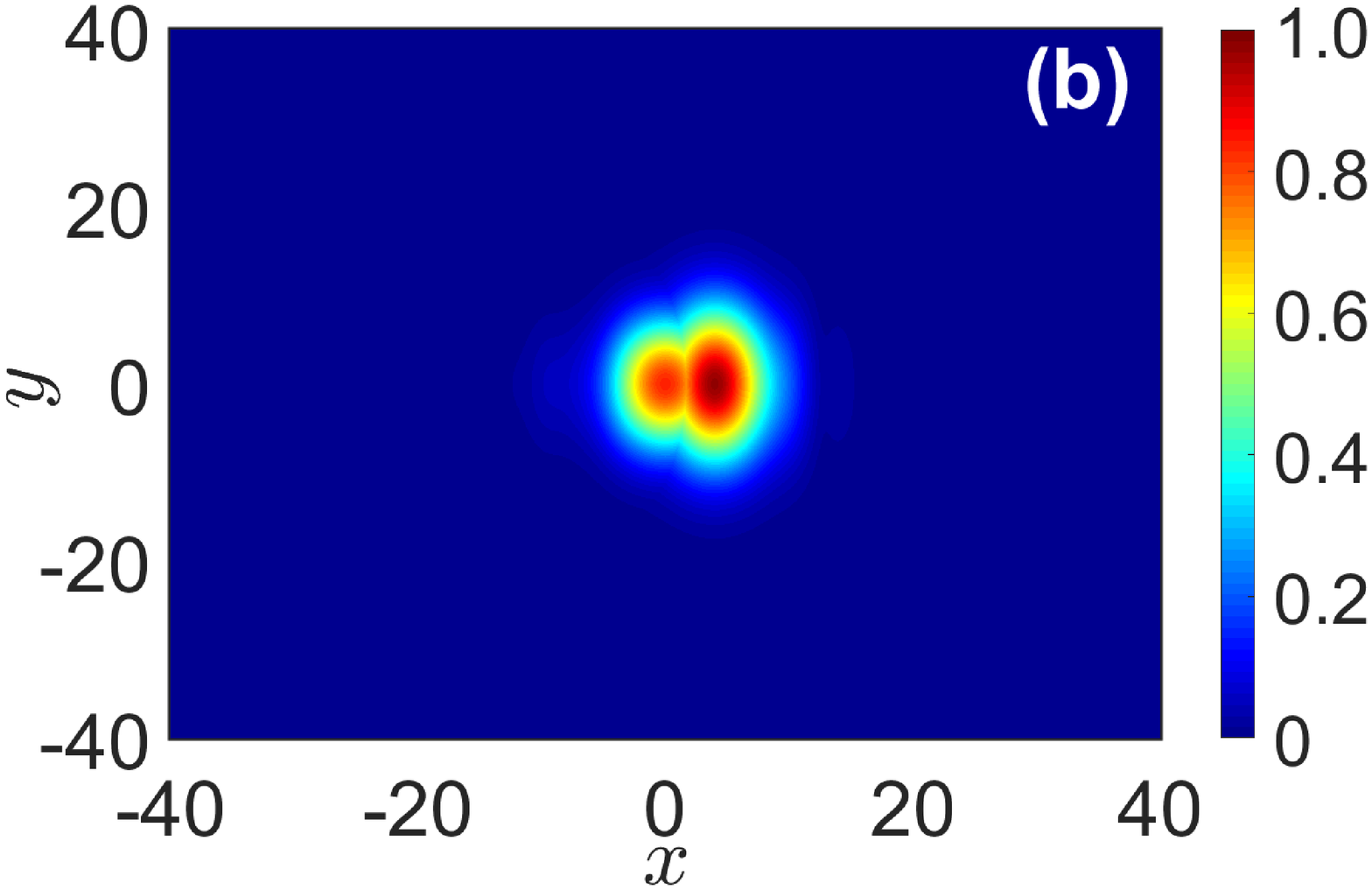} \\
\epsfxsize=10cm  \epsffile{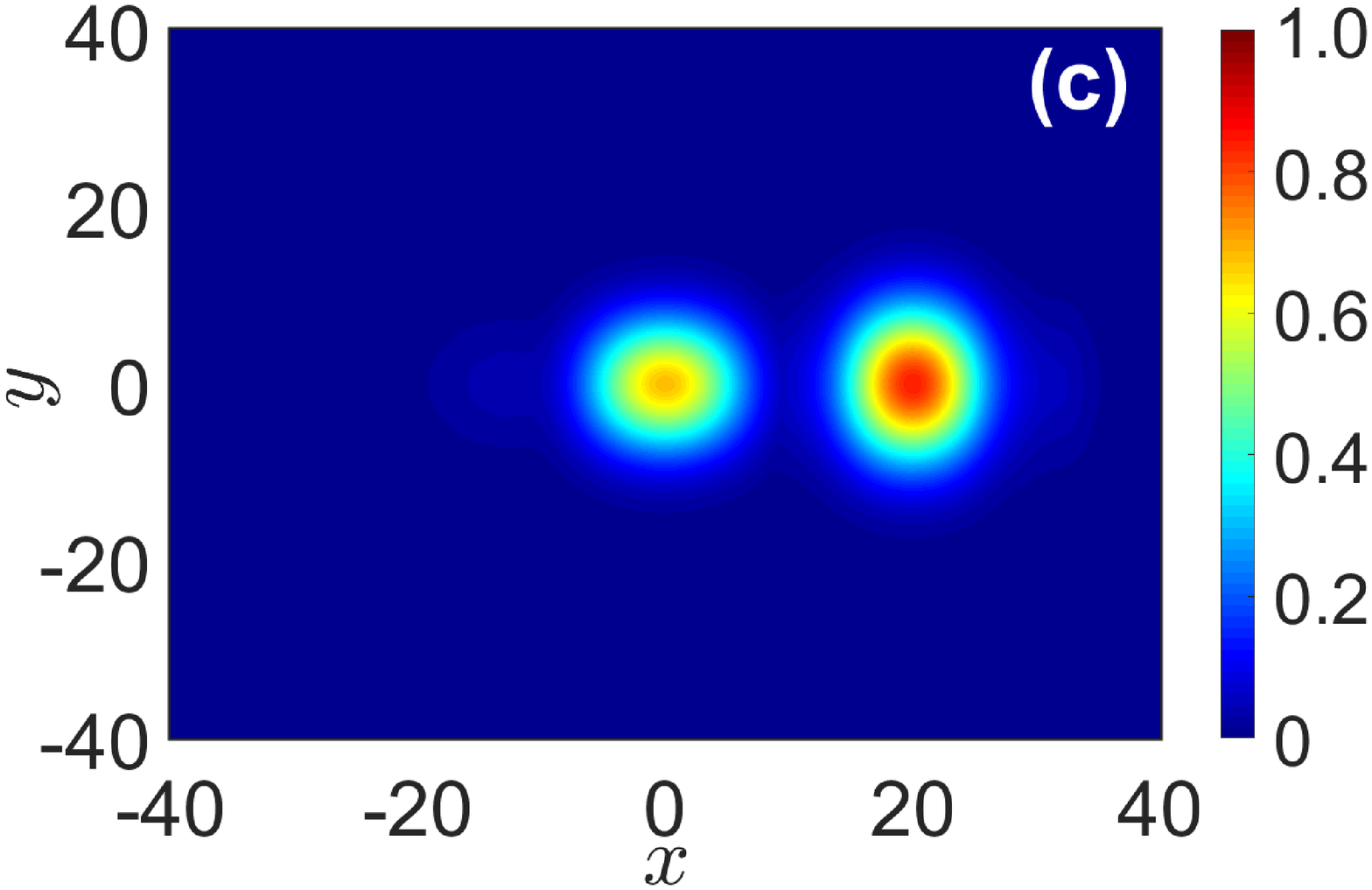} 
\end{center}
\caption{(Color online) 
Contour plots of the beam shapes $|\psi_{j}(x,y,z)|$ at $z=0$ (a), 
$z=z_{i}=2.4$ (b), and $z=z_{f}=4$ (c) in a fast collision between 
two Cauchy-Lorentz-Gaussian beams with parameter values 
$\epsilon_{3}=0.01$ and $d_{11}=10$. 
The plots represent the beam shapes obtained by numerical solution 
of Eq. (\ref{lp1}) with the initial condition (\ref{lp62}).}
\label{fig3}
\end{figure}

\begin{figure}[ptb]
\begin{center}
\epsfxsize=10cm \epsffile{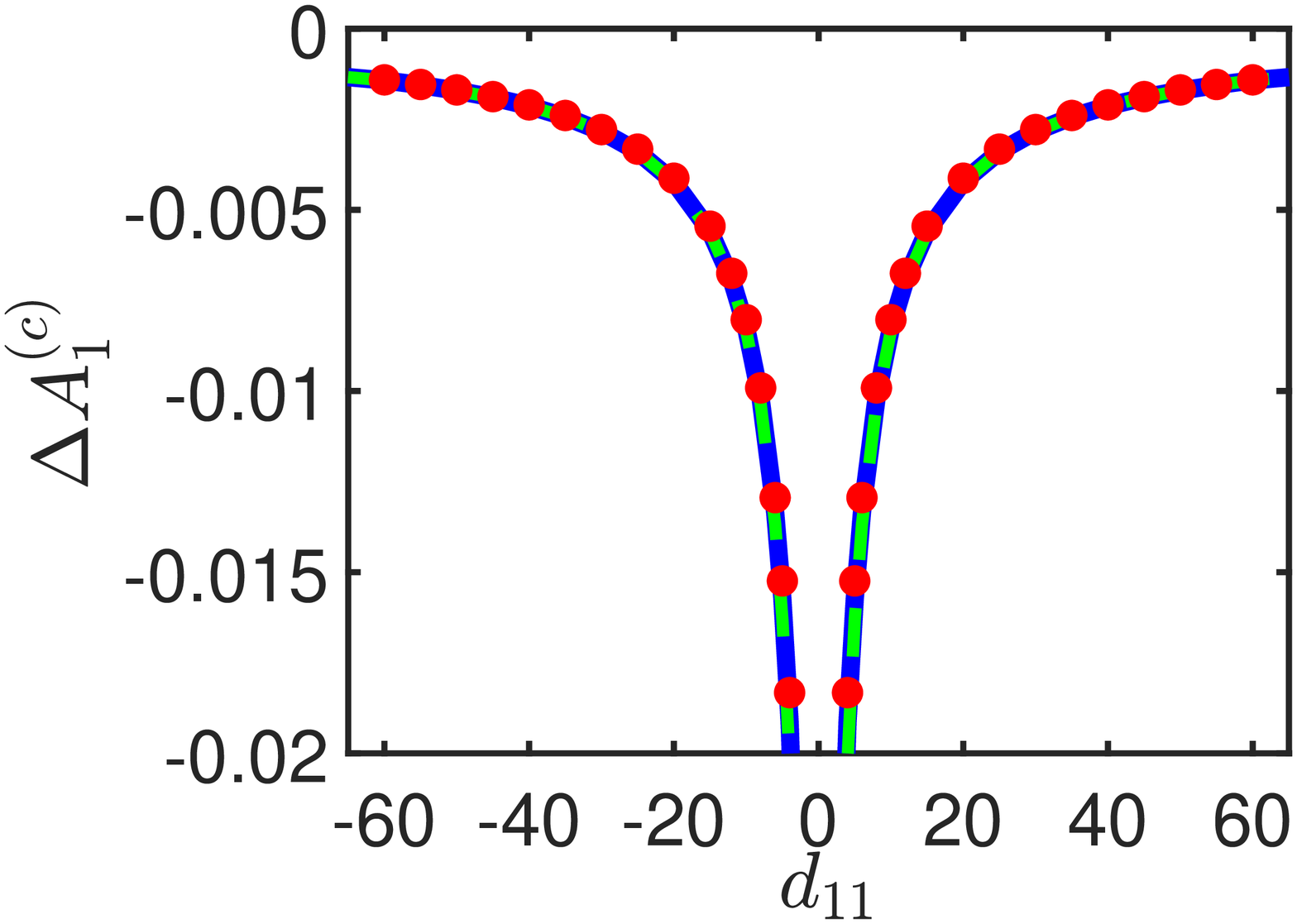}
\end{center}
\caption{(Color online) 
Dependence of the collision-induced amplitude shift of beam 1 
$\Delta A_{1}^{(c)}$ on the beam-steering coefficient $d_{11}$ 
in a fast collision between two Cauchy-Lorentz-Gaussian beams for $\epsilon_{3}=0.01$. 
The red circles represent the result obtained by numerical simulations 
with Eq. (\ref{lp1}) with the initial condition (\ref{lp62}). 
The solid blue and dashed green curves represent the perturbation 
theory predictions of Eqs. (\ref{lp64}) and (\ref{lp66}), respectively.}
\label{fig4}
\end{figure}

We now turn to discuss the simulations results for beams, 
whose tails exhibit slow decay in the longitudinal direction. 
For such beams it is unclear if the sharp-peak approximation, 
which is used in the derivation of Eq. (\ref{lp10}) from Eq. (\ref{lp8}), 
is valid. Therefore, in this case, the numerical simulations of the 
two-beam collision serve as an important check of the perturbation 
theory predictions. We use the Cauchy-Lorentz-Gaussian beams 
of Eq. (\ref{lp62}) as prototypical examples for beams, whose 
tails exhibit slow (power-law) decay with increasing distance 
from the beam center. Figure \ref{fig3} shows the initial beam shapes 
$|\psi_{j}(x,y,0)|$, and the beam shapes $|\psi_{j}(x,y,z)|$ obtained 
in the simulation with $d_{11}=10$ at $z=z_{i}=2.4$, and at 
$z=z_{f}=4$. We observe that the beams experience broadening 
and develop extended tails due to diffraction. In addition, the maximum 
values of $|\psi_{j}(x,y,z)|$ decrease with increasing $z$, mainly due 
to diffraction. The latter decrease is especially noticeable for beam 1. 
This can be explained by noting that $W_{10}^{(x)} < W_{20}^{(x)}$, 
and as a result, diffraction-induced beam broadening and generation of 
extended tails are stronger for beam 1 compared with beam 2. 
The $d_{11}$ dependence of $\Delta A_{1}^{(c)}$ obtained in 
the simulations is shown in Fig. \ref{fig4} together with the analytic predictions 
$\Delta A_{1}^{(c)}$ and $\Delta A_{1}^{(c)(s)}$ of Eqs. (\ref{lp64}) 
and  (\ref{lp66}). The agreement between the simulations result for 
$\Delta A_{1}^{(c)}$ and the perturbation theory prediction of 
Eq. (\ref{lp64}) is very good despite the beam broadening and the 
generation of extended beam tails. In particular, the value of the 
relative error $E_{r}^{(1)}$ is smaller than $2.8\%$ for 
$10 \le |d_{11}| \le 60$ and smaller than $5.1\%$ for $4 \le |d_{11}| <10$. 
Note that these values are comparable to the values of $E_{r}^{(1)}$ 
for collisions between Gaussian beams. Thus, based on the results of our 
simulations, we conclude that the longitudinal part in the expression for 
$\Delta A_{1}^{(c)}$, $c_{p2}^{(x)}W_{20}^{(x)}$, is indeed universal 
in the sense that it is not sensitive to the exact form of the initial beam shapes. 
We also note that the value of $E_{r}^{(2)}$ is smaller than $4.3\%$ for 
$10 \le |d_{11}| \le 60$ and smaller than $9.2\%$ for $4 \le |d_{11}| <10$. 
Thus, the deviation of the $d_{11}$ dependence of $\Delta A_{1}^{(c)}$ 
from the $1/|d_{11}|$ scaling is not significant for $|d_{11}| \ge 10$ 
for the parameter values used in our simulations. 

\subsection{Fast collisions between partially overlapping beams} 
\label{simu_overlap}

Another important property of a complete two-beam collision 
is related to the relative location of the beam centers at the 
collision distance $z_{c}$. When the $y$ coordinates of the beams 
are equal at $z_{c}$, $y_{1}(z_{c})=y_{2}(z_{c})$, we say that the 
beams are fully overlapping at $z_{c}$. In contrast, when the 
$y$ coordinates of the beams are not equal at $z_{c}$, 
$y_{1}(z_{c}) \ne y_{2}(z_{c})$, we say that the 
beams are only partially overlapping at $z_{c}$. 
It is clear that complete collisions between two partially 
overlapping beams exist only in spatial dimension higher 
than 1, since in spatial dimension 1, the two beams 
are always fully overlapping at $z_{c}$ in a complete fast collision. 
It is therefore interesting to employ the perturbation theory for 
studying the effect of the partial overlap between the colliding 
beams in 2D on the collision-induced amplitude shift. 
This problem is of further interest, since we can use it for checking 
the perturbation theory prediction for the transverse 
part in the expression for the amplitude shift in a nontrivial setup. 
Thus, in the current subsection, we investigate the dynamics of the 
amplitude shift in fast collisions between partially overlapping beams 
both analytically and by numerical simulations with Eq. (\ref{lp1}).

To demonstrate the effects of partial overlap on the collision-induced 
amplitude shift, we consider an initial condition in the form of two 
Gaussian beams with different initial values of the $y$ coordinates 
of the beam centers, $y_{10} \ne y_{20}$. For simplicity and without 
loss of generality, we assume that the initial beam widths satisfy 
$W_{10}^{(x)}=W_{20}^{(x)} \equiv W_{0}^{(x)}$ and 
$W_{10}^{(y)}=W_{20}^{(y)} \equiv W_{0}^{(y)}$. Therefore, 
the initial condition for the two-beam collision problem is given by: 
\begin{eqnarray}&&
\psi_{j}(x,y,0)=A_{j}(0)\exp \left[-\frac{(x-x_{j0})^{2}}{2W^{(x)2}_{0}}
-\frac{(y-y_{j0})^{2}}{2W^{(y)2}_{0}} + i\alpha_{j0}\right], 
\label{lp71}
\end{eqnarray}  
for $j=1,2$. Since the initial condition is separable for both beams, 
we can use Eq. (\ref{lp46}) for calculating the collision-induced 
amplitude shift. For this initial condition, $c_{p1}^{(y)}=c_{p2}^{(x)}=\pi^{1/2}$, 
and the functions $G_{j}^{(y)}$ are given by Eq. (\ref{lp63})
with $W_{10}^{(y)}=W_{20}^{(y)}=W_{0}^{(y)}$. 
Substituting these expressions into Eq. (\ref{lp46}) and 
integrating with respect to $y$, we obtain: 
\begin{eqnarray}&&
\Delta A_{1}^{(c)}=
-\frac{(2\pi)^{1/2}\epsilon_{3} A_{1}(z_{c}^{-})A_{2}^{2}(z_{c}^{-})}{|d_{11}|}
\nonumber \\&&
\times 
\frac{W_{0}^{(x)}W_{0}^{(y)2}}
{(W_{0}^{(y)4} + 4z_{c}^{2})^{1/2}}
\exp\left[-\frac{W_{0}^{(y)2}(y_{20}-y_{10})^{2}}{2(W_{0}^{(y)4} 
+ 4z_{c}^{2})} \right]. 
\label{lp72}     
\end{eqnarray} 
Thus, the effect of partial beam overlap on the amplitude shift 
is contained in the transverse part of the expression for $\Delta A_{1}^{(c)}$: 
\begin{eqnarray} &&
\!\!\!\!\!\!\!\!\!
\mbox{transverse factor}=
\frac{W_{0}^{(y)2}}{2^{1/2}
(W_{0}^{(y)4} + 4z_{c}^{2})^{1/2}}
\exp\left[-\frac{W_{0}^{(y)2}(y_{20}-y_{10})^{2}}{2(W_{0}^{(y)4} 
+4z_{c}^{2})} \right].
\nonumber \\&&
\label{lp73}
\end{eqnarray}                
We see that $\Delta A_{1}^{(c)}$ is a Gaussian function of 
the separation between the beam centers at $z_{c}$, $y_{20}-y_{10}$. 
The width of the Gaussian function is equal to $2^{1/2}W^{(y)}(z_{c})$, 
where $W^{(y)}(z_{c})=(W_{0}^{(y)2} + 4z_{c}^{2}/W_{0}^{(y)2})^{1/2}$ 
is the width of both beams in the transverse direction at $z_{c}$.

It is unclear if the approximations used by the perturbation theory 
hold when the separation between the beam centers at $z_{c}$ is relatively 
large. For this reason, it is important to check the predictions of 
Eqs. (\ref{lp72}) and (\ref{lp73}) by numerical solution of Eq. (\ref{lp1}). 
We take on this important numerical investigation by carrying out 
simulations with Eq. (\ref{lp1}) and by measuring the dependence of 
$\Delta A_{1}^{(c)}$ on $y_{20}-y_{10}$. 
For brevity, we describe the results of these simulations briefly 
without showing the corresponding figures. The physical parameter 
values are $\epsilon_{3}=0.01$ and $d_{11}=20$. 
The initial values of the beam parameters in Eq. (\ref{lp71}) are 
$A_{j}(0)=1$, $\alpha_{j0}=0$, $x_{10}=0$, $x_{20}=-20$, $y_{10}=0$, $W_{0}^{(x)}=4$, 
$W_{0}^{(y)}=5$, and the value of $y_{20}$ is varied in the interval 
$-10 \le y_{20} \le 10$. The final propagation distance is $z_{f}=2$, 
and the beam centers are well separated at $z=0$ and at $z=z_{f}$. 
In addition to $\Delta A_{1}^{(c)}$, we measure the 
relative error (in percentage) $|\Delta A_{1}^{(c)(num)}-\Delta A_{1}^{(c)(th)}|
\times 100/|\Delta A_{1}^{(c)(th)}|$, where $\Delta A_{1}^{(c)(th)}$ 
is given by Eq. (\ref{lp72}). The agreement between the numerical simulations 
result and the perturbation theory prediction is very good. 
More specifically, the relative error is smaller than $7.9\%$ in 
the entire interval $-10 \le y_{20} \le 10$. 
Thus, our numerical simulations confirm the prediction of Eqs. (\ref{lp72}) 
and (\ref{lp73}) for Gaussian dependence of $\Delta A_{1}^{(c)}$ 
and its transverse factor on $y_{20}-y_{10}$. Furthermore, 
these simulations demonstrate that the perturbation method 
of subsections \ref{lp_general_IC} and \ref{lp_separable_IC} is 
applicable for fast collisions between partially overlapping beams, 
even when the beam centers are relatively far from each other at the 
collision distance $z_{c}$.

\subsection{Dependence of the amplitude shift on the orientation angle 
between the beams} 
\label{simu_angle}

Another phenomenon that exists only in spatial dimension 
higher than 1 is associated with direction dependent 
collision-induced effects, i.e., with effects that exist  
due to some anisotropy in the system. 
In particular, we are interested in studying the effects of anisotropy 
in the initial condition. In a simple setup, this anisotropy 
can be characterized by a single angle, e.g., the angle 
$\theta_{0}$ between a ``preferred'' direction in the 
initial condition and the $x$ axis. To illustrate this situation, 
consider the case where the initial shape of beam 1 $|\psi_{1}(x,y,0)|$ 
is wider along one direction that we denote by $x'$, and narrower 
along the perpendicular direction that we denote by $y'$. 
We can then define the angle $\theta_{0}$, as the angle that 
the $x'$ axis forms with the $x$ axis of our coordinate system. 
Thus, $\theta_{0}$ is the angle between the relative velocity 
vector (between the beam centers) and the $x'$ axis of beam 1. 
In addition, if beam 2 is circularly symmetric, or is elongated 
along the $x$ or the $y$ axes, then $\theta_{0}$ can also be 
regarded as the orientation angle between the two beams. 
An important question about the collision dynamics in this 
anisotropic setup concerns the dependence of the collision-induced 
amplitude shift $\Delta A_{1}^{(c)}$ on the orientation angle $\theta_{0}$. 
In the current subsection, we address this important question by 
both analytic calculations and numerical simulations with Eq. (\ref{lp1}).

\begin{figure}[ptb]
\begin{center}
\epsfxsize=10cm \epsffile{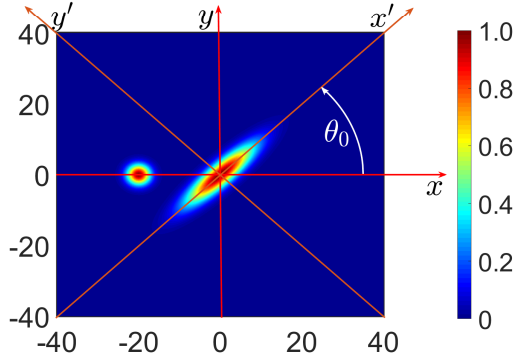}
\end{center}
\caption{(Color online) 
A contour plot of the initial beam shapes $|\psi_{j}(x,y,0)|$ 
for the anisotropic initial condition of Eqs. (\ref{lp82}) and (\ref{lp84}).
In this example, the initial beam widths are $W_{10}^{(x)}=8$, 
$W_{10}^{(y)}=2$, and $W_{20}=2$, and the orientation angle 
is $\theta_{0}=\pi/4$.} 
\label{fig_add1}
\end{figure}

We consider the following anisotropic collision setup, which consists 
of two initially well separated Gaussian beams. In this setup, beam 1 
is elongated along its $x'$ axis, which forms an angle $\theta_{0}$ 
with the $x$ axis, while beam 2 is circular in the $xy$ plane. 
Figure \ref{fig_add1} shows the contour plot of $|\psi_{j}(x,y,0)|$ 
for this initial condition in the case where $\theta_{0}=\pi/4$. 
The initial condition can be written as: 
\begin{eqnarray}&&
\!\!\!\!\!\!
\psi_{1}'(x',y',0)=A_{1}(0)\exp \left[-\frac{x^{\prime 2}}{2W^{(x)2}_{10}}
-\frac{y^{\prime 2}}{2W^{(y)2}_{10}} + i\alpha_{10}\right],
\label{lp81}
\end{eqnarray}  
and 
\begin{eqnarray}&&
\!\!\!\!\!\!
\psi_{2}(x,y,0)=A_{2}(0)\exp \left[-\frac{(x-x_{20})^{2}}{2W^{2}_{20}}
-\frac{y^{2}}{2W^{2}_{20}} + i\alpha_{20}\right], 
\label{lp82}
\end{eqnarray}   
where $\psi_{1}'(x',y',z)$ denotes the electric field of beam 1 
in the $(x',y',z)$ coordinate system, $W_{10}^{(x)} > W_{10}^{(y)}$, 
and 
\begin{eqnarray}&&
x' = x\cos\theta_{0}+ y\sin\theta_{0},
\nonumber \\&&
y' = y\cos\theta_{0} - x\sin\theta_{0}. 
\label{lp83}
\end{eqnarray}  
Substituting relation (\ref{lp83}) into Eq. (\ref{lp81}), 
we obtain: 
\begin{eqnarray}&&
\psi_{1}(x,y,0)=A_{1}(0)\exp\left[-B_{1}x^2 - B_{2}y^2 - B_{3}xy 
+ i\alpha_{10}\right],
\label{lp84}
\end{eqnarray}     
where 
\begin{eqnarray}&&
B_{1}=\frac{\cos^2 \theta_0}{2W_{10}^{(x)2}} + 
\frac{\sin^2 \theta_0}{2W_{10}^{(y)2}}, 
\nonumber
\end{eqnarray}     
\begin{eqnarray}&&
B_{2}=\frac{\sin^2 \theta_0}{2W_{10}^{(x)2}} + 
\frac{\cos^2 \theta_0}{2W_{10}^{(y)2}}, 
\nonumber
\end{eqnarray}     
and  
\begin{eqnarray}&&
B_{3}=\left(\frac{1}{W_{10}^{(x)2}} - 
\frac{1}{W_{10}^{(y)2}} \right)
\sin \theta_0 \cos \theta_0. 
\nonumber
\end{eqnarray}    
Notice that the initial condition for beam 1 is not separable 
in the $(x,y,z)$ coordinate system. Therefore, the investigation described 
in the current subsection also provides an example for 
collision-induced dynamics in a collision with a nonseparable 
initial condition.

The initial condition for beam 1 is nonseparable, and therefore 
we need to calculate $\Delta A_{1}^{(c)}$ by using the general 
expression, which is given by Eq. (\ref{lp13}). It is straightforward 
to show that for the current setup, $C_{p1}=\pi W_{10}^{(x)}W_{10}^{(y)}$. 
Furthermore, since the initial condition for beam 2 is separable, 
we can use Eq. (\ref{lp42}), where $c_{p2}^{(x)}=\pi^{1/2}$.  
Substitution of these relations into Eq. (\ref{lp13}) yields 
\begin{eqnarray} &&
\!\!\!\!
\Delta A_{1}^{(c)}=-\frac{2\epsilon_{3}
A_{1}(z_{c}^{-}) A_{2}^{2}(z_{c}^{-})}{\pi^{1/2}|d_{11}|}
\frac{W_{20}}{W_{10}^{(x)}W_{10}^{(y)}}
\nonumber \\&&
\times
\!\int_{-\infty}^{\infty} \!\!\!\!\! dy \, G_{2}^{(y)2}(y,z_{c})
\!\int_{-\infty}^{\infty} \!\!\!\!\! dx 
\;\tilde\Psi_{10}^{2}(x,y,z_{c}). 
\label{lp85}
\end{eqnarray}         
Since diffraction is isotropic, the unperturbed linear propagation 
equation for beam 1 in the $(x',y',z)$ coordinate system has the same form 
as in the $(x,y,z)$ coordinate system. 
Thus, the unperturbed propagation equation for $\tilde\psi'_{10}$ is: 
\begin{eqnarray}&&
\!\!\!\!\!\!\!
i\partial_z\tilde\psi'_{10} + \partial_{x'}^{2}\tilde\psi'_{10}
+ \partial_{y'}^{2}\tilde\psi'_{10}=0.
\label{lp86}
\end{eqnarray}    
Therefore, we can calculate $\tilde\Psi_{10}(x,y,z)$ by solving 
Eq. (\ref{lp86}) with the initial condition (\ref{lp81}) in the $(x',y',z)$ 
coordinate system, and by using Eq. (\ref{lp83}) to express the 
solution in the $(x,y,z)$ coordinate system. 
The solution of Eq. (\ref{lp86}) with the Gaussian initial condition (\ref{lp81})  
is described in Appendix \ref{appendD}. Using Eqs. (\ref{appD_8})-(\ref{appD_11}) in this 
Appendix, we obtain: 
\begin{eqnarray}&&
\tilde\Psi'_{10}(x',y',z)=
\frac{W_{10}^{(x)}W_{10}^{(y)}}
{(W_{10}^{(x)4} + 4z^{2})^{1/4}(W_{10}^{(y)4} + 4z^{2})^{1/4}}
\nonumber \\&&
\times
\exp\left[-\frac{W_{10}^{(x)2}x'^{2}}{2(W_{10}^{(x)4} + 4z^{2})} 
-\frac{W_{10}^{(y)2}y'^{2}}{2(W_{10}^{(y)4} + 4z^{2})}\right].
\label{lp87}
\end{eqnarray}   
Using the transformation relations (\ref{lp83}) in Eq. (\ref{lp87}),   
and using Eq. (\ref{lp63}) with $W_{20}^{(y)}=W_{20}$ 
for $G_{2}^{(y)}(y,z_{c})$, we obtain: 
\begin{eqnarray}&&
\!\!\!\!\!\!\!\!\!\!\!\!\!\!\!
G_{2}^{(y)2}(y,z_{c})\tilde\Psi_{10}^{2}(x,y,z_{c})=
\frac{W_{10}^{(x)2}W_{10}^{(y)2}W_{20}^{2}}
{(W_{10}^{(x)4} + 4z_{c}^{2})^{1/2}(W_{10}^{(y)4} + 4z_{c}^{2})^{1/2}
(W_{20}^{4} + 4z_{c}^{2})^{1/2}}
\nonumber \\&&
\times
\exp\left[-b_{1}^{2}x^2 - 2b_{2}xy - b_{3}^{2}y^2 \right],  
\label{lp88}
\end{eqnarray}
where 
\begin{eqnarray}&&
b_{1} = 
\left(\frac{W_{10}^{(x)2} \cos^2\theta_0}{W_{10}^{(x)4} + 4z_{c}^{2}}
 +\frac{W_{10}^{(y)2} \sin^2\theta_0}{W_{10}^{(y)4} + 4z_{c}^{2}}\right)^{1/2},  
\label{lp89}
\end{eqnarray}
\begin{eqnarray}&&
b_{2} = 
\left(\frac{W_{10}^{(x)2}}{W_{10}^{(x)4} + 4z_{c}^{2}}
 -\frac{W_{10}^{(y)2}}{W_{10}^{(y)4} + 4z_{c}^{2}}\right)
\sin\theta_{0} \cos\theta_{0},  
\label{lp90}
\end{eqnarray}
and 
\begin{eqnarray}&&
b_{3} = 
\left(\frac{W_{20}^{2}}{W_{20}^{4} + 4z_{c}^{2}}
+\frac{W_{10}^{(x)2} \sin^2\theta_0}{W_{10}^{(x)4} + 4z_{c}^{2}}
+\frac{W_{10}^{(y)2} \cos^2\theta_0}{W_{10}^{(y)4} + 4z_{c}^{2}}\right)^{1/2}. 
\label{lp91}
\end{eqnarray}
Substituting Eq. (\ref{lp88}) into Eq. (\ref{lp85}) and carrying out the double 
integration, we obtain the following expression for the collision-induced 
amplitude shift:  
\begin{eqnarray}&&
\Delta A_{1}^{(c)}=
-\frac{2\pi^{1/2}\epsilon_{3} A_{1}(z_{c}^{-})A_{2}^{2}(z_{c}^{-})}{|d_{11}|}
\nonumber \\&&
\times 
\frac{W_{10}^{(x)}W_{10}^{(y)}W_{20}^{3}}
{(W_{10}^{(x)4} + 4z_{c}^{2})^{1/2}(W_{10}^{(y)4} + 4z_{c}^{2})^{1/2}
(W_{20}^{4} + 4z_{c}^{2})^{1/2}
(b_{1}^{2}b_{3}^{2} - b_{2}^{2})^{1/2}},
\label{lp92}     
\end{eqnarray} 
where
\begin{eqnarray}&&
b_{1}^{2}b_{3}^{2} - b_{2}^{2}=
\frac{W_{20}^{2}}{W_{20}^{4} + 4z_{c}^{2}}
\left(\frac{W_{10}^{(x)2} \cos^2\theta_0}{W_{10}^{(x)4} + 4z_{c}^{2}}
+\frac{W_{10}^{(y)2} \sin^2\theta_0}{W_{10}^{(y)4} + 4z_{c}^{2}}\right)
\nonumber \\&&
+ \frac{W_{10}^{(x)2}W_{10}^{(y)2}}
{(W_{10}^{(x)4} + 4z_{c}^{2})
(W_{10}^{(y)4} + 4z_{c}^{2})}. 
\label{lp93}     
\end{eqnarray} 
We see that even the relatively simple anisotropic setup of the two-beam collision 
considered in the current subsection leads to a nontrivial dependence of 
$\Delta A_{1}^{(c)}$ on the orientation angle $\theta_{0}$. This nontrivial 
dependence on $\theta_{0}$ can also be associated with the nonseparable nature 
of the initial condition for beam 1.

\begin{figure}[ptb]
\begin{center}
\epsfxsize=10cm  \epsffile{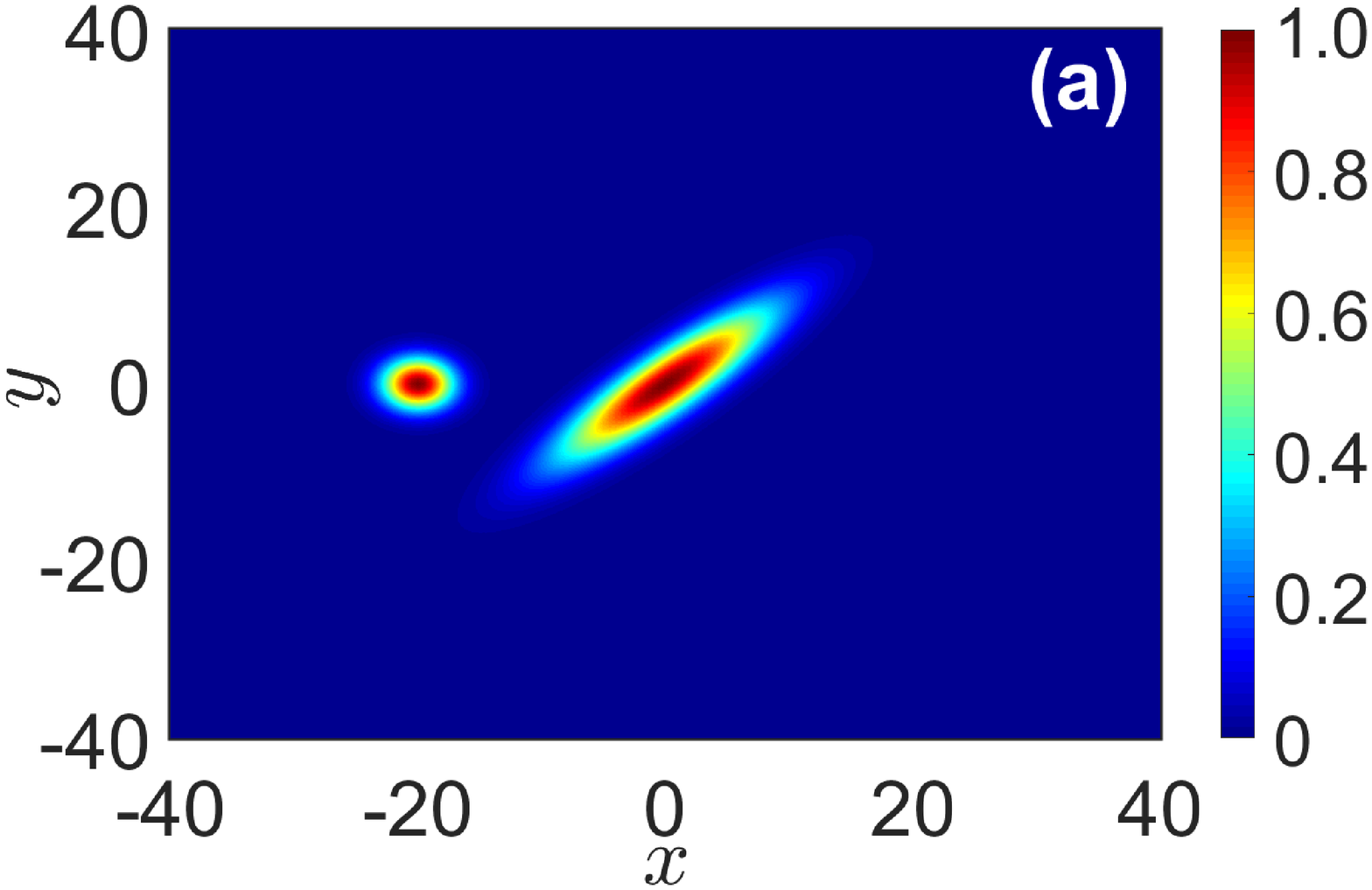} \\
\epsfxsize=10cm  \epsffile{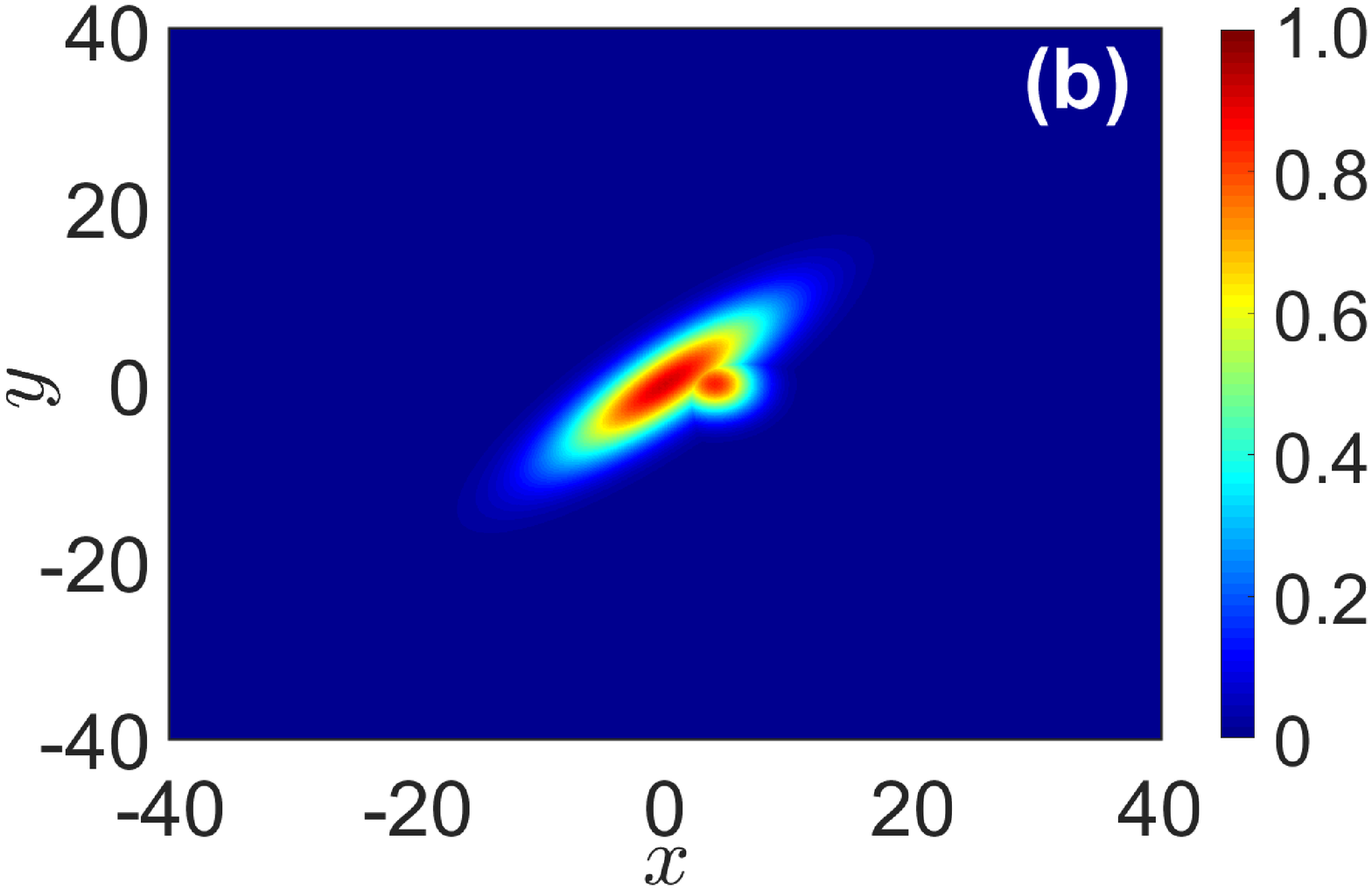} \\
\epsfxsize=10cm  \epsffile{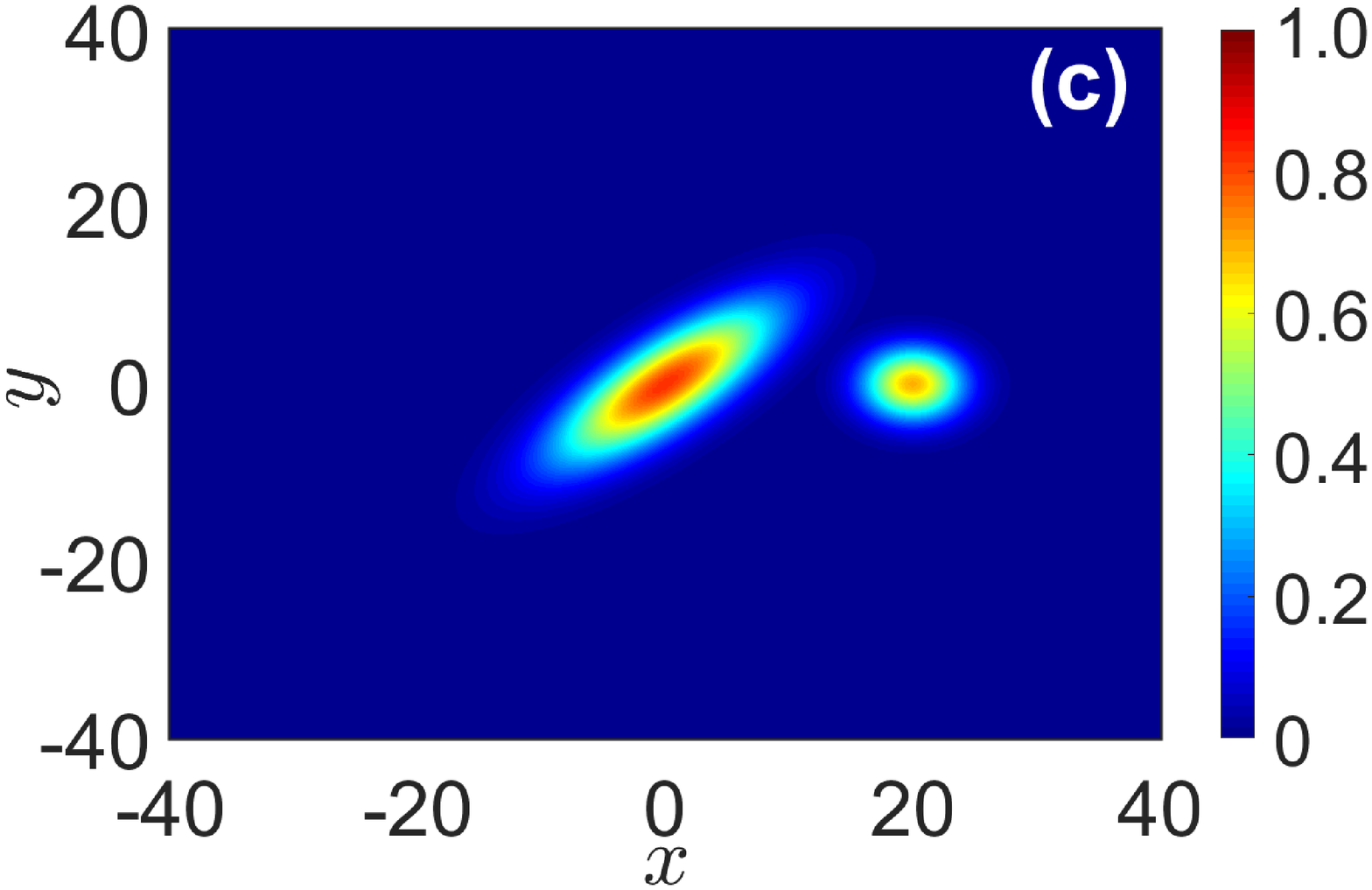} 
\end{center}
\caption{(Color online) 
Contour plots of the beam shapes $|\psi_{j}(x,y,z)|$ at $z=0$ (a), 
$z=z_{i}=1.2$ (b), and $z=z_{f}=2$ (c) in a fast collision between 
two Gaussian beams with the anisotropic initial condition of 
Eqs. (\ref{lp82}) and (\ref{lp84}). The plots represent the beam shapes 
obtained by numerical solution of Eq. (\ref{lp1}) with parameter values 
$\epsilon_{3}=0.01$ and $d_{11}=20$. The orientation angle is $\theta_{0}=\pi/4$.}
\label{fig5}
\end{figure}

\begin{figure}[ptb]
\begin{center}
\epsfxsize=10cm \epsffile{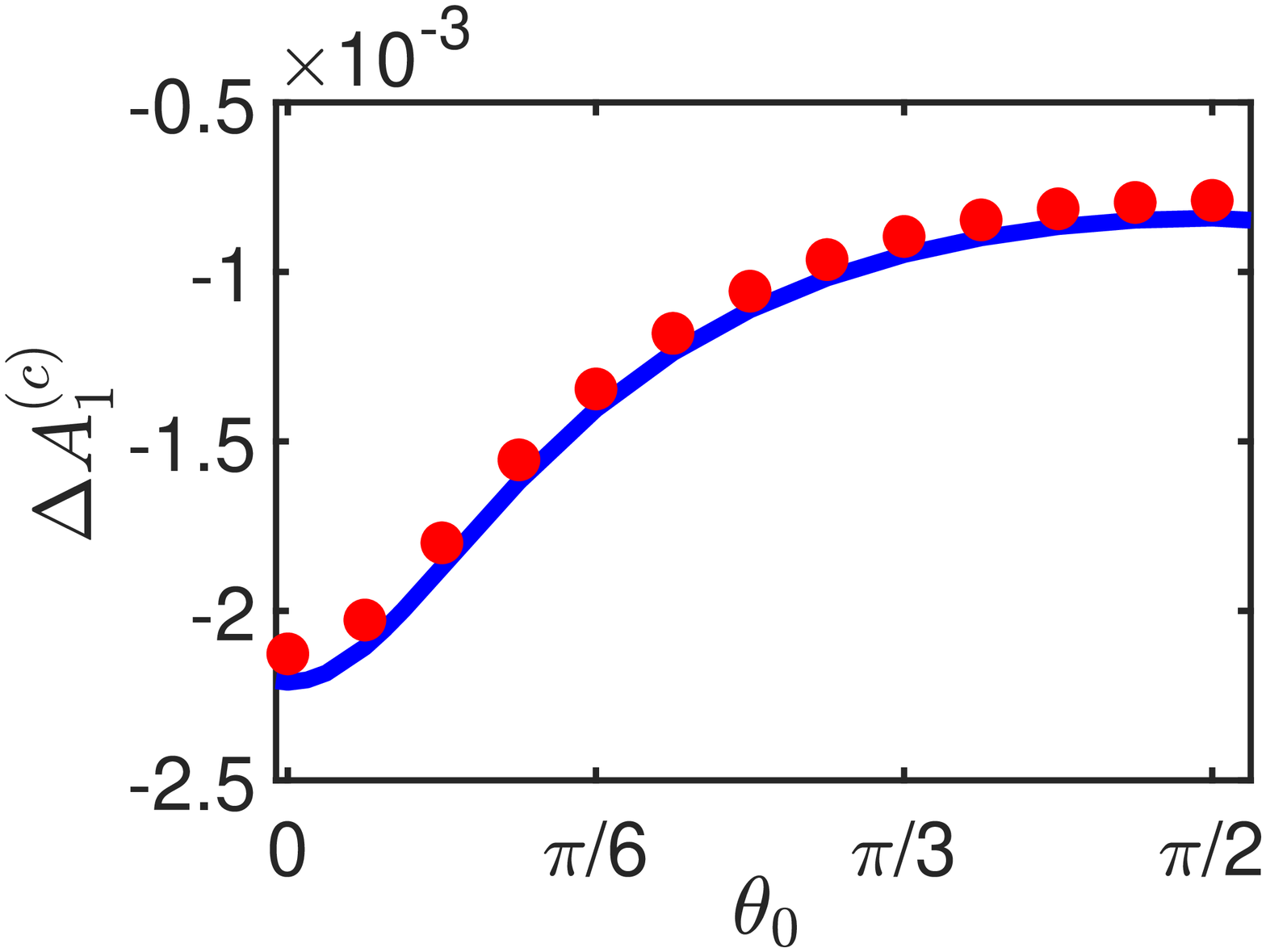}
\end{center}
\caption{(Color online) 
The collision-induced amplitude shift of beam 1 $\Delta A_{1}^{(c)}$ 
vs the orientation angle $\theta_{0}$ in a fast collision between two 
Gaussian beams with the anisotropic initial condition of 
Eqs. (\ref{lp82}) and (\ref{lp84}). The red circles represent 
the result obtained by numerical simulations with Eq. (\ref{lp1}).  
The solid blue curve represents the perturbation theory 
prediction of Eq. (\ref{lp92}).} 
\label{fig6}
\end{figure}

We check the predictions of Eq. (\ref{lp92}) for the dependence of the 
collision-induced amplitude shift on $\theta_{0}$ by carrying out numerical 
simulations with Eq. (\ref{lp1}) with the initial condition of 
Eqs. (\ref{lp82}) and (\ref{lp84}). The simulations are carried out for 
$\theta_{0}$ values in the interval $0 \le \theta_{0} \le \pi/2$. 
The physical parameter values are $\epsilon_{3}=0.01$ and $d_{11}=20$. 
The initial values of the beam parameters in Eqs. (\ref{lp82}) and (\ref{lp84}) 
are $A_{j}(0)=1$, $\alpha_{j0}=0$, $x_{20}=-20$, $W_{10}^{(x)}=8$, 
$W_{10}^{(y)}=2$, and $W_{20}=2$. The final propagation distance 
is $z_{f}=2$, and therefore, the beam centers are well separated at $z_{f}$. 
Figure \ref{fig5} shows the initial beam shapes 
$|\psi_{j}(x,y,0)|$, and the beam shapes $|\psi_{j}(x,y,z)|$ obtained 
in the numerical simulation with $\theta_{0}=\pi/4$ at the intermediate 
distance $z_{i}=1.2$, and at $z_{f}=2$. We observe that both beams experience 
broadening due to diffraction but do not develop extended tails. The dependence 
of $\Delta A_{1}^{(c)}$ on the orientation angle $\theta_{0}$ is shown in 
Fig. \ref{fig6}. We observe very good agreement between the perturbation theory  
prediction and the numerical simulations result. In particular, the numerical 
simulations confirm the expectation that for the chosen parameter values, 
the value of $|\Delta A_{1}^{(c)}|$ would be larger for smaller orientation angles, 
since in this case, beam 2 traverses through the wider part of beam 1. 
Furthermore, the relative error (in percentage) 
$|\Delta A_{1}^{(c)(num)}-\Delta A_{1}^{(c)(th)}|\times 100/|\Delta A_{1}^{(c)(th)}|$ 
is smaller than $6.3\%$ in the entire interval $0 \le \theta_{0} \le \pi/2$. 
Therefore, the numerical simulations with Eq. (\ref{lp1}) confirm the perturbation 
theory prediction for a nontrivial dependence of $\Delta A_{1}^{(c)}$ on 
$\theta_{0}$ due to the anisotropic (and nonseparable) nature of the initial condition.

\subsection{Collision-induced change in the beam shape} 
\label{simu_reshaping}


It was shown in subsection \ref{lp_general_IC} that the two-beam 
collision in the presence of weak cubic loss leads to a change of 
the beam shape in the transverse direction. 
In contrast, it was shown in Refs. \cite{PNH2017B,NHP2020} 
that within the leading order of the perturbative 
calculation for the one-dimensional case, 
the beam shape is preserved during a fast two-beam collision in the 
presence of cubic loss. Thus, the collision-induced change in 
the beam shape in the transverse direction is clearly 
a collisional effect that exists only in spatial dimension higher than 1. 
In the current subsection, we investigate this effect in detail for 
a concrete two-beam setup by both analytic calculations and 
numerical simulations.


To enable a more accurate comparison between the perturbation theory predictions 
and the numerical simulations, we assume that the effects of the optical 
medium's cubic loss on single-beam propagation are negligible compared 
with cubic loss effects on inter-beam interaction. This situation can be realized, 
for example, in certain semiconductors, 
in which two-photon absorption (2PA) associated with 
the simultaneous absorption of two photons with the same wavelength (degenerate 2PA) 
is much weaker than 2PA associated with the simultaneous absorption of two photons with 
different wavelengths (nondegenerate 2PA) \cite{Hagan2002,Hagan2011,Rauscher97}.   
Under this assumption, the dynamics of the two-beam collision is described by 
the following perturbed linear propagation model, in which the perturbation terms 
are only due to the effects of weak cubic loss on two-beam interaction:  
\begin{eqnarray}&&
\!\!\!\!\!\!\!
i\partial_z\psi_{1} + \partial_{x}^{2}\psi_{1}
+ \partial_{y}^{2}\psi_{1}=
-2i\epsilon_{3}|\psi_{2}|^2\psi_{1},
\nonumber \\&&
\!\!\!\!\!\!\!
i\partial_z\psi_{2}+id_{11}\partial_{x}\psi _{2}
+\partial_{x}^2\psi_{2} + \partial_{y}^{2}\psi_{2} =
-2i\epsilon_{3}|\psi_{1}|^2\psi_{2}.   
\!\!\!\!\!\!\!\!
\label{lp101}
\end{eqnarray}                                    
Similar to Eq. (\ref{lp1}), we assume that the cubic loss coefficient satisfies 
$0<\epsilon_{3} \ll 1$. We consider the change in the beam shape in a collision between two 
Gaussian beams as a concrete example. Therefore, the initial condition for the 
collision problem is given by Eq. (\ref{lp61}). 
This choice of the initial condition allows us to obtain an explicit analytic 
expression for the collision-induced change in the shape of beam 1 $\phi_{1}$ 
in the post-collision interval.


Since the initial condition (\ref{lp61}) is separable for both beams, 
we can calculate $\phi_{1}$ in the post-collision interval 
by employing Eq. (\ref{lp58}). In addition, since the effects of 
cubic loss on single-beam propagation are negligible, we can replace 
$A_{j}(z_{c}^{-})$ by $A_{j}(0)$ everywhere in the calculation. 
Therefore, the coefficient $\tilde a_{1}$, which is defined in 
Eq. (\ref{lp53}), takes the form \cite{a1zc}:
\begin{eqnarray} &&
\tilde a_{1}=
2\pi^{1/2}\epsilon_{3}A_{1}(0) A_{2}^{2}(0)
W_{20}^{(x)}/|d_{11}|, 
\label{lp111}
\end{eqnarray}        
where $c_{p2}^{(x)}=\pi^{1/2}$ is used.  
The function $g_{1}^{(x)}(x,z)$ in Eq. (\ref{lp58}) is obtained 
by the solution of the unperturbed linear propagation equation 
with the initial condition (\ref{lp61}). This function is 
given by Eqs. (\ref{appD_4}) and (\ref{appD_6}) in Appendix \ref{appendD}.     
Additionally, in Appendix \ref{appendC} we show that the inverse Fourier transform of  
$\hat g_{12}^{(y)}(k_{2},z_{c})\exp[-i k_{2}^{2}(z-z_{c})]$ 
is given by:  
\begin{eqnarray}&&
{\cal F}^{-1}\left(\hat g_{12}^{(y)}(k_{2},z_{c})
\exp[-i k_{2}^{2}(z-z_{c})]\right)=
\nonumber \\&&
\frac{W_{10}^{(y)}W_{20}^{(y)2}\exp\left[-q_{1}(z_{c})y^{2}/R_{1}^{4}(z,z_{c}) 
+i\chi_{1}^{(y)}(y,z)\right]}
{(W_{10}^{(y)4} + 4z_{c}^{2})^{1/4}
(W_{20}^{(y)4} + 4z_{c}^{2})^{1/2}R_{1}(z,z_{c})}, 
\label{lp120}     
\end{eqnarray}  
where $q_{1}(z_{c})$, $R_{1}(z,z_{c})$, and $\chi_{1}^{(y)}(y,z)$ are given by 
Eqs. (\ref{lp116}), (\ref{lp121}), and (\ref{lp122}), respectively.   
Substituting Eqs. (\ref{lp111}), (\ref{lp120}), (\ref{appD_4}), and (\ref{appD_6}) into 
Eq. (\ref{lp58}), we obtain the following expression for $\phi_{1}(x,y,z)$ 
in the post-collision interval: 
\begin{eqnarray}&&
\phi_{1}(x,y,z)= 
\frac{\tilde a_{1}W_{10}^{(x)}W_{10}^{(y)}W_{20}^{(y)2}}
{(W_{10}^{(x)4} + 4z^{2})^{1/4}
(W_{10}^{(y)4} + 4z_{c}^{2})^{1/4}
(W_{20}^{(y)4} + 4z_{c}^{2})^{1/2}R_{1}(z,z_{c})}
\nonumber \\&&
\times
\exp\left[-\frac{W_{10}^{(x)2} x^{2}}{2(W_{10}^{(x)4} + 4z^{2})}
-\frac{q_{1}(z_{c})y^{2}}{R_{1}^{4}(z,z_{c})} 
+i\chi_{1}^{(tot)}(x,y,z)\right],  
\label{lp123}     
\end{eqnarray}   
where the total phase factor $\chi_{1}^{(tot)}$ is given by: 
\begin{equation}
\chi_{1}^{(tot)}(x,y,z)= 
\chi_{10}^{(x)}(x,z) + \chi_{1}^{(y)}(y,z) + \alpha_{10} + \pi. 
\label{lp124}
\end{equation}       
Comparing Eqs. (\ref{lp124}) and (\ref{appD_11}),  
we observe that the beam's phase factor in the post-collision 
interval is different from the phase factor of the unperturbed 
beam. We therefore define the collision-induced change in the 
beam's phase factor by: 
\begin{equation}
\Delta\chi_{1}^{(tot)}(x,y,z)= 
\chi_{10}(x,y,z) - \chi_{1}^{(tot)}(x,y,z). 
\label{lp124_B}
\end{equation}     
Using Eqs. (\ref{lp124}) and (\ref{appD_11}), we obtain 
\begin{equation}
\Delta\chi_{1}^{(tot)}(y,z)=
\chi_{10}^{(y)}(y,z)-\chi_{1}^{(y)}(y,z)-\pi, 
\label{lp124_C} 
\end{equation}
where $\chi_{10}^{(y)}$ and $\chi_{1}^{(y)}$ are 
given by Eqs. (\ref{appD_7}) and (\ref{lp122}). Thus, only the 
dependence of the phase factor on the transverse 
coordinate is changed by the collision, while the 
dependence on the longitudinal coordinate remains unchanged. 
This change in the $y$ dependence of the phase factor is a 
result of the change in the beam's shape inside the collision 
interval, which leads to a change in the y-dependence of 
$\phi_{1}(x,y,z)$  in the post-collision interval [see detailed 
discussions following Eqs. (\ref{lp44}) and (\ref{lp58})].


The collision-induced change in the beam shape can be characterized by 
the fractional intensity reduction factor $\Delta I_1^{(r)}$, 
which is defined by: 
\begin{eqnarray}&&
\Delta I_1^{(r)}(x,y,z)=
\frac{\tilde I_{1}(x,y,z)-I_{1}(x,y,z)}{\tilde I_{1}(x,y,z)}=
1-\frac{I_{1}(x,y,z)}{A_{1}^{2}(0) \tilde \Psi_{10}^{2}(x,y,z)},
\label{lp102}
\end{eqnarray}
where $I_{1}(x,y,z)=|\psi_{1}(x,y,z)|^{2}$ is the intensity of beam 1 at 
$(x,y,z)$ in the presence of cubic loss, and 
$\tilde I_{1}(x,y,z)=A_{1}^{2}(0) \tilde \Psi_{10}^{2}(x,y,z)$ is the intensity 
of beam 1 in the absence of cubic loss. Thus, $\Delta I_1^{(r)}$ measures 
the ratio between the intensity decrease of beam 1, which is induced solely 
by the effects of cubic loss on the collision, and the intensity 
of beam 1 in the unperturbed single-beam propagation problem. 
We obtain the perturbation theory prediction for the fractional 
intensity reduction factor by expressing $\Delta I_1^{(r)}$  
in terms of $\phi_{1}$. Using Eq. (\ref{lp_add3}) and keeping 
terms up to order $\epsilon_{3}/d_{11}$, we arrive at  
\begin{eqnarray}&&
I_{1} \simeq A_{1}^{2}(0) |\tilde\psi_{10}|^{2}
+A_{1}(0)\left(\tilde \psi_{10}\phi_{1}^{*} 
+\tilde \psi_{10}^{*}\phi_{1}\right). 
\label{lp103}
\end{eqnarray} 
Substitution of Eq. (\ref{lp103}) together with the relation 
$|\tilde \psi_{10}|=\tilde \Psi_{10}$ into Eq. (\ref{lp102})
yields:  
\begin{eqnarray}&&
\Delta I_{1}^{(r)}= 
-\frac{\tilde \psi_{10}\phi_{1}^{*} + \tilde \psi_{10}^{*}\phi_{1}}
{A_{1}(0)\tilde \Psi_{10}^{2}}.
\label{lp104}
\end{eqnarray}
Using Eq. (\ref{lp_add4}) and the relation $\phi_{1}=|\phi_{1}|\exp[i\chi_{1}^{(tot)}]$, 
we obtain: 
\begin{equation}
\tilde \psi_{10}\phi_{1}^{*} + \tilde \psi_{10}^{*}\phi_{1}= 
2\tilde \Psi_{10}|\phi_{1}|\cos\left[\Delta \chi_{1}^{(tot)}\right].
\label{lp105}
\end{equation}
Therefore, we can also express $\Delta I_1^{(r)}$ as: 
\begin{eqnarray}&&
\Delta I_{1}^{(r)}= 
-\frac{2|\phi_{1}|\cos\left[\Delta \chi_{1}^{(tot)}\right]}
{A_{1}(0)\tilde \Psi_{10}}.
\label{lp106}
\end{eqnarray}
For a separable initial condition, $\Delta \chi_{1}^{(tot)}$ 
is given by Eq. (\ref{lp124_C}). In addition, the $x$   
dependences of $\tilde \Psi_{10}$ and of the leading order of $|\phi_{1}|$ 
are identical. As a result, in this case, the dependence on $x$ cancels out on 
the right hand side of Eq. (\ref{lp106}), and $\Delta I_1^{(r)}$ becomes 
independent of $x$. Therefore, for a separable initial condition, 
the expression for $\Delta I_1^{(r)}$ takes the form: 
\begin{eqnarray}&&
\Delta I_{1}^{(r)}(y,z)= 
-\frac{2|\phi_{1}(x,y,z)|\cos\left[\chi_{10}^{(y)}(y,z)-\chi_{1}^{(y)}(y,z)-\pi \right]}
{A_{1}(0)\tilde \Psi_{10}(x,y,z)}.  
\label{lp107}
\end{eqnarray}
The fractional intensity reduction factor for the collision setup 
considered in the current subsection is given by Eq. (\ref{lp107}), 
where $\phi_{1}$, $\tilde \Psi_{10}$, $\chi_{10}^{(y)}$, 
and $\chi_{1}^{(y)}$ are given by Eqs. (\ref{lp123}), 
(\ref{appD_12}), (\ref{appD_7}), and (\ref{lp122}), respectively. 
We note that the effects of the collision-induced change in the 
beam's phase factor are included in Eq. (\ref{lp107}) via the 
dependence of the expression on the right hand side of this equation 
on $\cos\left[\chi_{10}^{(y)}(y,z)-\chi_{1}^{(y)}(y,z)-\pi \right]$. 
We will see in the next paragraphs that these effects can lead to 
{\it negative} values of $\Delta I_1^{(r)}$ in certain intervals of $y$, 
that is, to a localized {\it increase} in the intensity of beam 1 relative 
to the intensity in the unperturbed single-beam propagation problem.


\begin{figure}[ptb]
\begin{center}
\epsfxsize=10cm  \epsffile{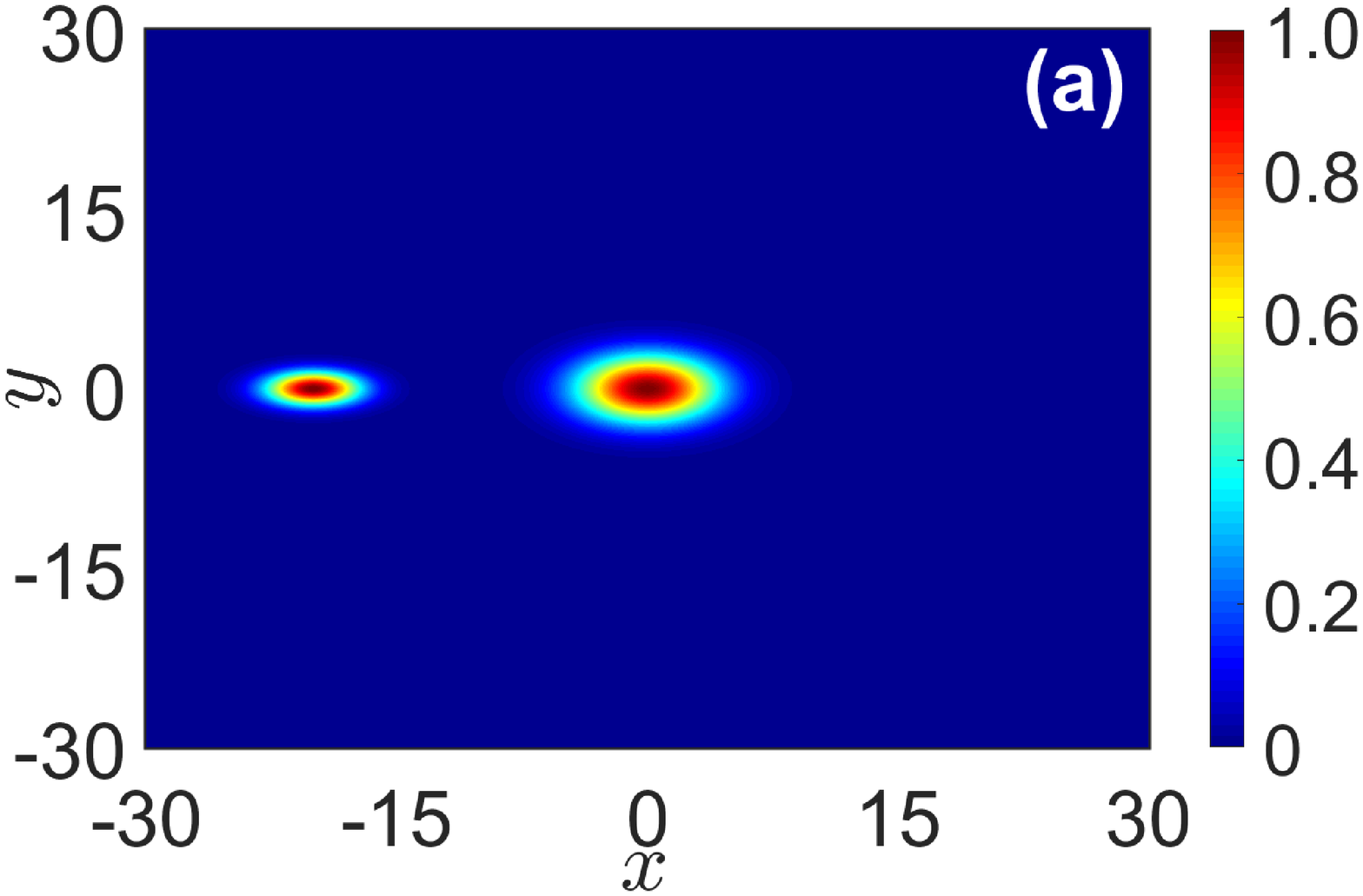} \\
\epsfxsize=10cm  \epsffile{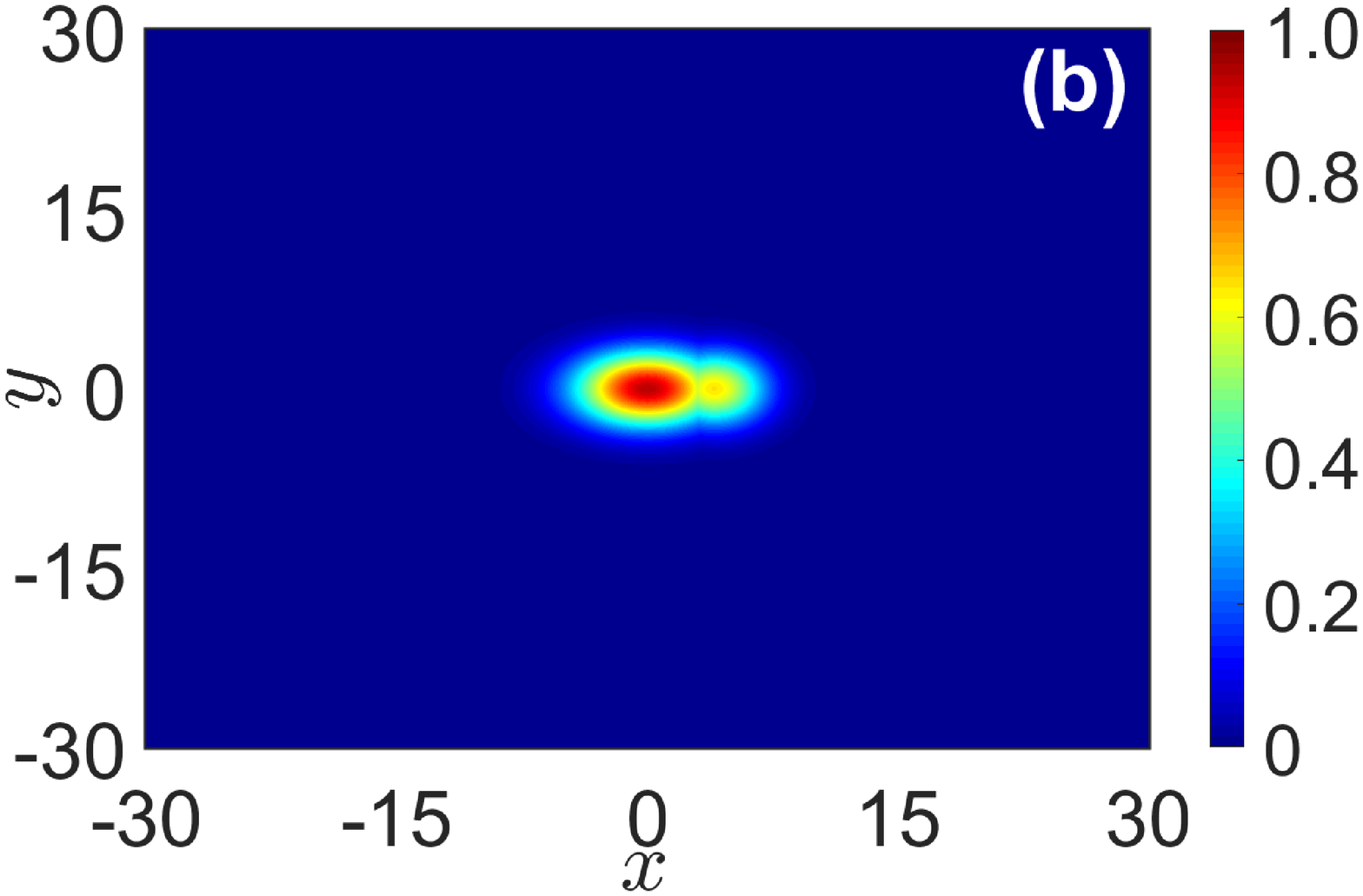} \\
\epsfxsize=10cm  \epsffile{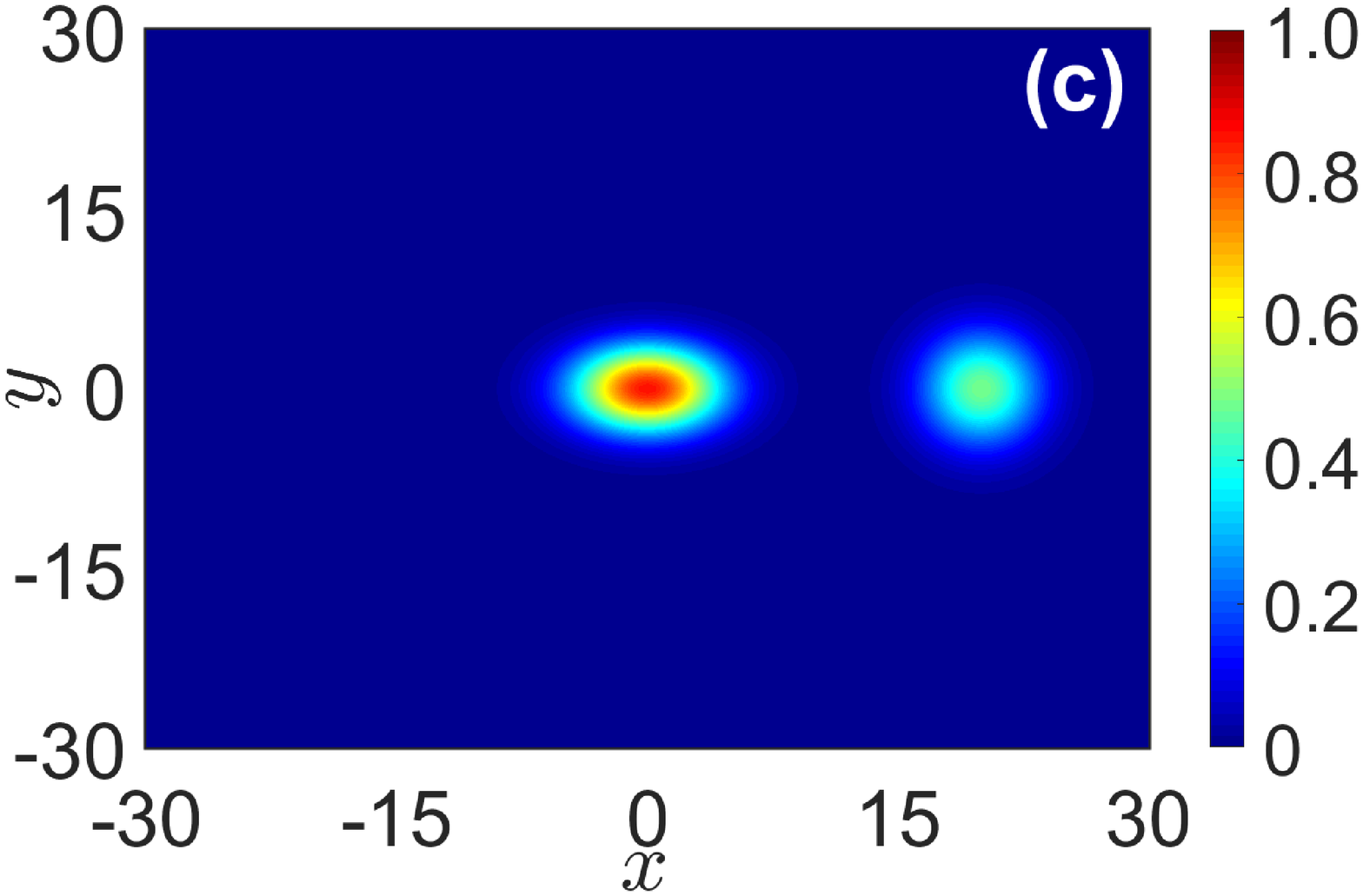} 
\end{center}
\caption{(Color online) 
Contour plots of the beam shapes $|\psi_{j}(x,y,z)|$ at $z=0$ (a), 
$z=z_{i}=0.96$ (b), and $z=z_{f}=1.6$ (c) in a fast collision between 
two Gaussian beams with parameter values $\epsilon_{3}=0.01$ and 
$d_{11}=25$. The plots represent the beam shapes obtained 
by numerical solution of Eq. (\ref{lp101}) with the initial 
condition (\ref{lp61}).}
\label{fig7}
\end{figure}

\begin{figure}[ptb]
\begin{center}
\epsfxsize=10cm \epsffile{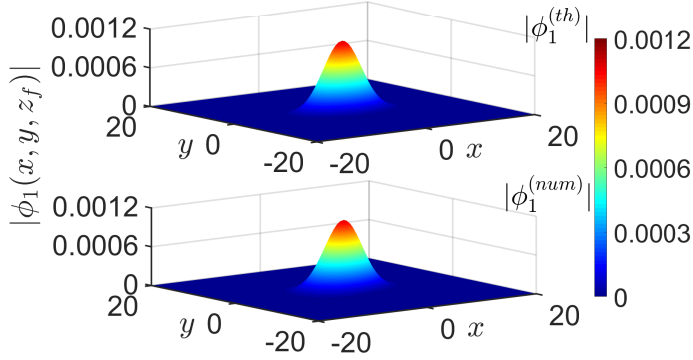}
\end{center}
\caption{(Color online) 
The collision-induced change in the shape of beam 1 $|\phi_{1}(x,y,z_{f})|$ 
at $z_{f}=1.6$ in a fast two-beam collision with parameter values  
$\epsilon_{3}=0.01$ and $d_{11}=25$.   
Top: the perturbation theory prediction of Eq. (\ref{lp123}). 
Bottom: the result obtained by numerical solution of Eq. (\ref{lp101}).} 
\label{fig_add3}
\end{figure}

We check the perturbation theory predictions for the collision-induced 
change in the beam shape by extensive numerical simulations 
with Eq. (\ref{lp101}). The simulations are carried out with $\epsilon_{3}=0.01$ 
and with $d_{11}$ values that are varied in the intervals $4 \le |d_{11}| \le 60$. 
The parameter values of the initial condition (\ref{lp61}) are $A_{j}(0)=1$,  
$\alpha_{j0}=0$, $x_{20}=\pm 20$, $W_{10}^{(x)}=3$, $W_{10}^{(y)}=2$, 
$W_{20}^{(x)}=2$, and $W_{20}^{(y)}=1$. 
The final propagation distance is $z_{f}=2z_{c}=-2x_{20}/d_{11}$, 
and therefore, the beam centers are well separated at $z_{f}$.  
Figure \ref{fig7} shows the initial beam shapes $|\psi_{j}(x,y,0)|$, 
and the beam shapes $|\psi_{j}(x,y,z)|$ obtained in the simulation 
with $d_{11}=25$ at $z_{i}=0.96>z_{c}$, and at $z_{f}=1.6$. We observe 
that both beams experience significant broadening due to diffraction. 
Figure \ref{fig_add3} shows the collision-induced change in the 
shape of beam 1 obtained in the simulation with $d_{11}=25$ 
at $z=z_{f}$ $|\phi_{1}^{(num)}(x,y,z_{f})|$.     
The perturbation theory prediction $|\phi_{1}^{(th)}(x,y,z_{f})|$, 
which is obtained by Eq. (\ref{lp123}) is also shown. 
The agreement between the simulation result and the perturbation 
theory prediction is very good both near the beam's maximum and at 
the tails. We quantify the difference between $|\phi_{1}^{(num)}(x,y,z)|$ 
and $|\phi_{1}^{(th)}(x,y,z)|$ by defining the relative error (in percentage)  
$E_{r}^{(|\phi_{1}|)}(z)$ in the following manner:  
\begin{eqnarray}&&
E_{r}^{(|\phi_{1}|)}(z)=
100 \times
\left [\int dx\int dy \, |\phi_{1}^{(th)}(x,y,z)|^{2} \right ]^{-1/2}
\nonumber \\&&
\times
\left\{\int dx\int dy 
\left[\;\left|\phi_{1}^{(th)}(x,y,z) \right| - 
\left|\phi_{1}^{(num)}(x,y,z) \right| \; 
\right]^2 \right\}^{1/2},   
\label{lp108}
\end{eqnarray}       
where the integration is carried out over the entire domain in the $xy$ 
plane, which is used in the numerical simulation. We find that the value 
of $E_{r}^{(|\phi_{1}|)}(z_{f})$ for $d_{11}=25$ is $7.0\%$, in 
accordance with the good agreement between simulation and theory 
that is observed in Fig.  \ref{fig_add3}.

\begin{figure}[ptb]
\begin{center}
\epsfxsize=10cm  \epsffile{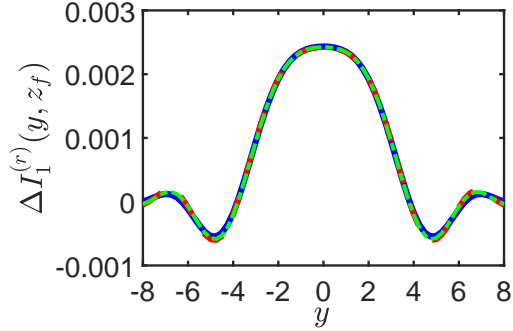} 
\end{center}
\caption{(Color online) 
The $y$ dependence of the fractional intensity reduction factor for beam 1  
at $z=z_{f}$ $\Delta I_{1}^{(r)}(y,z_{f})$ in a two-beam collision with 
parameter values $\epsilon_{3}=0.01$ and $d_{11}=25$. 
The solid blue curve corresponds to the perturbation theory prediction 
of Eqs. (\ref{lp107}) and (\ref{lp123}). The other two curves represent 
the results obtained by numerical solution of Eq. (\ref{lp101}). 
The dashed red curve is obtained by averaging $\Delta I_{1}^{(r)}(x,y,z_{f})$  
over the $x$-interval $[-2,2]$. The dashed-dotted green curve is obtained by using 
the numerically computed value of $\Delta I_{1}^{(r)}(0,y,z_{f})$.}
\label{fig8}
\end{figure}

\begin{figure}[ptb]
\begin{center}
\epsfxsize=10cm  \epsffile{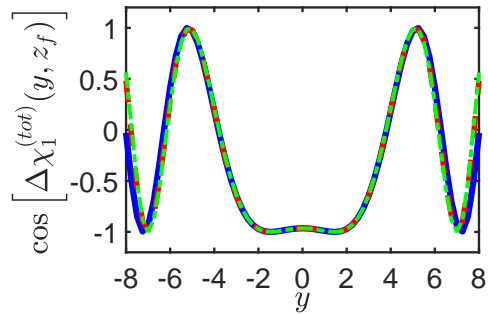} 
\end{center}
\caption{(Color online) 
The $y$ dependence of the cosine of the collision-induced 
change in the phase factor of beam 1 at $z=z_{f}$   
$\cos\,[\,\Delta\chi_{1}^{(tot)}(y,z_{f}) \,]$ 
in a two-beam collision with parameter values 
$\epsilon_{3}=0.01$ and $d_{11}=25$. 
The solid blue curve corresponds to the perturbation theory prediction, 
which is obtained by using Eqs. (\ref{lp124_C}), (\ref{appD_7}), and (\ref{lp122}). 
The other two curves represent the results 
obtained by numerical solution of Eq. (\ref{lp101}). 
The dashed red curve is obtained by averaging $\cos\,[\,\Delta\chi_{1}^{(tot)}(x,y,z_{f}) \,]$
over the $x$-interval $[-2,2]$. The dashed-dotted green curve is obtained by using 
the numerically computed value of $\cos\,[\,\Delta\chi_{1}^{(tot)}(0,y,z_{f}) \,]$.}
\label{fig_add4}
\end{figure}

We now turn to analyze the behavior of the fractional intensity 
reduction factor as a function of $y$. Figure \ref{fig8} shows 
the $y$ dependence of $\Delta I_{1}^{(r)}$ at $z=z_{f}$ obtained 
in the numerical simulation with $d_{11}=25$ \cite{Delta_I_1}. 
The perturbation theory prediction of Eqs. (\ref{lp107}) and 
(\ref{lp123}) is also shown. The agreement between the simulations 
result and the theoretical prediction is very good. 
Based on this result and on similar results that are obtained 
with other choices of the physical parameter values we conclude 
that the perturbation theory correctly captures the spatial 
distribution of the intensity reduction in fast two-beam collisions 
in the presence of weak cubic loss. 
We also point out that according to Eqs. (\ref{lp107}) and (\ref{lp123}),  
$\Delta I_{1}^{(r)}(y,z_{f})$ attains negative values at 
intermediate values of $y$, e.g., in the intervals 
$3.85 \le |y| \le 6.25$ (for $d_{11}=25$). This prediction 
is confirmed by the numerical simulation. In particular, 
the numerically obtained $\Delta I_{1}^{(r)}(y,z_{f})$ 
attains negative values in the intervals $3.8 \le |y| \le 6.1$, 
in very good agreement with the result of Eqs. (\ref{lp107}) and (\ref{lp123}).       
The negative values of $\Delta I_{1}^{(r)}(y,z_{f})$ correspond to a localized 
increase in the intensity of beam 1 relative to the intensity in the 
unperturbed single-beam propagation problem. According to the perturbation 
theory, these values are a result of the collision-induced change in the 
phase factor of beam 1, which affects the value of $\Delta I_{1}^{(r)}(y,z_{f})$ 
via its dependence on $\cos\,[\,\Delta\chi_{1}^{(tot)}(y,z_{f}) \,]$ 
[see Eqs. (\ref{lp107}) and (\ref{lp124_C})]. 
To check if this is indeed the case, we compare 
the numerical simulation result for the $y$ dependence of 
$\cos\,[\,\Delta\chi_{1}^{(tot)}(y,z_{f}) \,]$ 
with the perturbation theory prediction, 
which is obtained by using Eqs. (\ref{lp124_C}), (\ref{appD_7}), 
and (\ref{lp122}). This comparison is shown in Fig. \ref{fig_add4}. 
We observe good agreement between the results of the perturbative 
calculation and the numerical simulation for this quantity. 
In particular, the perturbation theory result for 
$\cos\,[\,\Delta\chi_{1}^{(tot)}(y,z_{f}) \,]$ attains positive values 
in the intervals $3.85 \le |y| \le 6.25$, while the simulation result 
attains positive values in the intervals $3.8 \le |y| \le 6.1$.   
These intervals are the same as the ones, in which the values 
of the theoretically and numerically obtained 
$\Delta I_{1}^{(r)}(y,z_{f})$ are negative. 
We therefore conclude that the relative localized intensity increase 
for beam 1 at intermediate $y$ values is indeed a result of the 
collision-induced change in the phase factor of this beam.

\begin{figure}[ptb]
\begin{center}
\epsfxsize=10cm  \epsffile{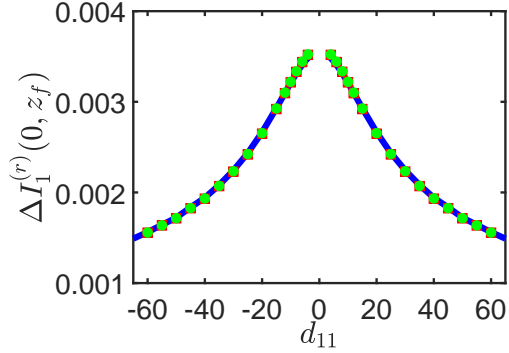} 
\end{center}
\caption{(Color online) 
The $d_{11}$ dependence of the fractional intensity reduction factor 
for beam 1 at $y=0$ and $z=z_{f}$, $\Delta I_{1}^{(r)}(0,z_{f})$,  
in a fast two-beam collision with $\epsilon_{3}=0.01$. 
The solid blue curve corresponds to the perturbation theory prediction 
of Eqs. (\ref{lp107}) and (\ref{lp123}). The other two curves are  
obtained from the numerical solution of Eq. (\ref{lp101}). 
The red squares represent the result obtained by averaging $\Delta I_{1}^{(r)}(x,0,z_{f})$  
over the $x$-interval $[-2,2]$. The green circles represent 
the result obtained by using the numerically computed value 
of $\Delta I_{1}^{(r)}(0,0,z_{f})$.}
\label{fig9}
\end{figure}

Finally, we study the dependence of the fractional intensity reduction factor 
on the value of the beam-steering coefficient by measuring   
$\Delta I_{1}^{(r)}(0,z_{f})$ as a function of $d_{11}$.    
Figure \ref{fig9} shows the dependence of $\Delta I_{1}^{(r)}(0,z_{f})$ 
on $d_{11}$ obtained in the numerical simulations together 
with the theoretical prediction of Eqs. (\ref{lp107}) and (\ref{lp123}). 
The agreement between the simulations result and the perturbation theory 
prediction is excellent over the entire interval of $d_{11}$ values. 
More specifically, the relative error in the approximation of 
$\Delta I_{1}^{(r)}(0,z_{f})$ (in percentage),   
which is defined by $|\Delta I_{1}^{(r)(num)}(0,z_{f}) - \Delta I_{1}^{(r)(th)}(0,z_{f})| 
\times 100 / |\Delta I_{1}^{(r)(th)}(0,z_{f})|$, is smaller than $0.6\%$ for 
$10 \le |d_{11}| \le 60$, and smaller than $1.2\%$ for $4 \le |d_{11}| <10$. 
We also checked the dependence of $\Delta A_{1}^{(c)}$ on $d_{11}$, 
and obtained very good agreement between the perturbation theory 
prediction and the numerical simulations result (similar to what 
is shown in Figs. \ref{fig2} and \ref{fig4}).      
Based on these results and on the results shown in Figs. 
\ref{fig_add3} - \ref{fig_add4} we conclude that the perturbation theory 
of subsections \ref{lp_general_IC} and \ref{lp_separable_IC} enables 
accurate calculation of both the change in the beam shapes and the dynamics 
of the beam amplitudes in fast two-beam collisions in the presence of 
weak cubic loss.

\section{Fast collisions between pulsed-beams in spatial dimension 3}
\label{lp_3D}

\subsection{Introduction}
\label{lp_3D_model} 

We consider the dynamics of collisions between two pulsed-beams 
in a three-dimensional linear optical medium with weak cubic loss. 
Similar to sections \ref{lp_2D} and \ref{simu}, we assume that 
the pulsed-beams propagate along the $z$ axis, and that 
the propagation is accurately described by the paraxial approximation. 
For each value of $z$, the distribution of the optical field 
depends on the spatial coordinates $x$ and $y$, and on time $t$. 
Therefore, using the terminology that was introduced in section 
\ref{lp_model}, the spatial dimension is 3, and the propagation 
is described by a $(3+1)$-dimensional propagation model.

We take into account the effects of first and second-order dispersion, 
isotropic diffraction, and weak cubic loss, and neglect beam-steering. 
Since beam-steering is neglected, the relative velocity vector 
between the beam centers lies along the $t$ axis in the $txy$ space. 
Therefore, the dynamics of the collision is described by the 
following weakly perturbed linear propagation model: 
\begin{eqnarray}&&
\!\!\!\!\!\!\!\!\!\!\!\!\!\!
i\partial_{z}\psi_{1} + \partial^{2}_{t} \psi_{1} 
+ d_{2}\partial^{2}_{x} \psi_{1} + d_{2}\partial^{2}_{y} \psi_{1}
= -i\epsilon_{3}|\psi_{1}|^{2}\psi_{1} - 2i\epsilon_{3}|\psi_{2}|^{2}\psi_{1},
\nonumber\\&&
\!\!\!\!\!\!\!\!\!\!\!\!\!\!
i\partial_{z}\psi_{2} + id_{11}\partial_{t}\psi_{2}
+ \partial^{2}_{t} \psi_{2} + d_{2}\partial^{2}_{x} \psi_{2}
+  d_{2}\partial^{2}_{y} \psi_{2}
= -i\epsilon_{3}|\psi_{2}|^{2}\psi_{2} -2i\epsilon_{3}|\psi_{1}|^{2}\psi_{2},
\nonumber\\&&
\label{lp140}
\end{eqnarray}
where $\psi_{j}$ are proportional to the electric fields of the beams, 
$x$, $y$, and $z$ are the spatial coordinates, and $t$ is time \cite{Dimensions2,sign_beta_2}.   
In Eq. (\ref{lp1}), $d_{11}$ is the first-order dispersion coefficient, 
$d_{2}$ is the diffraction coefficient, and $\epsilon_{3}$ 
is the cubic loss coefficient. The term $id_{11}\partial_{t}\psi_{2}$ 
describes the effects of first-order dispersion, the terms $\partial^{2}_{t} \psi_{j}$ 
describe the effects of second-order dispersion, and $d_{2}\partial_{x}^{2}\psi_{j}$ 
and $d_{2}\partial_{y}^{2}\psi_{j}$ describe the effects of isotropic diffraction. 
Additionally, the terms $-i\epsilon_{3}|\psi_{j}|^{2}\psi_{j}$ and  
$-2i\epsilon_{3}|\psi_{k}|^{2}\psi_{j}$ describe intra-beam and inter-beam effects 
due to weak cubic loss.

We consider collisions between pulsed-beams with general initial shapes 
and with tails that decrease sufficiently fast, such that the values of 
the integrals $\int_{-\infty}^{\infty} dt  \int_{-\infty}^{\infty} dx 
\int_{-\infty}^{\infty} dy \, |\psi_{j}(t,x,y,0)|^{2}$ are finite. 
We assume that the initial pulsed-beams can be characterized by the 
following parameters. (1) The initial amplitudes $A_{j}(0)$. 
(2) The initial widths of the pulsed-beams along the $t$, $x$, and $y$ axes, 
$W_{j0}^{(t)}$, $W_{j0}^{(x)}$, and $W_{j0}^{(y)}$. 
(3) The initial positions of the beam centers $(t_{j0}, x_{j0}, y_{j0})$.   
(4) The initial phases $\alpha_{j0}$. Thus, the initial electric fields 
can be written as: 
\begin{eqnarray} &&
\psi_{j}(t,x,y,0)=A_{j}(0)h_{j}(t,x,y)\exp(i\alpha_{j0}),  
\label{lp141}
\end{eqnarray} 
where $h_{j}(t,x,y)$ is real-valued. We are equally interested in the 
important case, where the initial electric fields of the two 
pulsed-beams are separable. i.e., where each of the functions 
$\psi_{j}(t,x,y,0)$ is a product of three functions of $t$, $x$, 
and $y$ \cite{laser_modes}. In this case, the initial electric fields 
can be expressed as:
\begin{eqnarray} &&
\psi_{j}(t,x,y,0)=A_{j}(0) h_{j}^{(t)}[(t-t_{j0})/W_{j0}^{(t)}]
h_{j}^{(x)}[(x-x_{j0})/W_{j0}^{(x)}]
\nonumber \\&&
\times 
h_{j}^{(y)}[(y-y_{j0})/W_{j0}^{(y)}]\exp(i\alpha_{j0}).  
\label{lp142}
\end{eqnarray}

We are interested in the collision-induced dynamics of complete 
fast collisions. We therefore obtain conditions on the 
physical parameter values for these collisions.   
As stated earlier, the relative velocity vector 
of the pulsed-beams centers lies along the $t$ axis in the $txy$ 
space. As a result, the conditions for a complete fast collision 
involve the initial widths of the pulsed-beams along the $t$ axis 
$W_{j0}^{(t)}$, as well as the initial and final values of the $t$ 
coordinate of the pulsed-beam centers $t_{j0}$ and $t_{j}(z_f)$, 
respectively. The conditions for a complete collision are obtained   
by requiring that the pulsed-beams are well-separated 
in time before and after the collision. 
This requirement yields the following inequalities:  
$|t_{20}-t_{10}| \gg W_{10}^{(t)}+W_{20}^{(t)}$ and 
$|t_{2}(z_f)-t_{1}(z_f)| \gg W_{1}^{(t)}(z_f)+W_{2}^{(t)}(z_f)$, 
where $W_{j}^{(t)}(z_f)$ are the pulsed-beam widths along the $t$ axis at $z=z_{f}$.     
The collision length $\Delta z_{c}$ is given by 
$\Delta z_{c}=2(W_{10}^{(t)}+W_{20}^{(t)})/|d_{11}|$.  
The assumption of a fast collision means that $\Delta z_{c}$ is 
much smaller than the length scale $z_{D}^{(min)}$, which is  
the smallest dispersion length or diffraction length in the problem.  
By definition, $z_{D}^{(min)}=\min \left\{ z_{d1}^{(t)}, z_{d2}^{(t)}, 
z_{D1}^{(x)}, z_{D2}^{(x)}, z_{D1}^{(y)}, z_{D2}^{(y)}\right\}$, 
where $z_{dj}^{(t)}$ are the dispersion lengths of the pulsed-beams, 
and $z_{Dj}^{(x)}$ and $z_{Dj}^{(y)}$ are the diffraction lengths 
along the $x$ and $y$ axes. Requiring that $\Delta z_{c} \ll z_{D}^{(min)}$, 
we obtain $2(W_{10}^{(t)}+W_{20}^{(t)}) \ll |d_{11}| z_{D}^{(min)}$, as 
the condition for a fast collision.

\subsection{The perturbation theory predictions for the collision-induced 
changes in the pulsed-beam shape and amplitude}
\label{lp_3D_calculation} 

The perturbative calculation of the collision-induced changes in the 
shapes and amplitudes of the pulsed-beams in spatial dimension 3 is very similar 
to the calculation that was presented in section \ref{lp_2D} for the 
two-dimensional case. For this reason and for brevity, we do not 
present the entire derivation of the equations for the collision-induced 
dynamics. Instead, we present only the two main results 
of the perturbative calculation, namely, the expressions for 
the collision-induced changes in the shape and amplitude of 
pulsed-beam 1, $\Delta\Phi_{1}(t,x,y,z_{c})$ and $\Delta A_{1}^{(c)}$. 
The notations for the physical quantities are the same as the ones 
used in section \ref{lp_2D}, apart from the fact that 
$\psi_{j}$, $\psi_{j0}$, $\tilde\psi_{j0}$, $\tilde\Psi_{j0}$, 
and $\chi_{j0}$ are now functions of $t$, $x$, $y$, and $z$.

We start by describing the results of the perturbative calculation 
for the general initial condition (\ref{lp141}). In this case, the 
collision-induced change in the shape of pulsed-beam 1 inside the 
collision interval is given by: 
\begin{eqnarray} &&
\!\!\!\!\!\!\!
\Delta\Phi_{1}(t,x,y,z_{c})\!=\!-\frac{2\epsilon_{3}
A_{1}(z_{c}^{-}) A_{2}^{2}(z_{c}^{-})}{|d_{11}|}\tilde\Psi_{10}(t,x,y,z_{c})
\!\!\int_{-\infty}^{\infty} \!\!\!\!\! d\tilde t \, \bar\Psi_{20}^{2}(\tilde t,x,y,z_{c}), 
\nonumber \\&&
\label{lp143}
\end{eqnarray} 
where $\tilde t=t-t_{20}-d_{11}z$. Additionally, the collision-induced 
change in the amplitude of pulsed-beam 1 is 
\begin{eqnarray} &&
\!\!\!\!
\Delta A_{1}^{(c)}=-\frac{2\epsilon_{3}
A_{1}(z_{c}^{-}) A_{2}^{2}(z_{c}^{-})}{C_{p1}|d_{11}|}
\nonumber \\&&
\times
\!\int_{-\infty}^{\infty} \!\!\!\!\! dt  
\!\int_{-\infty}^{\infty} \!\!\!\!\! dx 
\!\int_{-\infty}^{\infty} \!\!\!\!\! dy
\;\tilde\Psi_{10}^{2}(t,x,y,z_{c})
\!\int_{-\infty}^{\infty} \!\!\!\!\! d\tilde t \;\bar\Psi_{20}^{2}(\tilde t,x,y,z_{c}),  
\label{lp144}
\end{eqnarray}   
where $C_{p1}$ is given by: 
\begin{eqnarray}&&
C_{p1}= 
\!\int_{-\infty}^{\infty} \!\!\!\!\! dt 
\!\int_{-\infty}^{\infty} \!\!\!\!\! dx 
\!\int_{-\infty}^{\infty} \!\!\!\!\! dy
\;\tilde\Psi_{10}^{2}(t,x,y,0).  
\label{lp145}
\end{eqnarray}

Further insight into the collision dynamics is obtained when the 
initial condition is of the form (\ref{lp142}), which is completely 
separable for both beams. This case is also of special interest for 
practical reasons \cite{laser_modes}. In this case, the collisional 
change in the shape of pulsed-beam 1 in the collision interval 
is given by:  
\begin{eqnarray} &&
\!\!\!\!\!\!\!
\Delta\Phi_{1}(t,x,y,z_{c})\!=\!-\frac{2\epsilon_{3}
A_{1}(z_{c}^{-}) A_{2}^{2}(z_{c}^{-})}{|d_{11}|}
c_{p2}^{(t)} W_{20}^{(t)}
\nonumber \\&&
\times G_{1}^{(t)}(t,z_{c}) 
G_{1}^{(x)}(x,z_{c})G_{1}^{(y)}(y,z_{c})
G_{2}^{(x)2}(x,z_{c})G_{2}^{(y)2}(y,z_{c}). 
\label{lp146}
\end{eqnarray}    
The real-valued functions $G_{j}^{(t)}(t,z)$, $G_{j}^{(x)}(x,z)$, 
and $G_{j}^{(y)}(y,z)$ in Eq. (\ref{lp146}) are defined by:  
\begin{eqnarray}&&
\tilde\psi_{j0}(t,x,y,z)=G_{j}^{(t)}(t,z)G_{j}^{(x)}(x,z)G_{j}^{(y)}(y,z)
\nonumber \\&&
\times \exp\left\{i\left[\chi_{j0}^{(t)}(t,z) + 
\chi_{j0}^{(x)}(x,z) + \chi_{j0}^{(y)}(y,z) + \alpha_{0}\right] \right\},
\label{lp147}
\end{eqnarray}     
where $\tilde\psi_{j0}$ are the solutions to the unperturbed linear 
propagation equations with the separable initial condition (\ref{lp142}), 
and $\chi_{j0}^{(t)}(t,z)$, $\chi_{j0}^{(x)}(x,z)$, and $\chi_{j0}^{(y)}(y,z)$ 
are the (real-valued) phase factors. In addition, the coefficient 
$c_{p2}^{(t)}$ is given by the equation 
\begin{eqnarray}&&
\!\!\!\!\!\!\!\!
\int_{-\infty}^{\infty} \!\!\!\! dt \, G_{j}^{(t)2}(t,z)
=\int_{-\infty}^{\infty} \!\!\!\! dt \, G_{j}^{(t)2}(t,0)
=W_{j0}^{(t)}\int_{-\infty}^{\infty} \!\!\!\! ds \, h_{j}^{(t)2}(s)
=W_{j0}^{(t)}  c_{pj}^{(t)}.   
\nonumber \\&&
\label{lp148}
\end{eqnarray}        
Similar to the situation in the two-dimensional case, it follows 
from Eq. (\ref{lp146}) that the $t$ dependence of the pulsed-beam 
is not changed by the collision at all (within the leading order 
of the perturbative calculation).

The collision-induced change in the amplitude of pulsed-beam 1 
in the case of a completely separable initial condition 
is given by: 
\begin{eqnarray} &&
\!\!\!\!
\Delta A_{1}^{(c)}=-\frac{2\epsilon_{3}
A_{1}(z_{c}^{-}) A_{2}^{2}(z_{c}^{-})}{|d_{11}|}
\frac{c_{p2}^{(t)}W_{20}^{(t)}} 
{c_{p1}^{(x)}W_{10}^{(x)}c_{p1}^{(y)}W_{10}^{(y)}}
\nonumber \\&&
\times
\!\int_{-\infty}^{\infty} \!\!\!\!\! dx \, 
G_{1}^{(x)2}(x,z_{c})G_{2}^{(x)2}(x,z_{c})
\!\int_{-\infty}^{\infty} \!\!\!\!\! dy \, 
G_{1}^{(y)2}(y,z_{c})G_{2}^{(y)2}(y,z_{c}). 
\label{lp149}
\end{eqnarray}       
We observe that Eq. (\ref{lp149}) has the same form as 
Eq. (\ref{lp47}), where the longitudinal factor is now 
$c_{p2}^{(t)}W_{20}^{(t)}$ and the overall factor is 
$2\epsilon_{3}A_{1}(z_{c}^{-}) A_{2}^{2}(z_{c}^{-})/|d_{11}|$. 
This finding is in accordance with the expectation that the form 
(\ref{lp47}) is valid for a general spatial dimension when 
the initial condition is separable in the longitudinal direction 
for both beams. In addition, similar to the situation in spatial 
dimensions 1 and 2, the longitudinal factor $c_{p2}^{(t)}W_{20}^{(t)}$ 
is universal in the sense that it does not depend on the exact details 
of the initial pulsed-beam shapes and on the collision distance $z_{c}$.

\subsection{Numerical simulations for pulsed-beam collisions}
\label{lp_3D_simu} 

The predictions of the perturbative calculation in subsection 
\ref{lp_3D_calculation} are based on several simplifying assumptions. 
In particular, it is assumed that the pulsed-beams are sharply 
peaked in the $txy$ space throughout the propagation, and that as 
a result, the integration with respect to $z$ can be extended to 
$\pm \infty$. In addition, the explicit conditions for the validity 
of the approximations employed by the perturbative calculation are 
not known. For these reasons, it is important to check the predictions 
of the perturbative calculation by numerical simulations with the 
perturbed linear propagation model (\ref{lp140}).

As explained in the beginning of subsection \ref{lp_3D_calculation}, 
the perturbative calculation of the collision-induced dynamics in 
spatial dimension 3 is very similar to the calculation for spatial 
dimension 2. For this reason and for brevity, we do not present the 
results of the numerical simulations for all the collisional setups 
that were considered in section \ref{simu} in the two-dimensional case. 
Instead, we present only the simulations results for the collisional 
setup, which is used for checking the theoretical predictions for 
universality of the longitudinal part in the expression for 
$\Delta A_{1}^{(c)}$ (the setup considered in subsection 
\ref{simu_universality} for spatial dimension 2).

Similar to subsection \ref{simu_universality}, we choose two initial 
conditions with widely different pulsed-beam profiles in the longitudinal 
(temporal) direction. More specifically, the $t$ dependence of the 
pulsed-beams in the first initial condition is Gaussian, 
i.e., it has rapidly decaying tails. In contrast, the $t$ dependence 
of the pulsed-beams in the second initial condition is given by 
a generalized Cauchy-Lorentz distribution, i.e., it has slowly decaying 
tails, whose decay is characterized by a power-law. 
The initial profiles of the pulsed-beams in the transverse direction  
(that is, the initial dependence on $x$ and $y$) is taken as Gaussian, 
since this choice enables the explicit calculation of the integrals 
with respect to $x$ and $y$ on the right hand side of Eq. (\ref{lp149}). 
The numerical simulations with the two types of initial pulsed-beams, 
which posses widely different temporal (longitudinal) profiles, 
provide a careful test to the perturbation theory prediction  
for universal behavior of the longitudinal part in the expression  
for the collision-induced amplitude shift.

The initial condition for a collision between two Gaussian pulsed-beams 
is given by: 
\begin{eqnarray}&&
\!\!\!\!\!\!\!\!
\psi_{1}(t,x,y,0)=A_{1}(0)\exp\left[-\frac{t^{2}}{2W^{(t)2}_{10}}
-\frac{x^{2}}{2W^{(x)2}_{10}}
-\frac{y^{2}}{2W^{(y)2}_{10}} + i\alpha_{10} \right],
\nonumber \\&&
\!\!\!\!\!\!\!\!
\psi_{2}(t,x,y,0)=A_{2}(0)\exp\left[
-\frac{(t-t_{20})^{2}}{2W^{(t)2}_{20}}
-\frac{x^{2}}{2W^{(x)2}_{20}}
-\frac{y^{2}}{2W^{(y)2}_{20}} + i\alpha_{20} \right].  
\label{lp151}
\end{eqnarray}  
Additionally, the initial condition for a collision between two 
Cauchy-Lorentz-Gaussian pulsed-beams is given by: 
 \begin{eqnarray}&&
 \!\!\!\!\!\!\!\!
\psi_{1}(t,x,y,0)=A_{1}(0)\left[1 + \frac{2t^{4}}{W^{(t)4}_{10}} \right]^{-1}
\exp\left[-\frac{x^{2}}{2W^{(x)2}_{10}}
-\frac{y^{2}}{2W^{(y)2}_{10}} + i\alpha_{10} \right],
\nonumber \\&&
\!\!\!\!\!\!\!\!
\psi_{2}(t,x,y,0)=A_{2}(0)\left[1 + \frac{2(t-t_{20})^{4}}{W^{(t)4}_{20}} \right]^{-1}
\exp\left[-\frac{x^{2}}{2W^{(x)2}_{20}}
-\frac{y^{2}}{2W^{(y)2}_{20}} + i\alpha_{20} \right]. 
\nonumber \\&&
\label{lp152}
\end{eqnarray}    
The calculation of the collision-induced amplitude shift by using  
Eq. (\ref{lp149}) is similar to the one that was presented in 
subsection  \ref{simu_universality} for the two-dimensional 
problem. This calculation yields the following expression for 
$\Delta A_{1}^{(c)}$: 
\begin{eqnarray}&&
\Delta A_{1}^{(c)}=
-\frac{2b\epsilon_{3} A_{1}(z_{c}^{-})A_{2}^{2}(z_{c}^{-})}{|d_{11}|}
\frac{W_{10}^{(x)}W_{10}^{(y)}W_{20}^{(t)}W_{20}^{(x)2}W_{20}^{(y)2}}
{(W_{10}^{(x)2} + W_{20}^{(x)2})^{1/2}(W_{10}^{(y)2} + W_{20}^{(y)2})^{1/2}}
\nonumber \\&&
\times 
\frac{1}
{(W_{10}^{(x)2}W_{20}^{(x)2} + 4d_{2}^{2}z_{c}^{2})^{1/2}
(W_{10}^{(y)2}W_{20}^{(y)2} + 4d_{2}^{2}z_{c}^{2})^{1/2}},
\label{lp153}     
\end{eqnarray}    
where $b=\pi^{1/2}$ for Gaussian pulsed-beams, and $b=3\pi/2^{11/4}$ for 
Cauchy-Lorentz-Gaussian pulsed-beams. 
The longitudinal part of the expression for $\Delta A_{1}^{(c)}$, 
$c_{p2}^{(t)}W_{20}^{(t)}=bW_{20}^{(t)}$, is universal. 
On the other hand, the transverse part of the expression, 
which is given by 
\begin{eqnarray} &&
\!\!\!\!\!\!\!\!\!
\mbox{transverse factor}=
\frac{W_{10}^{(x)}W_{10}^{(y)}W_{20}^{(x)2}W_{20}^{(y)2}}
{(W_{10}^{(x)2} + W_{20}^{(x)2})^{1/2}
(W_{10}^{(y)2} + W_{20}^{(y)2})^{1/2}}
\nonumber \\&&
\times 
\frac{1}
{(W_{10}^{(x)2}W_{20}^{(x)2} + 4d_{2}^{2}z_{c}^{2})^{1/2}
(W_{10}^{(y)2}W_{20}^{(y)2} + 4d_{2}^{2}z_{c}^{2})^{1/2}}, 
\label{lp154}
\end{eqnarray}           
is clearly not universal. We define the quantity $\Delta A_{1}^{(c)(s)}$, 
which is used for measuring the deviation of the $d_{11}$ dependence 
of $\Delta A_{1}^{(c)}$ from the $1/|d_{11}|$ scaling, by a simple 
generalization of the definition in the two-dimensional case. 
For this purpose, we note that the collision distance $z_{c}$ satisfies 
$z_{c}=(t_{10}-t_{20})/d_{11}$. $\Delta A_{1}^{(c)(s)}$ is defined as 
the approximate expression for the collision-induced amplitude shift   
that is obtained by neglecting the terms 
$4d_{2}^{2}(t_{10}-t_{20})^2/d_{11}^{2}$ in the denominator 
of Eq. (\ref{lp153}). Therefore, $\Delta A_{1}^{(c)(s)}$ is given by: 
\begin{eqnarray}&&
\!\!\!\!\!\!\!\!\!
\Delta A_{1}^{(c)(s)}=
-\frac{2b\epsilon_{3} A_{1}(z_{c}^{-})A_{2}^{2}(z_{c}^{-})}{|d_{11}|}
\frac{W_{20}^{(t)}W_{20}^{(x)}W_{20}^{(y)}}
{(W_{10}^{(x)2} + W_{20}^{(x)2})^{1/2}
(W_{10}^{(y)2} + W_{20}^{(y)2})^{1/2}}. 
\nonumber \\&&
\label{lp155}     
\end{eqnarray}                                                                                                         
It is clear from the definition of $\Delta A_{1}^{(c)(s)}$ that the 
difference $|\Delta A_{1}^{(c)} - \Delta A_{1}^{(c)(s)}|$
measures the deviation of the $d_{11}$ dependence of $\Delta A_{1}^{(c)}$ 
from the $1/|d_{11}|$ scaling observed in spatial dimension 1.

We check the perturbation theory predictions for $\Delta A_{1}^{(c)}$ 
and for the universality of the longitudinal part in the expression 
for $\Delta A_{1}^{(c)}$ by numerical simulations with Eq. (\ref{lp140}) 
with the initial conditions (\ref{lp151}) and (\ref{lp152}). 
These initial conditions possess very different temporal (longitudinal) 
pulsed-beam profiles. In particular, the tails of the Gaussian pulsed-beams 
decay rapidly with increasing values of $|t|$ and $|t-t_{20}|$. 
For such pulsed-beams, the sharp-peak approximation that is used in the 
derivation of Eqs. (\ref{lp144}) and (\ref{lp149}) is expected to hold. 
In contrast, the tails of the Cauchy-Lorentz-Gaussian pulsed-beams are 
slowly decaying with increasing values of $|t|$ and $|t-t_{20}|$, 
and the decay is characterized by a power-law. It is unclear if the 
sharp-peak approximation is valid for pulsed-beams with such slowly decaying 
tails. Therefore, the numerical simulations with the initial conditions 
(\ref{lp151}) and (\ref{lp152}) provide a careful check for the validity 
of the perturbation theory approximations for pulsed-beams with widely 
different temporal (longitudinal) distributions. In this way, 
the simulations help to establish the universality of the 
longitudinal part in the expression for $\Delta A_{1}^{(c)}$.

Equation (\ref{lp140}) is numerically solved by the split-step method 
with periodic boundary conditions \cite{Agrawal2001,Yang2010}. Since we 
are interested in fast collisions, we perform the simulations for 
$d_{11}$ values in the intervals $4 \le |d_{11}| \le 60$. 
The other physical parameters values in Eq. (\ref{lp140}) 
are chosen as $\epsilon_{3}=0.01$ and $d_{2}=1.5$, 
as an example. The parameter values of the initial pulsed-beams are 
$A_{j}(0)=1$, $\alpha_{j0}=0$, $t_{20}=\pm 15$, $W_{10}^{(t)} = 2$, 
$W_{10}^{(x)} = 3$, $W_{10}^{(y)} = 4$, $W_{20}^{(t)} = 3$, 
$W_{20}^{(x)} = 4$, and $W_{20}^{(y)} = 5$. The final propagation 
distance is $z_{f}=2z_{c}=-2t_{20}/d_{11}$. We emphasize that results 
similar to the ones presented in the current subsection are obtained 
in simulations with other values of the physical parameters. 
For each initial condition, we compare the numerically obtained 
dependence of $\Delta A_{1}^{(c)}$ on $d_{11}$ with the theoretical 
predictions of Eqs. (\ref{lp153}) and (\ref{lp155}). We also describe 
the behavior of the relative errors in the approximation of 
$\Delta A_{1}^{(c)}$ (in percentage), $E_{r}^{(1)}$ and $E_{r}^{(2)}$, 
which were defined in subsection \ref{simu_universality}.

\begin{figure}[ptb]
\begin{center}
\epsfxsize=10cm  \epsffile{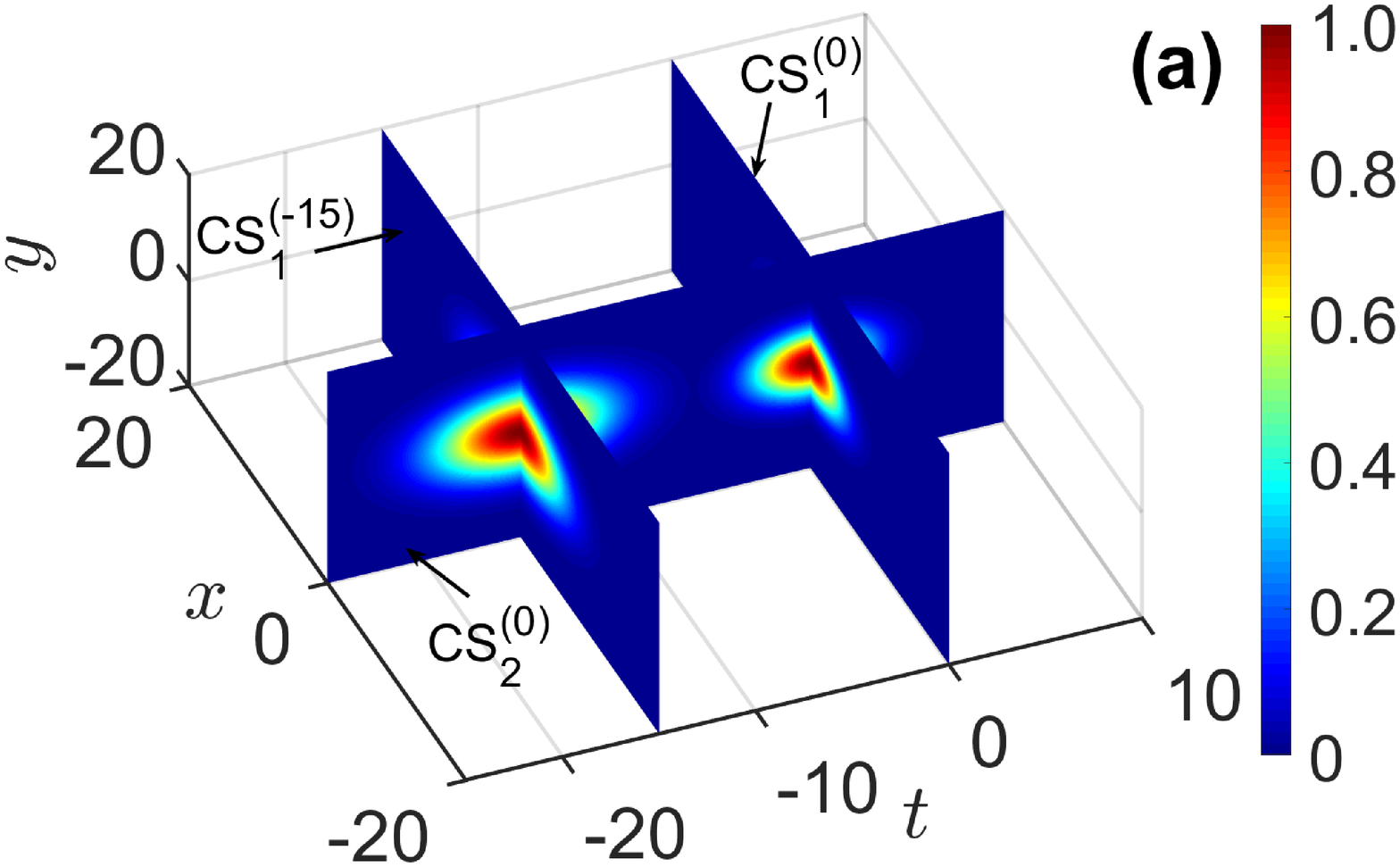} \\
\epsfxsize=10cm  \epsffile{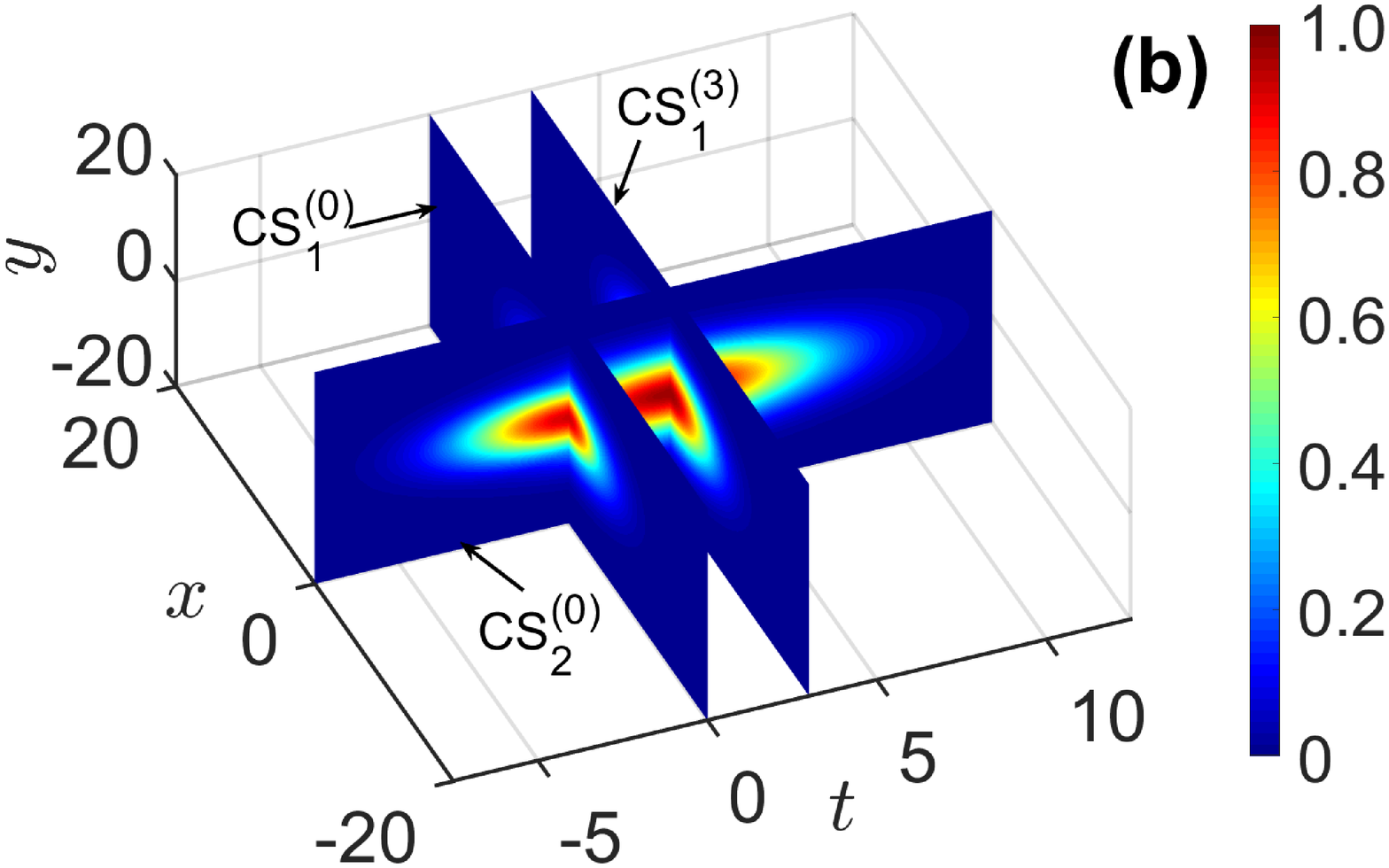} \\
\epsfxsize=10cm  \epsffile{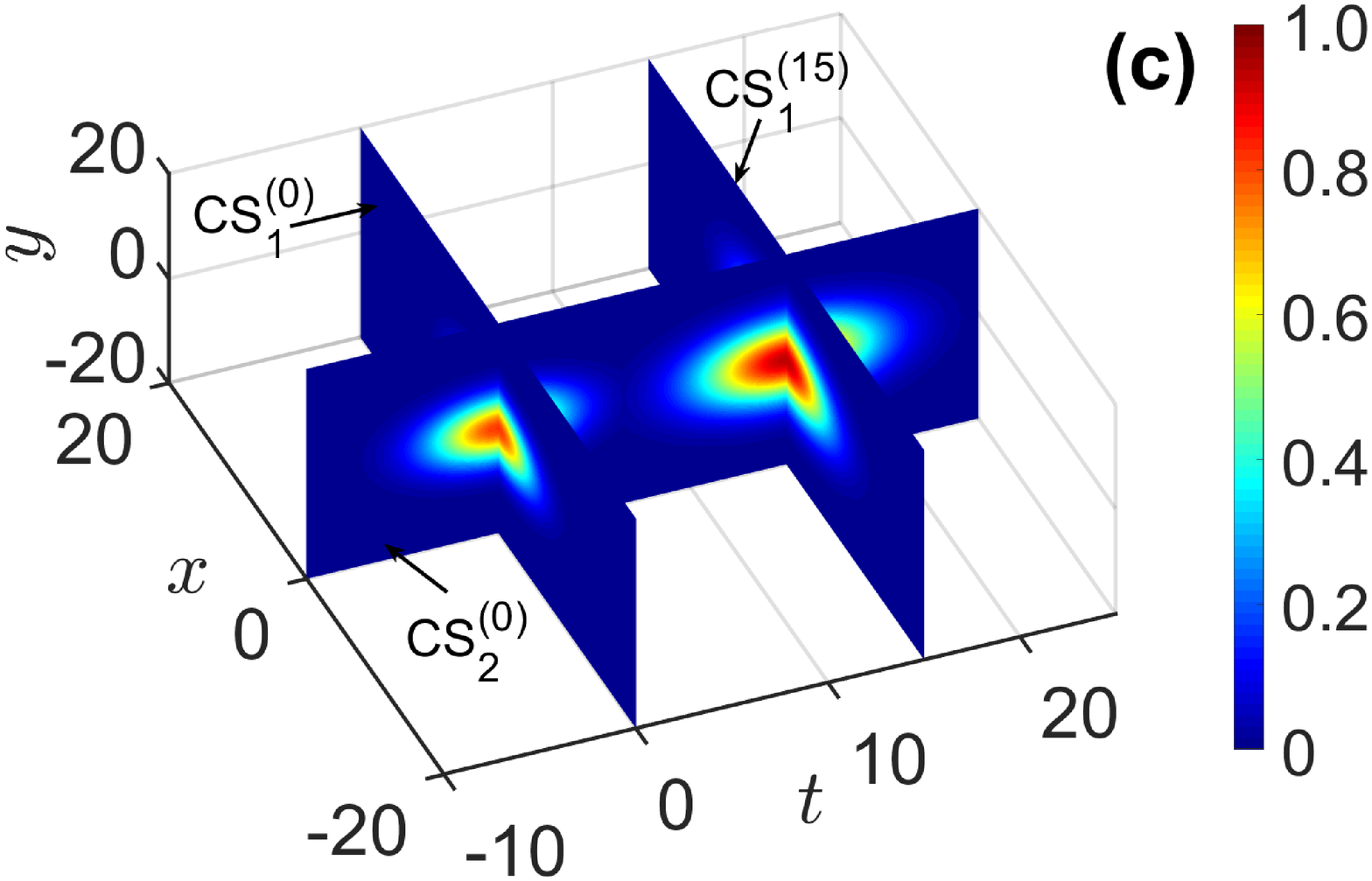} 
\end{center}
\caption{(Color online) 
Contour plots of the pulsed-beam shapes $|\psi_{j}(t,x,y,z)|$ 
on three planes in the $txy$ space at $z=0$ (a), 
$z=z_{i}=0.9$ (b), and $z=z_{f}=1.5$ (c) in a fast collision between 
two Gaussian pulsed-beams. The physical parameter values are 
$\epsilon_{3}=0.01$, $d_{2}=1.5$, and $d_{11}=20$.
The plots represent the beam shapes obtained by numerical solution 
of Eq. (\ref{lp140}) with the initial condition (\ref{lp151}).
The plane $x=0$ is denoted by $CS^{(0)}_{2}$, and the planes 
$t=t_{j}(z)$ with $j=1,2$ are denoted by $CS^{(t_{j}(z))}_{1}$.}
\label{fig11}
\end{figure}

\begin{figure}[ptb]
\begin{center}
\epsfxsize=10cm \epsffile{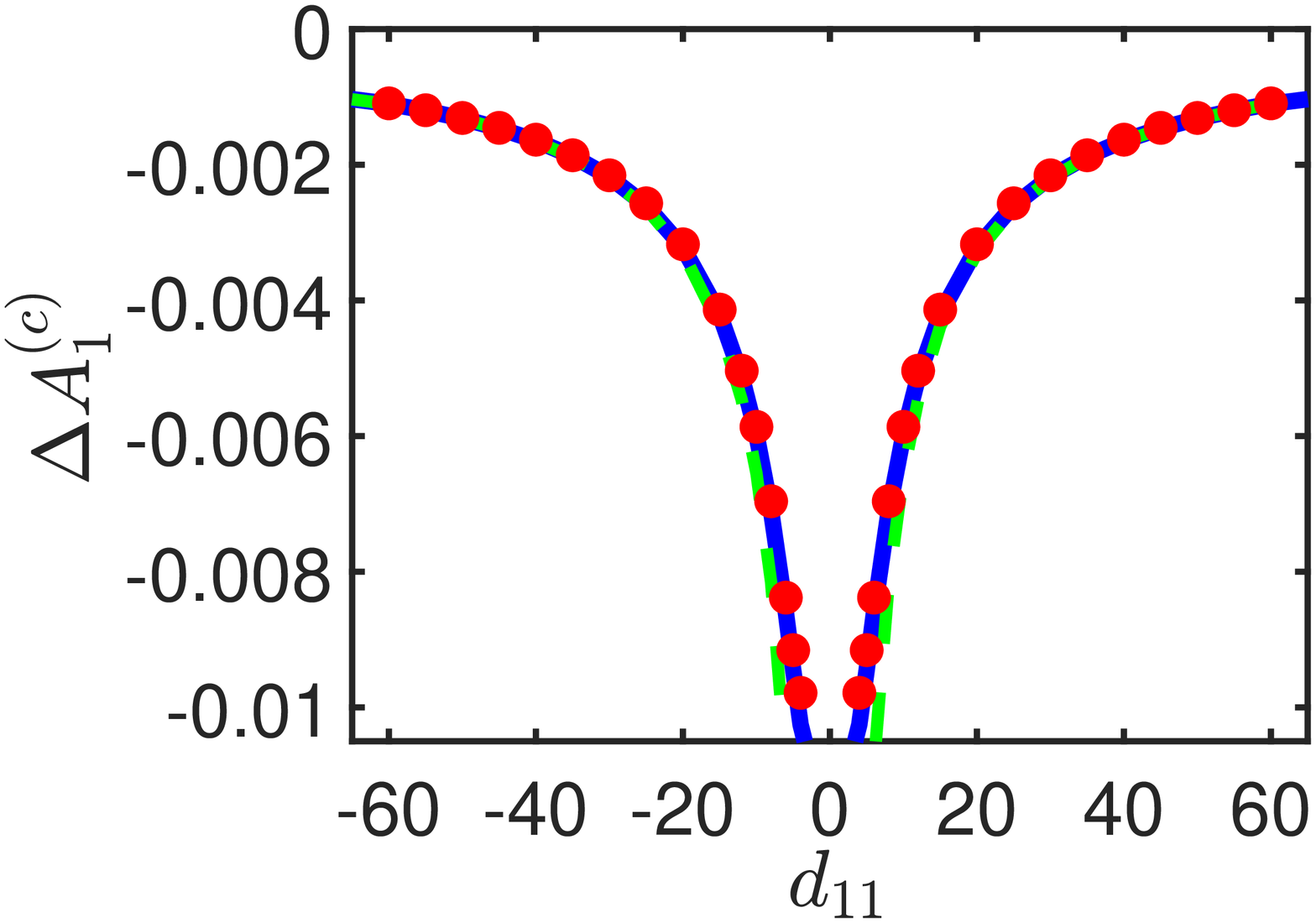}
\end{center}
\caption{(Color online) 
The collision-induced amplitude shift of pulsed-beam 1 
$\Delta A_{1}^{(c)}$ vs the first-order dispersion coefficient 
$d_{11}$ in a fast collision between two Gaussian pulsed-beams 
with parameter values $\epsilon_{3}=0.01$ and $d_{2}=1.5$. 
The red circles represent the result obtained by numerical 
simulations with Eq. (\ref{lp140}) with the initial condition (\ref{lp151}). 
The solid blue and dashed green curves represent the theoretical   
predictions of Eqs. (\ref{lp153}) and (\ref{lp155}), respectively.}
\label{fig12}
\end{figure}

We first discuss the simulations results for Gaussian pulsed-beams. 
Figure \ref{fig11} shows the values of $|\psi_{j}(t,x,y,z)|$ obtained 
in the simulation with $d_{11}=20$ at three specific planes (cross-sections) 
at the distances $z=0$, $z=z_{i}=0.9$, and $z=z_{f}=1.5$ \cite{zi_values}. 
At each distance, we choose the three cross-sections (planes) such that 
the main bodies of the pulsed-beams are shown clearly \cite{cross_sections}. 
In particular, one cross-section, which is denoted by $CS^{(0)}_{2}$, 
is located at the plane $x=0$. Additionally, the other two cross-sections 
are located at the planes $t=t_{j}(z)$, where $j=1,2$, $t_{j}(z)$ is the 
$t$ coordinate of the $j$th pulsed-beam's center, and $z$ can take the values 
$0$, $z_{i}$, or $z_{f}$. The latter cross-sections are denoted by 
$CS^{(t_{j}(z))}_{1}$, where $j=1,2$. It is seen that the pulsed-beams 
experience broadening due to the effects of second-order dispersion 
and diffraction. Additionally, the maximum values of  $|\psi_{j}(t,x,y,z)|$ 
decrease with increasing distance, mainly due to the broadening. 
The dependence of $\Delta A_{1}^{(c)}$ on $d_{11}$ obtained in the simulations 
is shown in Fig. \ref{fig12} together with the two theoretical predictions 
of Eqs. (\ref{lp153}) and (\ref{lp155}). The agreement between the simulations 
result and the prediction of Eq. (\ref{lp153}) is very good. More specifically, 
the relative error $E_{r}^{(1)}$ is smaller than $1.9\%$ for 
$10 \le |d_{11}| \le 60$ and smaller than $4.5\%$ for $4 \le |d_{11}| <10$. 
In addition, we observe good agreement between the simulations result and 
the more crude approximation $\Delta A_{1}^{(c)(s)}$ for large $|d_{11}|$ values, 
but there is a noticeable difference between the results for intermediate 
$|d_{11}|$ values. In particular, the relative 
error $E_{r}^{(2)}$ is smaller than $10.4\%$ for $10 \le |d_{11}| \le 60$ 
and smaller than $39.3\%$ for $4 \le |d_{11}| <10$. Thus, as expected from 
Eqs.  (\ref{lp153}) and (\ref{lp155}), the deviation of the $d_{11}$ dependence 
of $\Delta A_{1}^{(c)}$ from the $1/|d_{11}|$ scaling increases with 
decreasing value of $|d_{11}|$.

We now describe the results of the simulations for collisions 
between Cauchy-Lorentz-Gaussian pulsed-beams, which serve as 
an example for pulsed-beams with tails that exhibit slow 
(power-law) temporal decay. In this case, it is not clear 
if the sharp-peak approximation used in the perturbative 
calculation is valid. For brevity, we discuss the 
simulations results without showing the corresponding 
figures. The numerical simulation with $d_{11}=20$ shows 
that the pulsed-beams experience considerable broadening 
and develop extended tails due to second-order dispersion 
and diffraction. As a result, the maximum values of       
$|\psi_{j}(t,x,y,z)|$ decrease with increasing $z$. 
Furthermore, despite the broadening of the pulsed-beams, 
the agreement between the numerical simulations result for 
$\Delta A_{1}^{(c)}$ and the theoretical prediction of 
Eq. (\ref{lp153}) is very good. In particular, the relative 
error $E_{r}^{(1)}$ is less than $1.3\%$ for $10 \le |d_{11}| \le 60$ 
and less than $10.2\%$ for $4 \le |d_{11}| <10$. 
The values of $E_{r}^{(1)}$ are comparable to the values 
obtained for collisions between Gaussian pulsed-beams. 
Based on these findings and on similar results obtained with 
other parameter values, we conclude that the longitudinal part 
of the expression for $\Delta A_{1}^{(c)}$ is indeed universal, 
since it is not sensitive to the details of the initial pulsed-beams shapes. 
We also note that the relative error $E_{r}^{(2)}$ in 
the approximation of the amplitude shift by $\Delta A_{1}^{(c)(s)}$         
is smaller than $9.8\%$ for $10 \le |d_{11}| \le 60$ 
and smaller than $42.9\%$ for $4 \le |d_{11}| <10$. 
These values are significantly larger than the corresponding values 
of $E_{r}^{(1)}$. Thus, the agreement between the numerical result 
and the perturbation theory prediction for $\Delta A_{1}^{(c)}$ 
is significantly better than the agreement between the numerical 
result and the cruder approximation $\Delta A_{1}^{(c)(s)}$. 
Additionally, as in the case of Gaussian pulsed-beams, 
the deviation of the $d_{11}$ dependence of $\Delta A_{1}^{(c)}$ 
from the $1/|d_{11}|$ scaling increases with decreasing value 
of $|d_{11}|$.

\section{Conclusions}
\label{conclusions}

We studied the dynamics of fast collisions between two optical 
beams in linear optical media with weak cubic loss in spatial 
dimension higher than 1. For this purpose, we introduced a perturbation method, 
which generalizes the perturbation method developed in 
Refs. \cite{PNH2017B,NHP2020} for the one-dimensional case 
in three major ways. First, it extends the perturbative 
calculation from spatial dimension 1 to spatial dimension 2, 
and enables the extension of the calculation to a general 
spatial dimension in a straightforward manner. 
Second, it provides a perturbative calculation of 
the collision-induced dynamics of the beam shape both in the 
collision interval and outside of the collision interval. 
In contrast, the perturbative calculation in Refs. \cite{PNH2017B,NHP2020} 
was limited to the collision interval only. Third, it enables 
the discovery and analysis of several collision-induced effects, 
which exist only in spatial dimension higher than 1.

We used the generalized two-dimensional version of the perturbation 
method to obtain formulas for the collision-induced changes in 
the beam shapes and amplitudes. This was done both for a general 
initial condition and for the important case of an initial condition, 
which is separable for both beams. We found that for a general 
initial condition, the collision leads to a change in the beam shape 
in the direction transverse to the relative velocity vector 
between the beam centers. Additionally, we found that for a separable 
initial condition, the beam shape in the longitudinal direction is not 
changed by the collision within the leading order of the perturbative 
calculation. Moreover, we showed that for a separable initial condition, 
the longitudinal part in the expression for the collision-induced 
amplitude shift is universal, while the transverse part is proportional 
to the integral of the product of the beam intensities with respect to 
the transverse coordinate. We also demonstrated that the same behavior 
of the longitudinal and transverse parts of the expression for the 
amplitude shift exists in collisions between pulsed optical beams 
in spatial dimension 3.

We checked these predictions of the generalized perturbation method 
along with other predictions concerning the effects on the collision 
of partial beam overlap and anisotropy in the initial condition 
by extensive numerical simulations with the perturbed linear 
propagation model in spatial dimensions 2 and 3. The simulations 
in spatial dimension 2 were carried out for four different two-beam 
collision setups. These setups demonstrate four major collisional 
effects and properties that either exist only in 
spatial dimension higher than 1, or are qualitatively 
different from their one-dimensional counterparts. 
(1) The universality of the longitudinal part in the expression 
for the collision-induced amplitude shift. 
(2) The effect of partial beam overlap. 
(3) The effect of anisotropy in the initial condition. 
(4) The collision-induced change in the beam shape in the transverse direction. 
The prediction for universal behavior of the longitudinal part in the expression 
for the amplitude shift was also checked in spatial dimension 3 by 
numerical simulations of collisions between pulsed optical beams.

In all the simulation setups we obtained very good agreement between 
the perturbation theory and the numerical simulations. In particular, 
in setup (1), the simulations showed that the longitudinal part in the 
expression for the collision-induced amplitude shift is universal in 
the sense that it is not sensitive to the details of the initial beam 
shapes, and that this is true in both spatial dimensions 2 and 3. 
Additionally, the simulations in setup (2) verified the validity of 
the transverse part in the expression for the amplitude shift and 
demonstrated that the generalized perturbation method can be 
employed for fast collisions between partially overlapping beams. 
The simulations in the anisotropic setup [setup (3)] verified 
the complex dependence of the expression for the amplitude shift 
on the orientation angle $\theta_{0}$, which was predicted by the 
perturbation theory. We attributed this complex dependence to the nonseparable 
nature of the initial condition in the anisotropic setup. Moreover, the 
numerical simulations in setup (4) confirmed the perturbation theory predictions 
for the collision-induced change in the beam shape in the transverse direction. 
Based on the results of the latter simulations we concluded that the generalized 
perturbation method that we developed in the current paper enables accurate 
calculation of both the change in the beam shape and the dynamics of the 
beam amplitude in fast two-beam collisions in the presence of weak cubic loss.

In summary, our study extended the results of the 
previous works in Refs. \cite{PNH2017B,QMN2018,NHP2020} in two major ways. 
First, it generalized the perturbation method of Refs. \cite{PNH2017B,QMN2018,NHP2020} 
from spatial dimension 1 to a general spatial dimension. 
Second, it demonstrated a variety of collision-induced physical effects, 
which exist only for spatial dimension higher than 1. We point out 
that in another study, we developed a similar perturbation method 
for analyzing fast two-pulse collisions in systems described by 
linear diffusion-advection equations with weak quadratic loss 
in spatial dimension higher than 1 \cite{PHN2020}. 
Using the latter perturbation method and numerical simulations, 
we found that the collision-induced changes in pulse shapes and 
amplitudes in the perturbed linear diffusion-advection systems exhibit 
similar behavior to the one reported in the current paper. 
Thus, the perturbation methods developed in the current paper 
and in Ref. \cite{PHN2020} are very valuable tools for analyzing 
fast-collision dynamics in linear physical systems with weak 
nonlinear dissipation. Indeed, as described in the current paper 
and in Ref. \cite{PHN2020}, these methods enabled deep insight 
into many collision-induced effects in the perturbed linear 
systems in spatial dimensions 2 and 3. 
We also comment that detailed analytic results on collisions 
between pulse (or beam) solutions of linear or nonlinear evolution models 
in the presence of nonlinear dissipation in spatial dimension 
higher than 1 are quite scarce. Therefore, the current work 
and the work in Ref. \cite{PHN2020} also significantly extended  
the understanding of the more general high-dimensional problem 
of two-pulse (or two-beam) collisions in the presence of nonlinear dissipation.

\section*{Acknowledgments}
Q.M.N. and T.T.H. are supported by the Vietnam National Foundation 
for Science and Technology Development (NAFOSTED) under Grant No. 107.99-2019.340.

\appendix
\section{The relation between $\Delta A_{1}^{(c)}$ and $\Delta\Phi_{1}(x,y,z_{c})$}
\label{appendA}      
         
In this Appendix, we present the derivation of the relation (\ref{lp11}) 
between the collision-induced amplitude shift $\Delta A_{1}^{(c)}$ and the 
collision-induced change in the beam shape $\Delta\Phi_{1}(x,y,z_{c})$. 
This relation is used in subsection \ref{lp_general_IC} to obtain 
Eq. (\ref{lp13}) from Eq. (\ref{lp10}). The derivation is carried out 
for a collision in spatial dimension 2, but it can be generalized in 
a straightforward manner to spatial dimension $n$.

We first recall that the amplitude dynamics of a single beam propagating 
in the presence of diffraction and linear or nonlinear loss can be approximately 
determined by an energy balance equation of the form 
$\partial_{z}\int_{-\infty}^{\infty} \!\!\! dx 
\int_{-\infty}^{\infty} \!\!\! dy \, |\psi_{1}(x,y,z)|^{2}=...$, 
where the right hand side of the equation is determined by the type  
of the loss perturbation. [See, for example, Eqs. (\ref{appB_1}) and (\ref{appB_4}) 
in Appendix \ref{appendB}]. A fast collision in the presence of nonlinear loss  
at the distance $z=z_{c}$ leads to a jump in the value of $\int_{-\infty}^{\infty} \!\!\! dx 
\int_{-\infty}^{\infty} \!\!\! dy \, |\psi_{1}(x,y,z)|^{2}$ at $z=z_{c}$. 
Therefore, for a fast collision, the term $\partial_{z}\int_{-\infty}^{\infty} \!\!\! dx 
\int_{-\infty}^{\infty} \!\!\! dy \, |\psi_{1}(x,y,z)|^{2}$ 
in the equation that determines the amplitude dynamics can be replaced by:
\begin{eqnarray}&&
\Delta_{P}=
\int_{-\infty}^{\infty} \!\!\! dx 
\int_{-\infty}^{\infty} \!\!\! dy \, |\psi_{1}(x,y,z_{c}^{+})|^{2} 
\nonumber \\&&
- \int_{-\infty}^{\infty} \!\!\! dx 
\int_{-\infty}^{\infty} \!\!\! dy \, |\psi_{1}(x,y,z_{c}^{-})|^{2}. 
\label{appA_1}
\end{eqnarray}      
The derivation of the relation between $\Delta A_{1}^{(c)}$ and $\Delta\Phi_{1}(x,y,z_{c})$
is based on finding two expressions for $\Delta_{P}$, one involving $\Delta A_{1}^{(c)}$ 
and the other involving $\Delta\Phi_{1}(x,y,z_{c})$, and on equating the two expressions.

We note that in the limit of a fast collision, $\phi_{1}(x,y,z_{c}^{-}) \simeq 0$, 
and therefore, $\psi_{1}(x,y,z_{c}^{-}) \simeq \psi_{10}(x,y,z_{c}^{-})$. 
Using Eqs. (\ref{lp_add1}) and (\ref{lp_add4}), we find: 
$\psi_{1}(x,y,z_{c}^{-}) \simeq A_{1}(z_{c}^{-})
\tilde\Psi_{10}(x,y,z_{c})\exp[i\chi_{10}(x,y,z_{c})]$.    
We use the latter approximation for $\psi_{1}(x,y,z_{c}^{-})$ to evaluate 
the second integral on the right hand side of Eq. (\ref{appA_1}). 
This calculation yields: 
\begin{eqnarray}&&
\int_{-\infty}^{\infty} \!\!\! dx 
\int_{-\infty}^{\infty} \!\!\! dy \, |\psi_{1}(x,y,z_{c}^{-})|^{2}= 
C_{p1}A_{1}^{2}(z_{c}^{-}),
\label{appA_2}
\end{eqnarray}            
where $C_{p1}$ is given by Eq. (\ref{lp12}). Since $\phi_{1}(x,y,z_{c}^{-}) \simeq 0$, 
$\Delta\phi_{1}(x,y,z_{c})$ can be written as: 
$\Delta\phi_{1}(x,y,z_{c}) \simeq \phi_{1}(x,y,z_{c}^{+}) 
- \phi_{1}(x,y,z_{c}^{-}) \simeq \phi_{1}(x,y,z_{c}^{+})$.
Using this relation together with Eq. (\ref{lp2}) and the 
definition of $\psi_{10}$, we obtain:  
\begin{eqnarray}&&
\psi_{1}(x,y,z_{c}^{+})=\psi_{10}(x,y,z_{c})+\Delta\phi_{1}(x,y,z_{c}).
\label{appA_3}
\end{eqnarray} 
We now use Eqs. (\ref{appA_3}), (\ref{lp_add1}), and (\ref{lp_add4}) 
along with the definition of $\Delta\Phi_{1}$ to obtain  
\begin{eqnarray}&&
\int_{-\infty}^{\infty} \!\!\! dx 
\int_{-\infty}^{\infty} \!\!\! dy \, |\psi_{1}(x,y,z_{c}^{+})|^{2} = 
\nonumber \\&& 
\int_{-\infty}^{\infty} \!\!\! dx \int_{-\infty}^{\infty} \!\!\! dy \,
\left[A_{1}(z_{c}^{-})\tilde\Psi_{10}(x,y,z_{c})\!
+\!\Delta\Phi_{1}(x,y,z_{c})\right]^{2}.  
\label{appA_4}
\end{eqnarray}          
We expand the integrand on the right hand side of Eq.  (\ref{appA_4}), 
while keeping only the first two leading terms. We obtain 
\begin{eqnarray}&&
\int_{-\infty}^{\infty} \!\!\! dx 
\int_{-\infty}^{\infty} \!\!\! dy \, |\psi_{1}(x,y,z_{c}^{+})|^{2} 
\simeq C_{p1}A_{1}^{2}(z_{c}^{-})
\nonumber \\&& 
+ 2A_{1}(z_{c}^{-}) \int_{-\infty}^{\infty} \!\!\! dx 
\int_{-\infty}^{\infty} \!\!\! dy \,
\tilde\Psi_{10}(x,y,z_{c}) \Delta\Phi_{1}(x,y,z_{c}).  
\label{appA_5}
\end{eqnarray}     
Substitution of Eqs. (\ref{appA_2}) and (\ref{appA_5}) into Eq. (\ref{appA_1})  
yields the first expression for $\Delta_{P}$:
\begin{eqnarray}&&
\Delta_{P}=
2A_{1}(z_{c}^{-}) \int_{-\infty}^{\infty} \!\!\! dx 
\int_{-\infty}^{\infty} \!\!\! dy \,
\tilde\Psi_{10}(x,y,z_{c}) \Delta\Phi_{1}(x,y,z_{c}).  
\label{appA_6}
\end{eqnarray}     
The second expression for $\Delta_{P}$ is obtained by writing 
$\int_{-\infty}^{\infty} \!\!\! dx \int_{-\infty}^{\infty} \!\!\! dy \, 
|\psi_{1}(x,y,z_{c}^{+})|^{2}$ in terms of $\Delta A_{1}^{(c)}$ in the following manner:  
\begin{eqnarray}&&
\int_{-\infty}^{\infty} \!\!\! dx \int_{-\infty}^{\infty} \!\!\! dy \, 
|\psi_{1}(x,y,z_{c}^{+})|^{2}=
\left(A_{1}(z_{c}^{-}) +\Delta A_{1}^{(c)}\right)^{2}
\int_{-\infty}^{\infty} \!\!\! dx \int_{-\infty}^{\infty} \!\!\! dy \, 
\tilde\Psi_{10}^{2}(x,y,z_{c})
\nonumber \\&&
\simeq
C_{p1}A_{1}^{2}(z_{c}^{-})+2C_{p1}A_{1}(z_{c}^{-}) \Delta A_{1}^{(c)}.
\label{appA_7}
\end{eqnarray}  
Substitution of Eqs. (\ref{appA_2}) and (\ref{appA_7}) into Eq. (\ref{appA_1})  
yields the second expression for $\Delta_{P}$: 
\begin{eqnarray}&&
\Delta_{P}=2C_{p1}A_{1}(z_{c}^{-}) \Delta A_{1}^{(c)}.
\label{appA_8}
\end{eqnarray} 
Equating the right hand sides of Eqs. (\ref{appA_6}) and (\ref{appA_8}), 
we arrive at: 
\begin{eqnarray}&&
\!\!\!\!\!\!\!\!\!\!\!\!\!\!
\Delta A_{1}^{(c)}=C_{p1}^{-1}
\!\!\int_{-\infty}^{\infty} \!\!\!\!\! dx 
\!\int_{-\infty}^{\infty} \!\!\!\!\! dy
\;\tilde\Psi_{10}(x,y,z_{c})\Delta\Phi_{1}(x,y,z_{c}), 
\label{appA_9}
\end{eqnarray}       
which is  Eq. (\ref{lp11}). We point out that the relation (\ref{appA_9}) 
can be generalized to spatial dimension $n$ by replacing all the 
two-dimensional integrals in the equations in the current Appendix 
by $n$-dimensional integrals with respect to the $n$ spatial coordinates.

\section{Amplitude dynamics in the perturbed single-beam 
propagation problem}
\label{appendB}
In this Appendix, we derive the equation for the dynamics 
of the beam amplitudes in the perturbed single-beam propagation 
problem, i.e., for a single beam propagating in the presence 
of weak cubic loss. This equation is used for calculating 
the amplitude values in the approximate expressions for the 
$\psi_{j0}$. It is also used for calculating the values 
of $A_{j}(z_{c}^{-})$ in Eqs. (\ref{lp10}), (\ref{lp13}), (\ref{lp44}), and (\ref{lp46}) 
for $\Delta\Phi_{1}(x,y,z_{c})$ and $\Delta A_{1}^{(c)}$, 
and in other equations in section \ref{lp_2D}. 
We also show that the effects of weak linear loss can be 
incorporated in a straightforward manner in the equation for 
amplitude dynamics for single-beam propagation. Moreover, 
we show that the effects of weak linear loss do not change 
the form of the expressions for the collision-induced amplitude shifts.

Consider the propagation of a single beam in the presence of 
diffraction, beam steering, and weak cubic loss. The propagation 
is described by Eq. (\ref{lp3}) for beam 1 and by Eq. (\ref{lp4}) 
for beam 2. Employing energy balance calculations for these two 
equations, we obtain 
\begin{eqnarray}&&
\!\!\!\!\!\!\!\!\!\!\!\!
\partial_{z}\int_{-\infty}^{\infty} \!\!\!\!\! dx 
\int_{-\infty}^{\infty} \!\!\!\!\! dy \, |\psi_{j0}|^{2}\!=\!
-2\epsilon_{3}\int_{-\infty}^{\infty} \!\!\!\!\! dx \int_{-\infty}^{\infty} \!\!\!\!\! dy \,    
|\psi_{j0}|^{4}.
\label{appB_1}
\end{eqnarray}       
We now substitute the approximations to the $\psi_{j0}$, which are given by 
Eqs. (\ref{lp_add1})-(\ref{lp_add2}), into Eq. (\ref{appB_1}). This 
substitution yields the following equation for the $A_{j}$: 
\begin{eqnarray}&&
\!\!\!\!\!\!\!\!\!\!\!\!
C_{pj}\frac{dA_{j}^{2}}{dz}= 
-2\epsilon_{3}H_{4j}(z)A_{j}^{4} \,,  
\label{appB_2}
\end{eqnarray}     
where $H_{4j}(z)=\int_{-\infty}^{\infty} \!\!\!\!\! dx 
\int_{-\infty}^{\infty} \!\!\!\!\! dy \,\tilde\Psi_{j0}^{4}(x,y,z)$, 
$C_{p1}$ is given by Eq. (\ref{lp12}), and $C_{p2}$ is given by 
a similar equation, in which $\tilde\Psi_{10}^{2}(x,y,0)$ is replaced 
by $\tilde\Psi_{20}^{2}(x,y,0)$ on the right hand side.  
The solution of Eq. (\ref{appB_2}) on the interval $[0,z]$ is 
\begin{eqnarray}&&
\!\!\!\!\!\!\!\!\!\!\!\!
A_{j}(z)=\frac{A_{j}(0)}
{\left[1+2\epsilon_{3}\tilde H_{4j}(0,z)A_{j}^{2}(0)/C_{pj}\right]^{1/2}} \,,  
\label{appB_3}
\end{eqnarray}  
where $\tilde H_{4j}(0,z)=\int_{0}^{z} dz'\, H_{4j}(z')$.

It is straightforward to incorporate the effects of weak 
linear loss into the equation for the dynamics of the pulse amplitudes. 
In this case, single-beam propagation of beams 1 and 2 is described 
by Eqs. (\ref{lp3}) and (\ref{lp4}) with the terms $-i\epsilon_{1}\psi_{1}$ 
and $-i\epsilon_{1}\psi_{2}$ added on the right hand sides, where 
$0 < \epsilon_{1} \ll 1$ is the linear loss coefficient. 
Energy balance calculations for the two modified propagation equations 
yield the following equation 
\begin{eqnarray}&&
\!\!\!\!\!\!\!\!\!\!\!\!
\partial_{z}\int_{-\infty}^{\infty} \!\!\!\!\! dx 
\int_{-\infty}^{\infty} \!\!\!\!\! dy \, |\psi_{j0}|^{2}\!=\!
-2\epsilon_{1}\int_{-\infty}^{\infty} \!\!\!\!\! dx 
\int_{-\infty}^{\infty} \!\!\!\!\! dy \, |\psi_{j0}|^{2}
-2\epsilon_{3}\int_{-\infty}^{\infty} \!\!\!\!\! dx 
\int_{-\infty}^{\infty} \!\!\!\!\! dy \, |\psi_{j0}|^{4}.
\nonumber \\&&
\label{appB_4}
\end{eqnarray}     
Using the approximate expressions (\ref{lp_add1})-(\ref{lp_add2}) 
for the $\psi_{j0}$ in Eq. (\ref{appB_4}), we obtain 
\begin{eqnarray}&&
\!\!\!\!\!\!\!\!\!\!\!\!
C_{pj}\frac{dA_{j}^{2}}{dz}= 
-2\epsilon_{1}C_{pj}A_{j}^{2}
-2\epsilon_{3}H_{4j}(z)A_{j}^{4} \,.   
\label{appB_5}
\end{eqnarray}   
Equation (\ref{appB_5}) is a Bernoulli equation for $A_{j}^{2}(z)$. 
Its solution on the interval $[0,z]$ is given by: 
\begin{eqnarray}&&
\!\!\!\!\!\!\!\!\!\!\!\!
A_{j}(z)=\frac{A_{j}(0)e^{-\epsilon_{1}z}}
{\left[1+2\epsilon_{3}\tilde H_{4j}(0,z)A_{j}^{2}(0)/C_{pj}\right]^{1/2}} \,,  
\label{appB_6}
\end{eqnarray}  
where $\tilde H_{4j}(0,z)=\int_{0}^{z} dz'\, H_{4j}(z')e^{-2\epsilon_{1}z'}$.

We now show that the addition of the weak linear loss terms 
to Eq. (\ref{lp1}) does not alter the form of the expressions 
for the collision-induced changes in the beam shape and amplitude. 
For this purpose, we first note that the equation for $\phi_{1}$ 
in the leading order of the perturbative calculation 
is still Eq. (\ref{lp5}). As a result, the evolution     
of $\Phi_{1}$ in the collision interval is described by Eq. (\ref{lp7}).  
It follows that $\Delta\Phi_{1}(x,y,z_{c})$ and $\Delta A_{1}^{(c)}$ 
are still given by Eqs. (\ref{lp10}) and (\ref{lp13}). 
In addition, the evolution of $\phi_{1}$ in the post-collision interval 
is described by Eq. (\ref{lp21}), and as a result, $\phi_{1}$ is 
still given by Eq. (\ref{lp24}) in this interval. 
Thus, the addition of the weak linear loss terms to Eq. (\ref{lp1}) 
does not alter the form of the expressions for the collision-induced 
changes in the beam shape and amplitude in the leading order of the 
perturbative calculation. We point out that the weak linear loss 
affects the values of $\Delta\Phi_{1}(x,y,z_{c})$, $\Delta A_{1}^{(c)}$, 
and $\phi_{1}(x,y,z)$ only via the dependence of these quantities on the 
beam amplitudes. More specifically, in the absence of linear loss, 
these quantities are calculated with amplitude values that are given 
by Eq. (\ref{appB_3}), while in the presence of weak linear loss, 
these quantities are calculated with amplitude values that are given 
by Eq. (\ref{appB_6}).

\section{Derivation of Eq. (\ref{lp120})}
\label{appendC}
In this Appendix, we derive Eq. (\ref{lp120}) for the inverse Fourier transform of  
$\hat g_{12}^{(y)}(k_{2},z_{c})\exp[-i k_{2}^{2}(z-z_{c})]$
in the case where the initial condition for 
the collision problem is given by Eq. (\ref{lp61}).  
Equation (\ref{lp120}) is used in the calculation of 
$\phi_{1}(x,y,z)$ in the post-collision interval in 
subsection \ref{simu_reshaping}.

We first employ Eq. (\ref{lp54}) along with Eqs. (\ref{lp61}),  
(\ref{appD_5}), (\ref{appD_7}), and (\ref{appD_10}) to obtain 
an expression for the function $g_{12}^{(y)}(y,z_{c})$. 
We find:  
\begin{eqnarray}&&
g_{12}^{(y)}(y,z_{c})=
\frac{W_{10}^{(y)}W_{20}^{(y)2}}
{(W_{10}^{(y)4} + 4z_{c}^{2})^{1/4}
(W_{20}^{(y)4} + 4z_{c}^{2})^{1/2}}
\nonumber \\&&
\times 
\exp\left[-\tilde a_{2}^{2}(z_{c})y^{2}
-\frac{i}{2}\arctan\left(\frac{2z_{c}}{W_{10}^{(y)2}}\right)\right], 
\label{lp114}     
\end{eqnarray}  
where 
\begin{equation}
\tilde a_{2}^{2}(z_{c})= q_{1}(z_{c})+ iq_{2}(z_{c}), 
\label{lp115}
\end{equation}       
\begin{eqnarray} &&
\!\!\!\!\!\!\!\!\!\!\!\!\!\!
q_{1}(z_{c})=\frac{W_{10}^{(y)2}W_{20}^{(y)2} (2W_{10}^{(y)2} + W_{20}^{(y)2}) 
+ 4z_{c}^{2}(W_{10}^{(y)2} + 2W_{20}^{(y)2})}
{2(W_{10}^{(y)4} + 4z_{c}^{2}) (W_{20}^{(y)4} + 4z_{c}^{2})},
\nonumber \\&&
\label{lp116}
\end{eqnarray}
and 
\begin{eqnarray} &&
q_{2}(z_{c}) = -\frac{z_{c}}
{W_{10}^{(y)4} + 4z_{c}^{2}}. 
\label{lp117}
\end{eqnarray}
The Fourier transform of $g_{12}^{(y)}(y,z_{c})$ is: 
\begin{eqnarray}&&
\hat g_{12}^{(y)}(k_{2},z_{c})=
\frac{W_{10}^{(y)}W_{20}^{(y)2}(W_{10}^{(y)4} + 4z_{c}^{2})^{1/4}}
{\tilde a_{3}(z_{c})}
\nonumber \\&&
\times 
\exp\left[-\frac{k_{2}^{2}}{4\tilde a_{2}^{2}(z_{c})}
-\frac{i}{2}\arctan\left(\frac{2z_{c}}{W_{10}^{(y)2}}\right)\right], 
\label{lp118}     
\end{eqnarray}   
where  
\begin{equation}
\tilde a_{3}(z_{c})=
\left[2(W_{10}^{(y)4} + 4z_{c}^{2})
(W_{20}^{(y)4} + 4z_{c}^{2})\right]^{1/2}
\tilde a_{2}(z_{c}).  
\label{lp119}
\end{equation}  
Therefore, the inverse Fourier transform of  
$\hat g_{12}^{(y)}(k_{2},z_{c})\exp[-i k_{2}^{2}(z-z_{c})]$
is given by: 
\begin{eqnarray}&&
{\cal F}^{-1}\left(\hat g_{12}^{(y)}(k_{2},z_{c})
\exp[-i k_{2}^{2}(z-z_{c})]\right)=
\nonumber \\&&
\frac{W_{10}^{(y)}W_{20}^{(y)2}\exp\left[-q_{1}(z_{c})y^{2}/R_{1}^{4}(z,z_{c}) 
+i\chi_{1}^{(y)}(y,z)\right]}
{(W_{10}^{(y)4} + 4z_{c}^{2})^{1/4}
(W_{20}^{(y)4} + 4z_{c}^{2})^{1/2}R_{1}(z,z_{c})}, 
\label{lp120_a}     
\end{eqnarray}  
where 
\begin{equation}
R_{1}(z,z_{c})=
\left\{1 - 8q_{2}(z_{c})(z-z_{c}) 
+ 16[q_{1}^{2}(z_{c}) + q_{2}^{2}(z_{c})](z-z_{c})^{2}\right\}^{1/4},  
\label{lp121}
\end{equation}  
and 
\begin{eqnarray}&&
\chi_{1}^{(y)}(y,z)=
-\frac{1}{2}\arctan\left[\frac{2z_{c}}{W_{10}^{(y)2}}\right] 
-\frac{1}{2}\arctan\left[\frac{4q_{1}(z_{c})(z-z_{c})}{1 - 4q_{2}(z_{c})(z-z_{c})}\right]
\nonumber \\&&
-\left\{q_{2}(z_{c}) - 4[q_{1}^{2}(z_{c}) + q_{2}^{2}(z_{c})](z-z_{c})\right\}
\frac{y^{2}}{R_{1}^{4}(z,z_{c})}. 
\label{lp122}     
\end{eqnarray} 
Equation (\ref{lp120_a}) is Eq. (\ref{lp120}) of 
subsection \ref{simu_reshaping}.

\section{The solution of the unperturbed linear propagation equation 
with a Gaussian initial condition}
\label{appendD} 

In Section \ref{simu}, we extensively used the solution of the 
unperturbed linear propagation equation with a Gaussian initial 
condition as an example. We therefore present here a brief summary 
of the different forms of this solution.

We consider the unperturbed linear propagation equation 
\begin{eqnarray}&&
\!\!\!\!\!\!\!
i\partial_z\psi + \partial_{x}^{2}\psi
+ \partial_{y}^{2}\psi=0 
\label{appD_1}
\end{eqnarray}         
with the separable Gaussian initial condition 
\begin{eqnarray}&&
\psi(x,y,0)=A\exp \left[-\frac{(x-x_{0})^{2}}{2W^{(x)2}_{0}}
-\frac{(y-y_{0})^{2}}{2W^{(y)2}_{0}} + i\alpha_{0}\right]. 
\label{appD_2}
\end{eqnarray}  
The solution of Eq. (\ref{appD_1}) with the initial condition (\ref{appD_2}) 
can be written as: 
\begin{eqnarray}&&
\psi(x,y,z)=Ag^{(x)}(\tilde x,z)
g^{(y)}(\tilde y,z)\exp(i\alpha_{0}),
\label{appD_3}
\end{eqnarray}    
where $\tilde x = x - x_{0}$, $\tilde y = y - y_{0}$, 
\begin{eqnarray}&&
\!\!\!\!\!\!\!\!\!\!\!\!\!\!
g^{(x)}(\tilde x,z)= 
\frac{W_{0}^{(x)}}{(W_{0}^{(x)4} + 4z^{2})^{1/4}}
\exp\left[-\frac{W_{0}^{(x)2} \tilde x^{2}}{2(W_{0}^{(x)4} + 4z^{2})}
+i\chi_{0}^{(x)}(\tilde x,z)\right], 
\label{appD_4}
\end{eqnarray}
and  
\begin{eqnarray}&&
\!\!\!\!\!\!\!\!\!\!\!\!\!\!
g^{(y)}(\tilde y,z)= 
\frac{W_{0}^{(y)}}{(W_{0}^{(y)4} + 4z^{2})^{1/4}}
\exp\left[-\frac{W_{0}^{(y)2} \tilde y^{2}}{2(W_{0}^{(y)4} + 4z^{2})}
+i\chi_{0}^{(y)}(\tilde y,z)\right].
\label{appD_5}
\end{eqnarray} 
The phase factors $\chi_{0}^{(x)}(\tilde x,z)$ and $\chi_{0}^{(y)}(\tilde y,z)$ 
in Eqs. (\ref{appD_4}) and (\ref{appD_5}) are given by: 
\begin{eqnarray} &&
\chi_{0}^{(x)}(\tilde x,z)= 
-\frac{1}{2}\arctan\left(\frac{2z}{W_{0}^{(x)2}}\right) 
+\frac{\tilde x^{2}z}{W_{0}^{(x)4} + 4z^{2}}, 
\label{appD_6}
\end{eqnarray}       
and 
\begin{eqnarray} &&
\chi_{0}^{(y)}(\tilde y,z)= 
-\frac{1}{2}\arctan\left(\frac{2z}{W_{0}^{(y)2}}\right) 
+\frac{\tilde y^{2}z}{W_{0}^{(y)4} + 4z^{2}}. 
\label{appD_7}
\end{eqnarray}

One can also write the solution (\ref{appD_3}) in the form: 
\begin{eqnarray}&&
\psi(x,y,z)=A\Psi(x,y,z)\exp[i\chi_{0}(\tilde x,\tilde y,z)],
\label{appD_8} 
\end{eqnarray}      
where 
\begin{eqnarray}&&
\Psi(x,y,z)=G^{(x)}(\tilde x,z)G^{(y)}(\tilde y,z),
\label{appD_12} 
\end{eqnarray}      
\begin{eqnarray}&&
\!\!\!\!\!\!\!\!\!\!\!\!\!\!
G^{(x)}(\tilde x,z)= 
\frac{W_{0}^{(x)}}{(W_{0}^{(x)4} + 4z^{2})^{1/4}}
\exp\left[-\frac{W_{0}^{(x)2} \tilde x^{2}}{2(W_{0}^{(x)4} + 4z^{2})}\right], 
\label{appD_9}
\end{eqnarray}
\begin{eqnarray}&&
\!\!\!\!\!\!\!\!\!\!\!\!\!\!
G^{(y)}(\tilde y,z)= 
\frac{W_{0}^{(y)}}{(W_{0}^{(y)4} + 4z^{2})^{1/4}}
\exp\left[-\frac{W_{0}^{(y)2} \tilde y^{2}}{2(W_{0}^{(y)4} + 4z^{2})}\right], 
\label{appD_10}
\end{eqnarray} 
and 
\begin{equation}
\chi_{0}(\tilde x,\tilde y,z)= 
\chi_{0}^{(x)}(\tilde x,z) + \chi_{0}^{(y)}(\tilde y,z) + \alpha_{0}. 
\label{appD_11}
\end{equation}       
We also note that the solution of Eq. (\ref{appD_1}) with the term 
$id_{11}\partial_{x}\psi$ on its left hand side and with the initial 
condition (\ref{appD_2}) is given by Eqs. (\ref{appD_3})-(\ref{appD_7}) 
[or by Eqs. (\ref{appD_8})-(\ref{appD_11})]  
with $\tilde x = x - x_{0} - d_{11}z$, and $\tilde y = y - y_{0}$.

\section{Invariance of $\Delta A_{1}^{(c)}$ under rotations}
\label{appendE}                                            

In this Appendix, we show that the change in the coordinate system, 
in which we rotate the $x'$ and $y'$ axes by an angle $\Delta\theta$, 
such that in the new coordinate system the relative velocity vector 
between the beam centers lies on the $x$ axis, 
does not change the value of $\Delta A_{1}^{(c)}$. 
That is, the value of the collision-induced amplitude shift is 
invariant under this rotation transformation. 
This calculation provides the justification for choosing the 
direction of the relative velocity vector between the beam centers 
along the direction of the $x$ axis in sections \ref{lp_2D} 
and \ref{simu}.

We consider the fast two-beam collision problem in the $(x',y',z)$ 
coordinate system, in which the relative velocity vector 
$\mathbf{d_{1}'}=(d_{11}', d_{12}')$ does not 
lie on the $x'$ or $y'$ axes. We assume that 
$d_{1}'=|\mathbf{d_{1}'}| \gg 1$. The perturbed linear 
propagation model in the $(x',y',z)$ coordinate system is 
\begin{eqnarray}&&
\!\!\!\!\!\!\!
i\partial_z\psi'_{1} + \partial_{x'}^{2}\psi'_{1}
+ \partial_{y'}^{2}\psi'_{1}=
-i\epsilon_{3}|\psi'_{1}|^2\psi'_{1}
-2i\epsilon_{3}|\psi'_{2}|^2\psi'_{1},
\nonumber \\&&
\!\!\!\!\!\!\!
i\partial_z\psi'_{2} + id_{11}'\partial_{x'}\psi' _{2} 
+id_{12}'\partial_{y'}\psi' _{2}
+\partial_{x'}^2\psi'_{2} + \partial_{y'}^{2}\psi'_{2} = 
\nonumber \\&&
-i\epsilon_{3}|\psi'_{2}|^2\psi'_{2}
-2i\epsilon_{3}|\psi'_{1}|^2\psi'_{2},   
\label{appE_1}
\end{eqnarray}                                  
where $\psi'_{j}(x',y',z)$ is the electric field of the $j$th 
beam in this coordinate system. The initial condition is: 
\begin{eqnarray} &&
\psi'_{j}(x',y',0)=A_{j}(0)h'_{j}(x',y')\exp(i\alpha_{j0}),  
\label{appE_2}
\end{eqnarray} 
where $h'_{j}(x',y')$ is real-valued.

We assume that the solution to the unperturbed propagation equation 
\begin{eqnarray}&&
\!\!\!\!\!\!\!
i\partial_z\psi'_{1} + \partial_{x'}^{2}\psi'_{1}
+ \partial_{y'}^{2}\psi'_{1}=0
\label{appE_3}
\end{eqnarray} 
does not contain any fast dependence on $z$. 
In addition, we assume that the only fast dependence on 
$z$ in the solution to the unperturbed propagation equation 
\begin{eqnarray}&&
\!\!\!\!\!\!\! 
i\partial_z\psi'_{2} + id_{11}'\partial_{x'}\psi' _{2} 
+id_{12}'\partial_{y'}\psi' _{2}
+\partial_{x'}^2\psi'_{2} + \partial_{y'}^{2}\psi'_{2} = 0 
\label{appE_4}
\end{eqnarray}                      
is contained in factors of the form $x'-x'_{20}-d_{11}'z$ 
and $y'-y'_{20}-d_{12}'z$, where $(x'_{20},y'_{20})$ is the 
initial location of beam 2 in the $x'y'$ plane. Under these 
assumptions, we can use the perturbation method of subsection 
\ref{lp_general_IC} to show that within the leading order of the 
method, the equation for $\Phi_{1}'$ in the collision interval is 
\begin{eqnarray} &&
\partial_{z}\Phi'_{1}=
-2\epsilon_{3}\Psi_{20}^{\prime 2}\Psi'_{10}.
\label{appE_5}
\end{eqnarray}  
In addition, in a similar manner to the calculation in subsection 
\ref{lp_general_IC}, we can show that $\Delta\Phi'_{1}(x',y',z_{c})$ 
can be approximated by: 
\begin{eqnarray} &&
\!\!\!\!\!\!\!
\Delta\Phi'_{1}(x',y',z_{c})\!=\!-
2\epsilon_{3} A_{1}(z_{c}^{-}) A_{2}^{2}(z_{c}^{-})
\tilde\Psi'_{10}(x',y',z_{c})
\nonumber \\&&
\!\!\! \times
\int_{-\infty}^{\infty} \!\!\!\!\! dz' \, 
\bar\Psi_{20}^{\prime 2}(x'-x'_{20}-d_{11}'z',y'-y'_{20}-d_{12}'z',z_{c}).
\label{appE_6}
\end{eqnarray}
It follows that the collision-induced amplitude shift in the $(x',y',z)$ 
coordinate system is 
\begin{eqnarray} &&
\!\!\!\!
\Delta A_{1}^{\prime (c)}=-\frac{2\epsilon_{3}
A_{1}(z_{c}^{-}) A_{2}^{2}(z_{c}^{-})}{C'_{p1}}
\nonumber \\&&
\!\!\!\!\!\!\!\!
\times
\!\!\int_{-\infty}^{\infty} \!\!\!\!\! dx' 
\!\int_{-\infty}^{\infty} \!\!\!\!\! dy'
\;\tilde\Psi_{10}^{\prime 2}(x',y',z_{c})
\int_{-\infty}^{\infty} \!\!\!\!\! dz' \, 
\bar\Psi_{20}^{\prime 2}(x'-x'_{20}-d_{11}'z',y'-y'_{20}-d_{12}'z',z_{c}), 
\nonumber \\&&
\label{appE_7} 
\end{eqnarray}    
where 
\begin{eqnarray}&&
C'_{p1}= 
\!\!\int_{-\infty}^{\infty} \!\!\!\!\! dx' 
\!\int_{-\infty}^{\infty} \!\!\!\!\! dy'
\;\tilde\Psi_{10}^{\prime 2}(x',y',0).  
\label{appE_8} 
\end{eqnarray}

We now make a change of variables by going to the 
$(x,y,z)$ coordinate system, in which the relative 
velocity vector $\mathbf{d_{1}'}$ lies on the $x$ axis. 
The $(x,y,z)$ system is found by rotating the $x'$ and $y'$ 
axes by an angle $\Delta\theta = \arctan(d_{12}'/d_{11}')$ 
about the $z$ axis. The equations that define this change of 
variables are: 
\begin{eqnarray}&&
x' = x \cos\Delta\theta - y\sin\Delta\theta,
\nonumber \\&&
y' = x \sin\Delta\theta + y\cos\Delta\theta, 
\label{appE_9} 
\end{eqnarray}  
and 
\begin{equation}
\psi'_{j}(x',y',z) = \psi_{j}(x,y,z).  
\label{appE_10} 
\end{equation}  
It is straightforward to show that the perturbed linear propagation 
model in the $(x,y,z)$ coordinate system is Eq. (\ref{lp1}) and that $d_{11}=d_{1}'$.  
The initial condition for the collision problem is given by 
Eq. (\ref{lp_IC1}), where $h_{j}(x,y) = h'_{j}(x',y')$. 
We observe that the only large parameter in Eq. (\ref{lp1}) 
is $d_{11}$. Thus, the change of variables in 
Eqs. (\ref{appE_9}) and (\ref{appE_10}) does not change the 
properties of the fast dependence on $z$ of the solutions to 
the unperturbed propagation equations (\ref{appE_3}) and (\ref{appE_4}). 
This means that the solution to the unperturbed equation 
\begin{eqnarray}&&
\!\!\!\!\!\!\!
i\partial_z\psi_{1} + \partial_{x}^{2}\psi_{1}
+ \partial_{y}^{2}\psi_{1}=0 
\label{appE_13} 
\end{eqnarray}      
does not contain any fast dependence on $z$. 
In addition, the only fast dependence on $z$ in the 
solution to the equation 
\begin{eqnarray}&&
\!\!\!\!\!\!\!
i\partial_z\psi_{2}+id_{11}\partial_{x}\psi _{2}
+\partial_{x}^2\psi_{2} + \partial_{y}^{2}\psi_{2} = 0  
\label{appE_14}     
\end{eqnarray}      
is contained in factors of the form $x-x_{20}-d_{11}z$. 
If follows that we can employ the perturbation method of 
subsection \ref{lp_general_IC} to calculate the collision-induced 
amplitude shift $\Delta A_{1}^{(c)}$ in the $(x,y,z)$ system, 
and that $\Delta A_{1}^{(c)}$ is given by Eq. (\ref{lp13}), 
where $C_{p1}$ is given by Eq. (\ref{lp12}).

Let us show that the value of the amplitude shift $\Delta A_{1}^{\prime (c)}$ 
in Eq. (\ref{appE_7}) is equal to the value $\Delta A_{1}^{(c)}$ in Eq. (\ref{lp13}). 
For this purpose we note that the determinant of the Jacobian matrix 
for the transformation (\ref{appE_9}) is $|J|=1$.  
Using this together with Eqs. (\ref{lp12}), (\ref{appE_8}), and (\ref{appE_10}), 
we obtain $C'_{p1}=C_{p1}$. In addition, from Eq. (\ref{appE_10}) it follows 
that $\tilde\Psi'_{j0}(x',y',z_{c}) = \tilde\Psi_{j0}(x,y,z_{c})$.        
Furthermore, since the transformation in Eqs. (\ref{appE_9}) and (\ref{appE_10}) 
does not change the properties of the fast dependence on $z$ of the solutions to 
the unperturbed propagation equations, and since $d_{12}=0$, we obtain 
\begin{equation}
\bar\Psi'_{20}(x'-x'_{20}-d_{11}'z,y'-y'_{20}-d_{12}'z,z_{c}) = 
\bar\Psi_{20}(x-x_{20}-d_{11}z,y,z_{c}). 
\label{appE_15} 
\end{equation}                        
Using all the relations mentioned in the current paragraph 
in Eq. (\ref{appE_7}), we arrive at: 
\begin{eqnarray} &&
\!\!\!\!
\Delta A_{1}^{\prime (c)}=-\frac{2\epsilon_{3}
A_{1}(z_{c}^{-}) A_{2}^{2}(z_{c}^{-})}{C_{p1}|d_{11}|}
\nonumber \\&&
\times
\!\!\int_{-\infty}^{\infty} \!\!\!\!\! dx 
\!\int_{-\infty}^{\infty} \!\!\!\!\! dy
\;\tilde\Psi_{10}^{2}(x,y,z_{c})
\!\int_{-\infty}^{\infty} \!\!\!\!\! d\tilde x \;\bar\Psi_{20}^{2}(\tilde x,y,z_{c})
=\Delta A_{1}^{(c)}. 
\label{appE_16} 
\end{eqnarray}    
Thus, the value of $\Delta A_{1}^{(c)}$ is indeed invariant under rotation 
transformations in the $xy$ plane.

  

\begin{thebibliography}{}
\bibitem{spatial_dimension} In the current paper, we define 
the spatial dimension of the problem as the number of the 
coordinates on which the electric field depends for a given 
distance $z$. Therefore, in the three works in Refs.  
\cite{PNH2017B,QMN2018,NHP2020}, for example, the spatial 
dimension was 1, since for each value of $z$, 
the electric field was a function of one coordinate only.
\bibitem{Dimensions1} The dimensionless distance $z$ in Eq. (\ref{lp1})  is 
$z=Z/L_{D}$, where $Z$ is the dimensional distance, 
$L_{D}=x_{0}^{2}/\tilde d_{2}$ is the diffraction length
for a reference beam with width $x_{0}$ in the $x$ direction,  
and $\tilde d_{2}$ is the dimensional diffraction coefficient.  
The dimensionless coordinates $x$ and $y$ are $x=X/x_{0}$ and 
$y=Y/x_{0}$, where $X$ and $Y$ are the dimensional coordinates. 
$\psi_{j}=E_{j}/\sqrt{P_{0}}$, where $E_{j}$ is the 
electric field of the $j$th beam and $P_{0}$ is the peak power.   
The coefficients $d_{11}$ and $\epsilon_{3}$ are 
given by: $d_{11}=\tilde d_{11} x_{0}/\tilde d_{2}$ 
and $\epsilon_{3}=\tilde\rho_{3} P_{0} x_{0}^{2}/\tilde d_{2}$, 
where $\tilde d_{11}$ is the dimensional beam-steering coefficient, 
and $\tilde\rho_{3}$ is the dimensional cubic loss coefficient.
\bibitem{laser_modes} The separable initial condition is of special importance 
since it describes the electric fields that are produced by many 
lasers \cite{Siegman86,Kogelnik66}.
\bibitem{nD} A method similar to the one presented in the current subsection 
can be used for calculating the collision-induced effects in any spatial dimension 
$n$, where $n \ge 2$. In section \ref{lp_3D}, we carry out this calculation for 
spatial dimension 3.
\bibitem{zi_values} The values of $z_{i}$ are determined by:  
$z_{i}=z_{c}+r(z_{f}-z_{c})$, where $r=1/5$, as an example. 
Thus, $z_{i}$ is an intermediate distance that is larger than $z_{c}$, 
at which the collision is not yet completed.
\bibitem{a1zc} Since in subsection \ref{simu_reshaping} the coefficient $\tilde a_{1}$ 
depends on $A_{1}(0)$ and $A_{2}(0)$ [instead of on $A_{1}(z_{c}^{-})$ and 
$A_{2}(z_{c}^{-})$], we replace the notation $\tilde a_{1}(z_{c}^{-})$ by 
$\tilde a_{1}$ in this subsection.
\bibitem{Delta_I_1} Since the fractional intensity reduction factor obtained in 
the simulation $\Delta I_{1}^{(r)(num)}(x,y,z)$ shows weak dependence 
on $x$, we use two different methods to obtain $\Delta I_{1}^{(r)}(y,z)$ 
from the simulation result. In the first method, 
we calculate $\Delta I_{1}^{(r)}(y,z)$  by averaging 
$\Delta I_{1}^{(r)(num)}(x,y,z)$ over the $x$-interval $[-2,2]$, 
and in the second method, we use the value of $\Delta I_{1}^{(r)(num)}(0,y,z)$.
\bibitem{Dimensions2} The dimensionless distance in Eq. (\ref{lp140}) is 
$z=Z/(2L_{D})$, where $Z$ is the dimensional distance, 
$L_{D}=\tau_{0}^{2}/|\tilde\beta_{2}|$ is the dispersion length, 
$\tau_{0}$ is the temporal width of a reference pulsed-beam, 
and $\tilde\beta_{2}$ is the dimensional second-order dispersion coefficient. 
The dimensionless coordinates $x$ and $y$ are $x=X/x_{0}$ and $y=Y/x_{0}$, 
where $X$ and $Y$ are the dimensional coordinates, and $x_{0}$ is  
the width of a reference pulsed-beam along the $x$ axis. 
$\psi_{j}=E_{j}/\sqrt{P_{0}}$, where $E_{j}$ is the 
electric field of the $j$th beam and $P_{0}$ is the peak power. 
The coefficients $d_{11}$ and $d_{2}$ are given by: 
$d_{11}=2\tilde\beta_{1}\tau_{0}/|\tilde\beta_{2}|$ and 
$d_{2}=2\tilde d_{2} \tau_{0}^{2}/(|\tilde \beta_{2}|x_{0}^{2})$, 
where $\tilde\beta_{1}$ is the dimensional first-order dispersion coefficient, 
and $\tilde d_{2}$ is the dimensional diffraction coefficient. 
In addition, the coefficient $\epsilon_{3}$ is given by: 
$\epsilon_{3}=2P_{0}\tau_{0}^{2}\tilde\rho_{3}/|\tilde\beta_{2}|$, 
where $\tilde\rho_{3}$ is the dimensional cubic loss coefficient.
\bibitem{sign_beta_2} In writing Eq. (\ref{lp140}) we assume (without
loss of generality) $\mbox{sgn}(\tilde \beta_{2}) = -1$. The case  
$\mbox{sgn}(\tilde \beta_{2}) = 1$ can be treated in exactly the 
same manner.
\bibitem{cross_sections} We choose to present the pulsed-beam shapes 
using cross-sections, since this enables a clear presentation of the 
dynamics of the main bodies of the pulsed-beams. Indeed, in a conventional 
contour plot of the pulsed-beams shapes in dimension 3, the main bodies are 
typically obscured by the outer parts of the pulsed-beams (i.e., by the 
pulsed-beams tails).
\end{thebibliography}
\end{document}